\documentclass[11pt]{article}
\usepackage{graphicx}
\usepackage{amssymb}
\usepackage{epstopdf}
\usepackage{hyperref}
\usepackage{amsmath, amssymb, amsfonts}
\usepackage{color}
\usepackage{hyperref}

\def\ie{{\rm i.e.,\/}\ }
\def\etc{{\rm etc.\/}\ }
\def\cf{{\rm cf.\/}\ }

\newcommand{\nco}{\newcommand}
\nco{\one}{\ensuremath{\,\,\mathrm{l}\!\!\!1}} 
\nco{\NN}{\mathbb{N}}
\nco{\ZZ}{\mathbb{Z}}
\nco{\QQ}{\mathbb{Q}}
\nco{\RR}{\mathbb{R}}
\nco{\CC}{\mathbb{C}}
\nco{\HH}{\mathbb{H}}
\nco{\OO}{\mathbb{O}}
\nco{\red}{\color{red}}

\nco{\redend}{\normalcolor}
\nco{\colorend}{\normalcolor}

\nco{\rnc}{\renewcommand}
\rnc{\title}[1]{{\Large\bf\mbox{}\\\medskip#1\bigskip\medskip\\}}
\rnc{\author}[1]{{\large #1\smallskip\\}}
\nco{\address}[1]{{\em #1\medskip\\}}
\begin{document}
\begin{titlepage}
\vspace*{\fill}

\begin{center}
\title{The history of the universe is an elliptic curve}
\medskip
\author{Robert Coquereaux} 
\address{Centre de Physique Th\'eorique (CPT),\\ 
Aix Marseille Universit\'e, Universit\'e de Toulon, CNRS, CPT, UMR 7332, 13288 Marseille, France}
\bigskip\medskip

\begin{abstract}
\noindent  {
Friedmann-Lema\^{i}tre equations with contributions coming from matter, curvature, cosmological constant, and radiation, when written in terms of conformal time $u$ rather than in terms of cosmic time $t$, can be solved explicitly in terms of standard Weierstrass elliptic functions. The spatial scale factor, the temperature, the densities, the Hubble function, and almost all quantities of cosmological interest (with the exception of $t$ itself) are elliptic functions of $u$, in particular they are bi-periodic with respect to a lattice of the complex plane, when one takes $u$ complex. After recalling the basics of the theory, we use these explicit expressions, as well as the experimental constraints on the present values of density parameters (we choose for the curvature density a small value  in agreement with experimental bounds) to display the evolution of the main cosmological quantities for one real period $2 \omega_r$ of conformal time (the cosmic time $t$ ``never ends'' but it goes to infinity for a finite value $u_f <  2 \omega_r$ of $u$).
A given history of the universe, specified by the measured values of present-day densities, is  associated with a lattice in the complex plane, or with an elliptic curve, and therefore with two Weierstrass invariants $g_2, g_3$. 
Using the same experimental data we calculate the values of these invariants, as well as the associated modular parameter and the corresponding Klein $j$-invariant.
If one takes the flat case $k=0$, the lattice is only defined up to homotheties, and if one, moreover, neglects the radiation contribution, the $j$-invariant vanishes and the corresponding modular parameter $\tau$ can be chosen in one corner  of the standard fundamental domain of the modular group (equihanharmonic case: $\tau = exp(2 i \pi/3)$). Several exact -- \ie non-numerical -- results  of independent interest are  obtained in that case.}
\end{abstract}
\end{center}
\vspace*{20mm}

\vspace*{\fill }
%%{Version of \today } 19/05/2014
\end{titlepage}
%%%%%%%%

%%%%%%%%%%%%%%%%%%%%%%%%%%%%%%%%%%

\section{Introduction}

Friedmann-Lema\^{i}tre equation is studied in many places (articles, books, encyclopedias, etc.). The subject is hardly new and it is known that the general solution to this differential equation involves elliptic integrals \cite{Lemaitre}, but few people tried to obtain explicit expressions.
To our knowledge, the first article where such an analysis is performed in a detailed manner, keeping all the contributions coming from matter, curvature, cosmological constant, and radiation, is \cite{CoqueGros}. 
Explicit formulae for all range of values of the reduced cosmological constant, and assuming\footnote{This assumption was then made by a majority of people, but at the same time the cosmological constant was not fashioned and was almost always assumed to be zero in the cosmology community.}  $k=+1$  (the so-called ``closed case''), are given in the same reference.
This work was quickly followed by  \cite{Dabrowski}, where the same analysis was performed for the flat and open cases.
Fifteen years later or so (see \cite{cosmolCERNpaper} and the lectures \cite{cbpfCosmologycourse}), the same analysis was reconsidered, in the light of experiments showing that the cosmological constant was probably not zero, after all$\ldots$
A little bit more than thirty years after the first paper we find it useful to return to this problem since the range of cosmological parameters specifying our universe history has been made more precise thank's to recent experiments, and since the elliptic functions (essentially Weierstrass elliptic functions) that allow one to give explicit expressions for the quantities of interest have been made available, with a very good precision, in most computer packages. About this last point it seems that many people still prefer to perform simplifying assumptions or use techniques of numerical integration to describe the evolution of quantities of cosmological interest in terms of cosmic time. We think that using exact expressions is highly preferable, not only for conceptual reasons, but also for practical reasons: some features of the solutions (presence of local extrema, inflection points, behavior near the singularities, etc.) are almost obvious if one uses exact expressions, because of the well-known properties of the involved functions, but may be sometimes difficult to detect numerically. If one is interested in the evolution of the solutions in terms of conformal time, \ie "before" the Big Bang, of "after" the end of cosmic time (infinity),  the use of such expressions is, of course, required.

As it is clear from the very definition of elliptic functions, cosmological solutions of Friedmann equations are periodic in conformal time, they are even bi-periodic if the latter is allowed to take complex values.
This is by no means in contradiction with the fact that many solutions (in particular the one that seems to be dictated by experiments) describe a never-ending universe starting with a Big Bang, since the ``never-ending'' qualifier refers to the cosmic time variable, not to the conformal time variable.
Rather than trying to study the evolution of the spatial scale factor $a(t)$ as a function of cosmic time $t$, the starting point of the method of resolution is to notice that it is much better to give a parametric representation $(a(u), t(u))$ of $a(t)$, because $a(u)$, like most functions of cosmological interest (not the function $t(u)$), turns out to be an elliptic function of $u$, the conformal time.
 Actually, the resolution of Friedmann equations for the quantity $1/a$, the inverse of the scale factor, is almost immediate; this quantity is essentially equal to the temperature $\widetilde T$ of the Cosmic Microwave Blackbody (CMB) radiation
 (warning: in our paper the variable $T$ will denote another quantity that differs from $\widetilde T$ by a scale factor).
 
 The present article is supposed to be self-contained and does not require from the reader any familiarity with Friedmann equations. We nevertheless still refer to the article \cite{CoqueGros} for a general study of their analytic solutions, for all possible values of parameters, but here we shall  restrict our attention to those solutions that are compatible with the recent experiments.
 In particular the evolution of density parameters and of other quantities of cosmological interest, that are displayed in sec.~\ref{sec:evolutionofallquantities} as functions of conformal time, are determined by using the  experimental results for the present-day values of density parameters and Hubble constant.
 Section 2 is certainly standard, but it serves the purpose of specifying our notations. At the beginning of section 3, we remind the reader that the main features of the evolution of the temperature, as a function of conformal time, can be simply obtained by studying the classical motion of point in a potential. Then we give explicit analytical solutions,  with or without radiation (the latter case is of course simpler but hides interesting physical phenomena that show up in a neighborhood of $u=0$). In the same section we define the Weierstrass invariants of the universe. In section 4 we use the values of present-day densities coming from experiment (in particular from the Planck collaboration) to calculate and display the behavior of most quantities of cosmological interest in terms of conformal time. This is done by assuming a small (positive) value for the present value of the curvature density, compatible with the experimental bounds. This section ends with a table giving the values of various quantities at several important dates in the (conformal) history of the universe -- the fact, for instance, that the curvature density function has an extremum located in our past is a phenomenon that is often overlooked.
Special features of the case $k=0$ (flat case) are discussed in section 5. In that particular situation, the experimental data only allows one to determine up to scale the lattice in the complex plane associated with the universe history.
 In other words, the corresponding elliptic curve is only known up to isomorphism, but several exact results (values of periods, value of conformal time when $t\rightarrow \infty$, \etc) can nevertheless also be obtained in that case.
 Section 6 contains miscellaneous comments. Several properties of elliptic functions in relation with lattices, tori, elliptic curves, and modular considerations, are given in the appendix.

As it was already mentioned, the general method of resolution of Friedmann equation in terms of elliptic functions was discussed in \cite{CoqueGros}, but a good part of the discussion relating cosmology to modular considerations is only described in the present paper. In this respect, let us summarize some of the properties that will be discussed (see also the abstract):
to each universe history determined by the measurement of cosmological parameters one can associate an elliptic curve, or, equivalently, a lattice in the complex plane, or a torus with a complex structure.
If $k=\pm 1$, the measurement of all the densities (summing up to $1$) specifies in particular the Weierstrass invariants $g_2$ and $g_3$.
The period parallelogram can be chosen as a rhombus, symmetric with respect to the real axis.
Conversely, the history of the universe, described by the evolution of the temperature as a function of conformal time is fully specified by the Weierstrass invariants, together with today's date (the value of the conformal time ``now''),  and a dimensionful quantity defining the centimeter, for instance the Hubble constant.
If $k=0$, the lattice is only determined up to a complex homothety and the elliptic curve only up to isomorphism (the invariants $g_2, g_3$ are obtained up to scale).
If one furthermore neglects the effect of radiation, the curve is equianharmonic, which corresponds to a modular parameter equal to $exp(2 i \pi/3)$ sitting in the corner of the boundary of the standard fundamental domain for the modular group.
In all cases one can determine the value of the Klein $j$-invariant.

\section{Friedmann-Lema\^itre equations and elliptic functions}

\subsection{Natural variables for the Friedmann-Lema\^itre equations}

It is usual to model space-time as a four-dimensional smooth manifold endowed with a pseudo-riemannian structure specified by a metric.
Assuming that our space-time neighborhood is, in first approximation, homogeneous and isotropic, this metric is described, in local coordinates $(t, \chi, \theta, \phi)$, \ie  in the domain of a chart, by the line element $ds^{2}$.
There are three possibilities ($k=\pm1$ or $0$):
\[ ds^{2} = -dt^2 + a(t)^2 \, [d\chi^2 +  s(\chi)^2 \; (d\theta^2 + \sin^2\theta \, d\phi^2)] \]
with $s(\chi) = \sin(\chi)$ if $k=+1$, $s(\chi) = \sinh(\chi)$ if $k=-1$, and $s(\chi) = \chi$ if $k=0$. 

In conventional cosmology one assumes that the average local energy described by a rank-two tensor $T$,  the energy-momentum tensor, splits into three parts ($G$ denotes the Newton constant): 
 \begin{itemize} 
 \item  Vacuum contribution $\rho_{vac} =\frac{\Lambda}{ 8 \pi G}$ 
 	where ${\Lambda}$ is the cosmological constant, 
 \item  Radiation contribution $\rho_{rad}$ such that $a^{4}(t) 
 	\rho_{rad}(t) = const. = \frac{3}{ 8 \pi G} C_{r}$, 
 \item Averaged matter contribution  $\rho_{mat}$ such that $a^{3}(t) 
 	\rho_{mat}(t) = const. = \frac{3}{ 8 \pi G}C_{m}$ , 
 \end{itemize} 

Einstein's equations read $E_{\mu \nu} = 8 \pi G \, T_{\mu \nu}$ where $E$ is the Einstein tensor. 
The cosmological term described by $\Lambda$ is sometimes written explicitly on the lhs of those equations but here it is included as a part of $T$ itself, on the rhs.
The quantities $\Lambda$, $C_r$ and $C_m$ are constant, as well as $\rho_{vac}$, but $\rho_{rad}$ and $\rho_{mat}$ are, a priori, time-dependent quantities.
Assuming a Levi-Civita connection (no torsion), the Einstein tensor is determined from the metric alone, and Einstein's equations imply that 
the evolution of  $a(t)$ is governed by the Friedmann  equation:
\begin{equation}
\frac{1}{ a^{2}}{(\frac{da}{ dt})}^{2} = \frac{C_{r}}{  a^{4}} + \frac{C_{m}}{
a^{3}} - \frac{k}{ a^{2}} + \frac{\Lambda}{ 3}\ . 
\end{equation}
Lema\^\i tre \cite{Lemaitre} did, long ago, an analytic study of the solutions of Friedmann  equations, with a cosmological constant; 
his discussion, made in terms of $a$, the scale factor,  and $t$, the cosmic time, involves elliptic integrals.
In order to discuss these equations it is however convenient  \cite{CoqueGros} to use another set of variables, \ie
to introduce a {\sl conformal time} ${u}$, and a dimensionless {\sl reduced temperature} $T({u})$, defined as:
\begin{equation}
d {u} = \frac {d t}{ a}  \qquad \text{and} \qquad  T({u}) = \frac{1}{ \Lambda_{c}^{1/2} a({u})} \quad \text{with} \qquad \sqrt{\Lambda_{c}} = \frac{2}{ 3 C_{m}}
\label{eq:cosmictime_eq} 
\end{equation}
The variable ${u}$ is natural, both geometrically (it gives three-dimensional 
geodesic distances) and analytically, as we shall 
see below, since it is only when we express the cosmological quantities of interest
in terms of  $T({u})$ that these quantities can be written themselves as 
elliptic functions with respect to a particular lattice.

Intuitively,  this change of variables replaces the scale factor $a$, that measures the ``size'' of the spatial universe, by its inverse, a quantity proportional to the temperature of the cosmic microwave radiation (see later).  In a Big Bang cosmology,  $T$ is infinite at the Big Bang and decreases as the universe expands. This change of variables  replaces the cosmic time $t$ by a parameter ${u}$, the conformal time,  that measures geodesic distances in dimensionless units. The previous equations define ${u}$ only up to an additive constant;  in a Big Bang cosmology, it is natural to set ${u}=0$ at the Big Bang, so that ${u}$ gives the dimensionless distance along the trajectory of a photon that would have been emitted at the Big Bang. This is a perfectly natural way of measuring ``time''.  Rather than describing  the dynamics of the universe by a single function $a(t)$ of one variable, we therefore use the parametric equations $(T({u}), t({u}))$. 
The advantage is that the differential equation for $T({u})$ is very simple -- see below --  and can be integrated immediately in terms of standard elliptic functions. 
Very often -- and in particular, as we shall see later,  for the cosmological solutions of physical interest -- the function $t({u})$ approaches a logarithmic singularity when ${u}$ approaches a finite limit ${u}_f$.
In other words, when the cosmic time goes to infinity, the universe expands for ever (in terms of $t$) and cools down ($T \sim 1/a \rightarrow 0$) but the conformal time goes to ${u}_f$.  
So, even if the universe is spatially closed (let us say that it is $S^3$), an observer will never see the back of his head if the cosmology is such that ${u}_f$ is finite and smaller than $2 \pi$, even if this observer waits for an infinite (cosmic) time. 

In this article we use the natural system of units for which $\hbar = c = 1$. All quantities are therefore homogenous with $L^p$, for some integer $p$, where $L$ is a length (the cm, say). In particular 
the quantities $C_r$, $C_m$ and $\Lambda$ have dimensions $L^2, L, L^{-2}$ respectively, $a$ has dimension $L$ and $t$ has dimension $L^{-1}$. The Newton constant $G$ is homogeneous with $L^2$. Notice that $T$ and ${u}$ are {\sl dimensionless}.

With these new variables, Friedmann  equation becomes 
\begin{equation} 
(dT /d{u})^{2} = \alpha T^{4} + \frac{2}{ 3} T^{3} - k T^{2} + \frac{\lambda}{3} 
\label{eq:Friedmann} 
\end{equation} 
where 
\begin{equation} 
\lambda = {\Lambda \over \Lambda_{c}} \quad \text{and} \quad \alpha =  C_{r} \Lambda_{c} 
\label{def:lambdaandalpha}
\end{equation} 
are two constant dimensionless parameters: the {\sl reduced cosmological constant} and the {\sl reduced radiation parameter}.
We shall see later how these parameters can be extracted from the more standard densities used to describe experimental results.

The next section will deal with analytic solutions, but the direct link with the theory of elliptic functions should be already clear from the fact 
that, with our parametrization, the RHS of eq~\ref{eq:Friedmann} is a polynomial of degree four.

\subsection{Hubble function and density ``parameters'' }

As a rule we shall add an upper or lower index $o$ to denote the present-day value of the cosmological quantities. For instance $t_o$ is the age of the universe (using cosmic time), ${u}_0$ its age in terms of conformal time, $a_0$ is the present value of the scale factor, \etc

{\sl The Hubble function} describing the rate of expansion is  defined, as usual, by  $H = \dfrac{d\, Log(a)}{dt}$ and can be written, in  terms of the reduced temperature $T({u})$ as 
\begin{equation} 
H({u}) = - \Lambda_{c}^{1/ 2} \, {dT \over d{u}} 
\label{eq:HubbleDef} 
\end{equation}  
In natural units, most quantities used in cosmology are dimensionless, but the Hubble function is a dimensionful quantity homogenous to an inverse length, so we may consider that the present value of $H_o = H({u}_0)$, or rather of $(H_o)^{-1}$, defines what the centimeter is (today).

Notice that 
\begin{equation}
H^2 = \alpha \, \Lambda_c \, T^4 + 2/3 \,  \Lambda_c \, T^3 - k \, \Lambda_c \, T^2 + \Lambda/3
\label{eq:HubbleEq} 
\end{equation}  

Multiplying this equation by $1 / H^2$, one obtains the famous relation\footnote{$\Omega_r^o$ being non zero, but very small, it is sometimes dropped from equation \ref{eq:CosmicTriangle}, and  because $\Omega_K^o$ is experimentally compatible with $0$, it is also often dropped from many presentations. It is interesting to remember that thirty years ago, it was another parameter (namely $\Omega_\Lambda$) that was often forgotten from many presentations.\\} : 
 
\begin{equation} 
1 = \Omega_r + \Omega_m + \Omega_K + \Omega_\Lambda 
\label{eq:CosmicTriangle} 
\end{equation} 
 with 
\begin{eqnarray} 
\Omega_K = - k T^2 {\Lambda_c \over H^2}, \quad
 \Omega_m = {2\over 3}  T^3 {\Lambda_c \over H^2}, \quad
\Omega_\Lambda = {\lambda \over 3}{\Lambda_c \over H^2}, \quad
 \Omega_r = \alpha T^4 {\Lambda_c \over H^2} 
\label{eq:TheOmegas} 
\end{eqnarray} 

Equivalently,
\begin{eqnarray} 
\Omega_K  = - \frac{k}{a^2 H^2},  \quad
\Omega_m =  \frac{C_m}{ a^3 H^2},  \quad
\Omega_\Lambda = \frac{\Lambda}{3 H^2},  \quad
\Omega_r = \frac{C_r}{a^4 H^2}
\label{eq:TheusualOmegas} 
\end{eqnarray}

We remind the reader that those quantities\footnote{\label{note1} The densities $\Omega_r(u)$, $\Omega_m(u)$, $\Omega_\Lambda(u)$  were respectively called $\alpha_S(\tau)$, $\Omega(\tau)$ and $\lambda_S(\tau)$ in ref. \cite{CoqueGros}.}, although often called ``density parameters'',  are function of time (${u}$ or $t$). 
For this reason it is usual to introduce the notations
 $\Omega_m^o$, $\Omega_r^o$,  $\Omega_K^o$ and $\Omega_\Lambda^o$ to denote their present-day values. 
 Obviously, $\Omega_m$ and $\Omega_r$ are positive but $\Omega_\Lambda$ and $\Omega_K$ could be of both signs. 
 It is quite common to call $\Omega= \Omega_m + \Omega_r + \Omega_\Lambda$ the ``total density parameter''.
 Since $- \Omega_K =  \Omega - 1$ and since, for historical reasons, $\Omega_K$ is negative when $k=+1$ and vice-versa, one sees that $k = +1, 0$ or $-1$, respectively,  if $\Omega >1$, $=1$, or $<1$.

The {\sl constant} parameters $\Lambda$, $k$, $C_m$ and $C_r$ entering the original Friedmann equation are expressed as follows  in terms of the {\sl time dependent} density parameters:
\begin{equation} 
\Lambda = 3 H^2 \Omega_\Lambda, \quad k = - \text{sign}{(\Omega_K)}, \quad C_m = \frac{\Omega_m}{\vert \Omega_K \vert^{3/2} H}, \quad C_r = \frac{\Omega_r}{\vert \Omega_K \vert^{2} H^2}
\label{fromomegatoparameters}
\end{equation} 

In the closed case ($k=+1)$, rather than $C_m$ or $\Lambda_c$, one often uses the mass $M$
\begin{equation} 
M = 2 \pi^2 a^3 \rho_m = \frac{\pi}{2 G \sqrt{\Lambda_c}} = \frac{3 \pi}{4 G} C_m
\label{totalmass}
\end{equation} 
 to parametrize the matter contents of the universe.

If $k=\pm 1$, the {\sl constant} parameters $\alpha$ and $\lambda$,  introduced in equation (\ref{def:lambdaandalpha}) can be expressed as follows in terms of the {\sl time dependent} density parameters:
\begin{equation}
\alpha = \frac{4}{9} \frac{\Omega_r \vert \Omega_K \vert}{\Omega_m^2} \quad \lambda = \frac{27}{4} \frac{\Omega_\Lambda \Omega_m^2}{\vert \Omega_K^3 \vert}
\label{alphaandlambdaasfunctionofomega} 
\end{equation} 

If $k=0$, the measurements (now) of $\Omega_r$, $\Omega_m$ and $\Omega_\Lambda$ give no information on the individual values of $\alpha$ and $\lambda$, but they are nevertheless related --- at all times -- by the relation: 
\begin{equation}
\alpha^3 \lambda = \frac{16}{27} \; \frac {\Omega_r^3 \; \Omega_\Lambda}{\Omega_m^4}
\label{eq:lambdaalphacube} 
\end{equation}

\subsection{Temperature, units and dimensions}
The reason for calling $T$ a dimensionless ``reduced  temperature'' is that 
it is proportional to the temperature $\widetilde T$ of the black body 
radiation. Indeed, $\rho_{rad} = 4 \sigma {\widetilde T}^{4}$ where 
$\sigma$ is the Stefan-Boltzmann constant. Since $\rho_{rad} = {3 
\over 8 \pi G} \alpha \Lambda_{c} T^{4}$, one finds 
\begin{equation} 
\widetilde T^{4} = {3 \over 8 \pi G} {\alpha \Lambda_{c} \over 4 \sigma} T^{4} 
\label{eq:UsualTemperature} 
\end{equation} 
With $\hbar = c = 1$, the value of the Stefan-Boltzmann constant is  $\sigma = {\pi^2 k_B^4 / 60} = 59.8 \; cm^{-4}$, where $k_B$ is the Boltzmann constant.
Notice that $\widetilde T$ is in degrees Kelvin, hence dimensionless, but the Boltzmann constant $k_B$ (as $k_B \, \widetilde T$) is an energy, hence homogenous to an inverse length, as it should.
We call  $T_o$ and $\tilde T_o$ the {\sl present} values of the time-dependent quantities $T$ and $\tilde T$. 

As already recalled, in this system of units all quantities are either dimensionless or have a dimension which is some power of a length ($[cm]$).
We gather the relevant information as follows:
$$ t \sim a \sim C_{m} \sim [cm], \qquad G \sim C_{r} \sim [cm^2], \qquad H \sim k_{B} \sim {energy} \sim [cm^{-1}]$$
$$ \Lambda \sim \Lambda_{c} \sim [cm^{-2}], \qquad  \rho_{vac} \sim \rho_{rad} \sim \rho_{m} \sim \sigma \sim [cm^{-4}] $$

Finally we list the dimensionless quantities:
$$\Omega_{m}\sim \Omega_{\Lambda}\sim  \Omega_{K}\sim \Omega_{r}\sim q 
\sim T \sim {\tilde T} \sim u \sim \alpha \sim \lambda  \sim [cm^{0} = 1] $$

 Remember that $k,\alpha, \lambda, \Lambda, \Lambda_c$ are constant parameters. 

\section{Solutions}
\subsection{Qualitative Behavior of Solutions} 
\label{sec:qualitativebehavior}
\begin{figure} 
\includegraphics[scale=0.60]{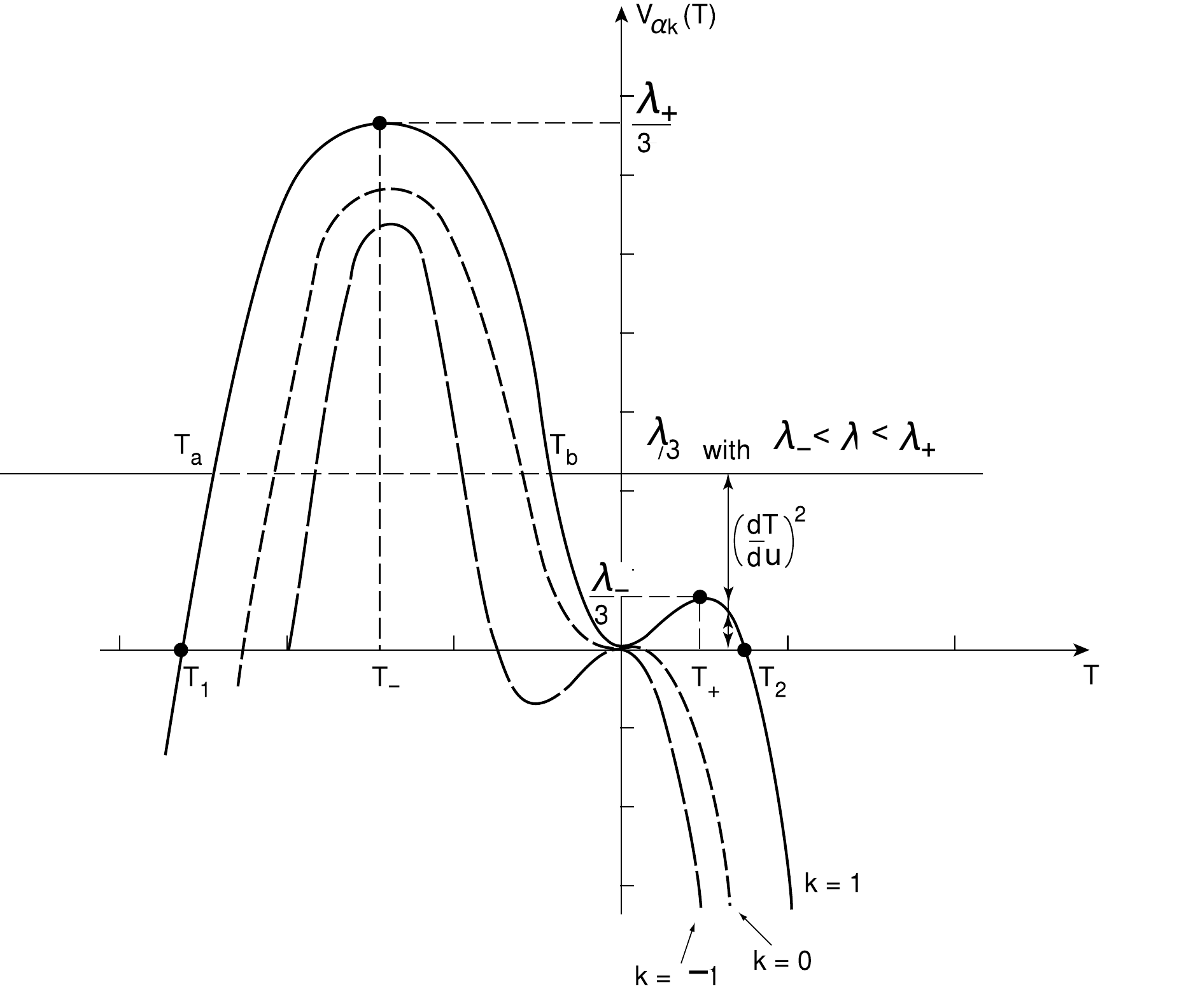} 
\caption{Potential for the associated mechanical system. Case $\alpha  \neq 0$} 
\label{potential} 
\end{figure}

Friedmann equation in $\{u,T\}$ variables (eq~\ref{eq:Friedmann}) can also be written 
\begin{equation} 
({dT \over du})^{2} + V_{\alpha, k}(T) = {\lambda \over 3} 
\quad
\text{with}
\quad
 V_{\alpha, k}(T) = -\alpha T^{4} -{2\over 3} T^{3} + k T^{2} 
\end{equation} 
This is the equation of a one-dimensional mechanical system with 
``coordinate'' T, potential $V_{\alpha, k}(T)$ (displayed in Fig 
\ref{potential}) and total energy $\lambda /3$. 
We shall also set  
\begin{equation} 
Q_{\alpha, k}(T) = -V_{\alpha, k}(T)+\lambda/3
\label{eq:Qpolynomial}
\end{equation} 
When no confusion arises we write $V= V_{\alpha, k}$,  $Q= Q_{\alpha, k}$.
Formally, $du = dT / \sqrt{Q}$.

The kinetic energy 
being non negative, the associated mechanical system describes a 
horizontal line in the $(V(T),T)$ plane but never penetrates under 
the curve $ V_{\alpha, k}(T)$ --this would correspond to $u$ 
imaginary. The length of the vertical line segment between a point 
belonging to the curve and  a point with same value of $T$ but 
belonging to the horizontal line $\lambda /3$ (on which the associated mechanical system moves) is a measure of $({dT / du})^{2}$.

\begin{figure} 
\includegraphics[scale=0.75]{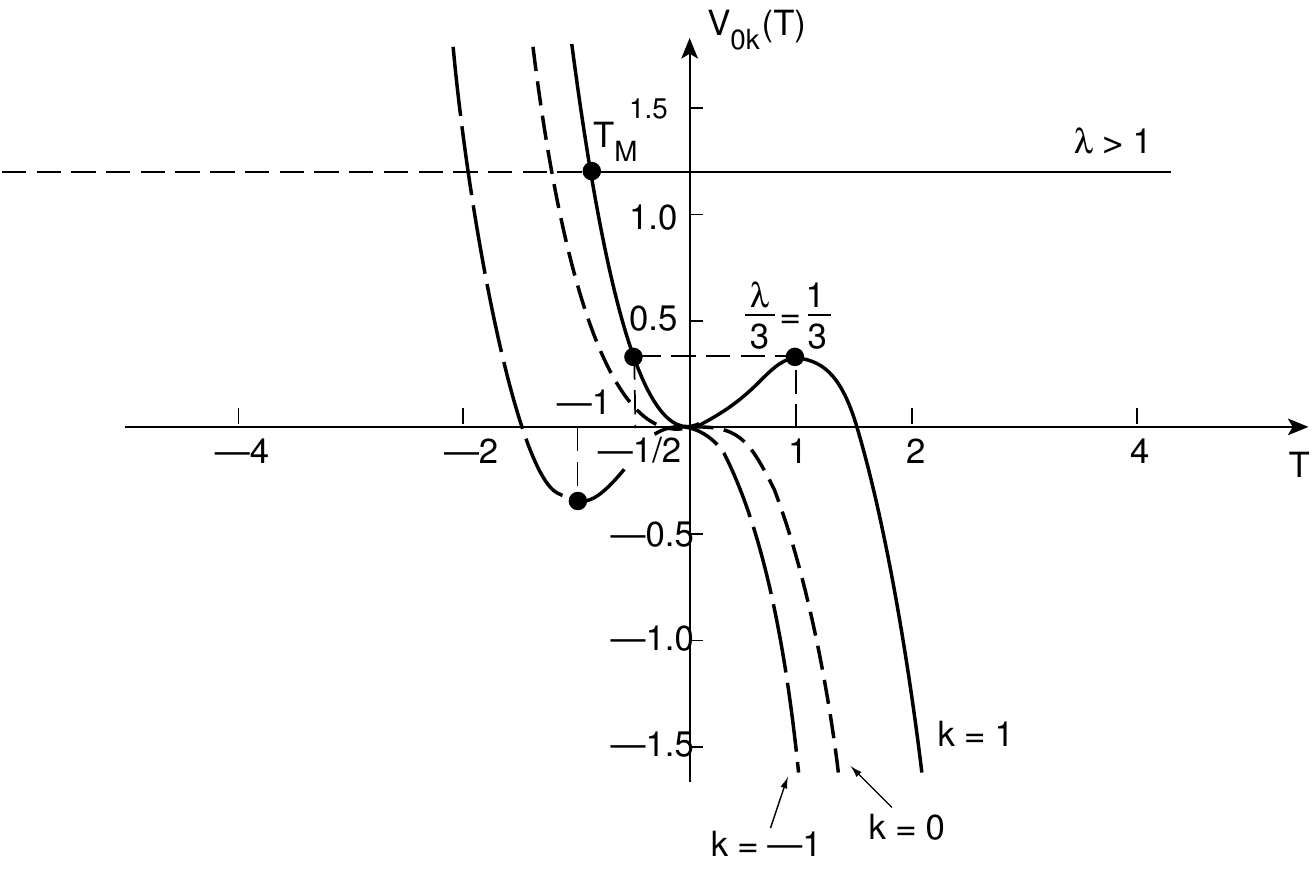} 
\caption{Potential for the associated mechanical system. Case $\alpha =0$} 
\label{potential0} 
\end{figure}

For a given value of $\alpha$, the radiation parameter, and for $k=\pm 
1$, the curve  $V_{\alpha, k}(T)$ 
has typically two bumps (two local maxima). For $k = 1$ (closed 
universe), the right maximum occurs for a positive value of 
$T$ and $V(T)$ whereas, for  $k=-1$ (open case), this maximum is 
shifted to $T=0$ 
and $V(T)=0$. For $k=0$ (flat case), the right  maximum 
disappears and we are left with an inflection point at $T=0, V(T)=0$. 
 If $\alpha = 0$, the curve $V(T)$ 
becomes a cubic (Fig. \ref{potential0}) and the LHS maximum disappears: it moves 
to $- \infty$ as $\alpha$ goes to $0$. 
 
Warning: figure  \ref{potential} gives only the qualitative features 
of the curve $V_{\alpha, k}(T)$. Indeed, for reasonable values of $\alpha$ 
and $\lambda$, \ie values compatible with  experimental 
constraints, the vertical coordinate of the left maximum should be at least 
$1000$ times higher than the vertical coordinate of the right maximum. 
Let us call  $\lambda_{\pm}/3$ the ordinates of the non-zero extrema of the potential $V_{\alpha, k}(T)$ (see fig. \ref{potential}).
For small values of $\alpha$, the values of these extrema (maxima if $k=+1$) are given by 
$\lambda_{-}\simeq k(1-3 \alpha k)$ and $\lambda_{+}\simeq {1\over 
16\alpha^{3}}(1 + 12 k \alpha)$; the former goes to $k$, and the latter to infinity when $\alpha$ goes to $0$.
Actually, if $k=0$, one has exactly $\lambda_{-}=0$ and $\lambda_{+}={1/ (16\alpha^{3})}$.
 
All the recent experimental results (see sec.~\ref{closedcase}) seem to agree on the fact that ${\Lambda}$ is non-zero and positive.
As a consequence, $\lambda$ is also positive. This will be assumed in the rest of this paper. 
At the end of sec.~\ref{closedcase} we will show, using the experimental values (or bounds) on the densities, that 
if $k=1$,   then $\lambda_{-} < \lambda < \lambda_{+}$, and if $k=0$ or $-1$, then  $0< \lambda < \lambda_{+}$.
Actually, if $k=0$ the right maximum of the potential disappears: ($\lambda_{-} =0$) and if $k=-1$,  the (right) maximum becomes a (left) minimum and moves to the non-physical region

 In other words, the associated mechanical system  moves along an horizontal line like the one 
displayed in fig.~\ref{potential}, with an ordinate  located between the two extrema of the potential $V_{\alpha, k}(T)$.
Typically, a given universe starts from the right of the picture (an infinite $T$ corresponding to the Big Bang) and moves to the left until it reaches the vertical axis ($T=0$). This  takes 
place in a finite conformal time $u_{f}$ but corresponds to a cosmic time $t$ going to infinity, so that, in the 
universe in which we live, ``History'' stops there. However, the 
solution can be continued for $T<0$ (a negative radius $a$) until the 
system bumps against $V(T)$ and goes back to right infinity; the 
system then jumps to left infinity, follows the same horizontal line (but now from left to right) till it 
bumps against $V(T)$ again, and comes back. This round trip of the 
associated mechanical system is done in a (conformal) time 
$2\omega_{r}$ -- a period of the corresponding elliptic function. 
If $k = 1$ there is an inflection point for $T(u)$ coming from the existence of a positive right 
maximum for the curve $V(T)$: the expansion speeds up anyway, but there is a time $u_{I}$ 
for which the rate of expansion vanishes. If $k=0$ the right maximum disappears and there is no inflection point for $T(u)$.
In the case $k=-1$ the inflection point of $T(u)$ moves to the non-physical region.

Since the radius $a$ is proportional to $1/T$, the discussion in 
terms of $a$ is of 
course different: the system 
starts with $a=0$ (Big Bang), and 
expands forever; as $\lambda >1$ the expansion speeds up in all three cases 
$k=\pm 1$, $k=0$.  
Let us stress the fact that only the first part of the motion of the associated mechanical system (from right-infinity to the intersection with the vertical axis) is physically relevant 
for the history of the universe in which we live. 
 
\begin{figure} 
 \begin{center}
\includegraphics*[scale=0.55]{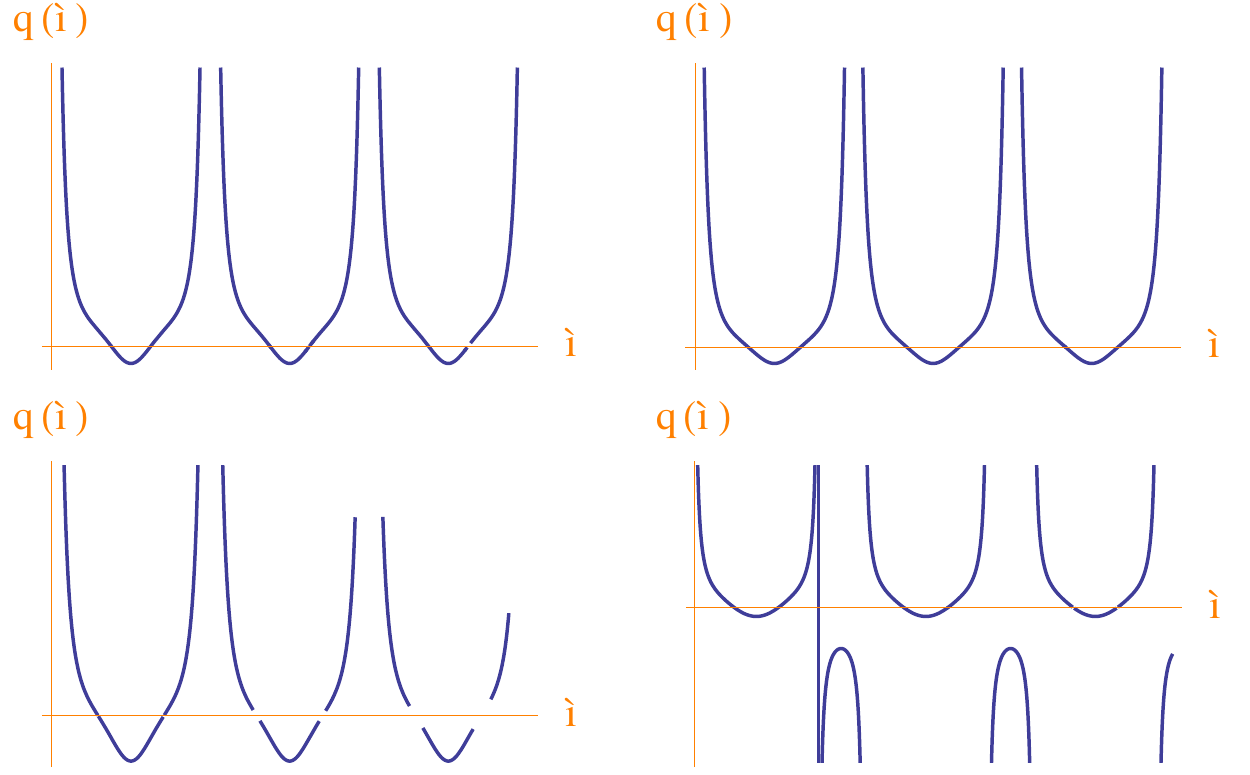} 
 \hspace{1.0cm}
\includegraphics*[scale=0.55]{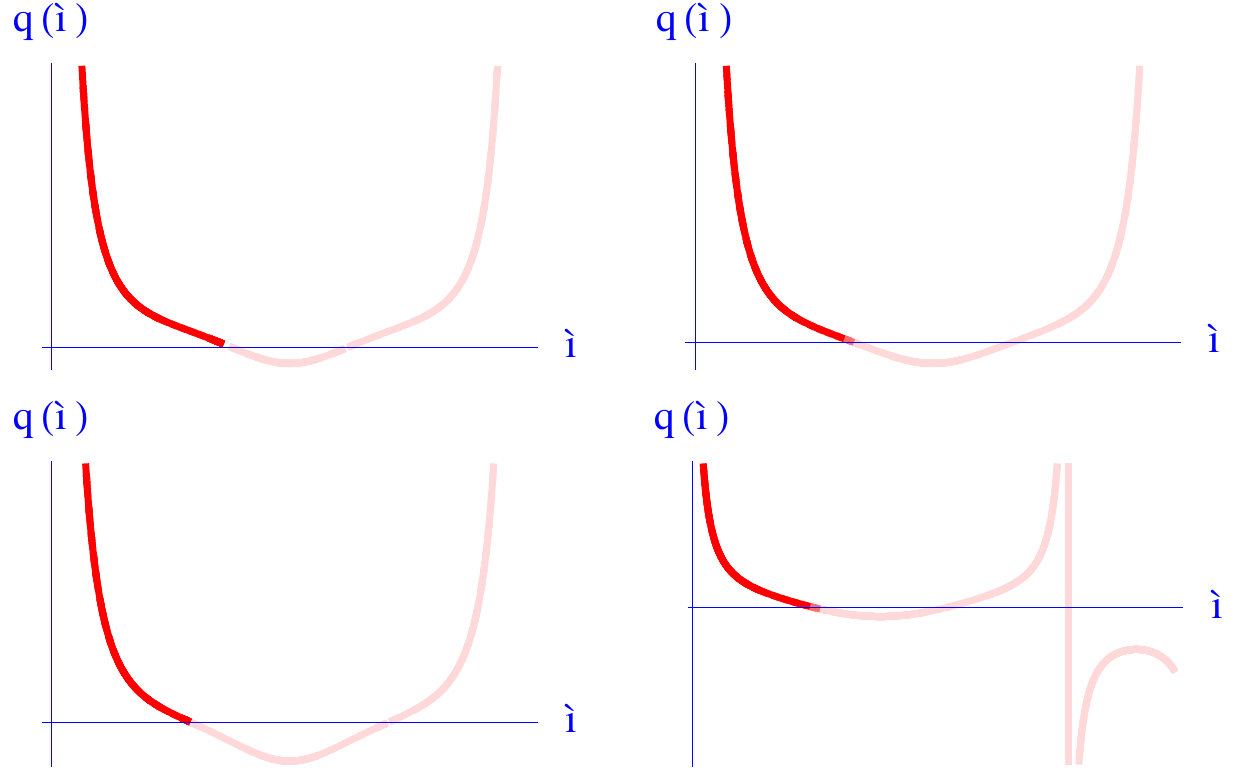} 
\end{center}
\caption{Qualitative behavior of $T(u)$. {\small Typical plots for cases (i) $(k=1, \alpha =0)$, assuming $ \lambda \geq 1$, (ii) 
$(k=0, \alpha =0)$, assuming $ \lambda \geq 0$, (iii) $(k=-1, \alpha =0)$, and case (iv) $\alpha \neq 0$, assuming $\lambda_{-} \leq \lambda \leq \lambda_{+}$. 
The function $T(u)$ is doubly periodic for complex $u$, and periodic, with period $2\omega_r$, if the conformal time $u$ is real (graphs on the left). 
Only the decreasing positive branch of $T(u)$,  in the first period, is ``physical'' \ie describes the history of our universe (graphs on the right).  
In the first three cases the inflection point occurs respectively for $T>0$, $T=0$, and $T<0$. 
We have the same type of behavior when $\alpha \neq 0$ but the curve also develops a negative connected branch (see details in fig.~\ref{qualitativeTbis}).}}
\label{qualitativeT} 
\end{figure}

 The above elementary discussion shows immediately that, as a function of the conformal time $u$, the behavior of the reduced temperature $T(u)$ is described by fig~\ref{qualitativeT}.
 Notice that, if $\alpha=0$ the connected negative branch(es) of this curve disappears.  
 Call $u_f$ the first positive zero of $T$, \ie the first positive value of $u$ for which $T(u_f)=0$. 
 Only that part of the curve corresponding to the interval $0 \leq u \leq u_f$ matters, for the universe in which we live. 
 %%%%%%
 %%%%%%% New qualitative 
  \begin{figure} 
 \begin{center}
\includegraphics*[scale=0.50]{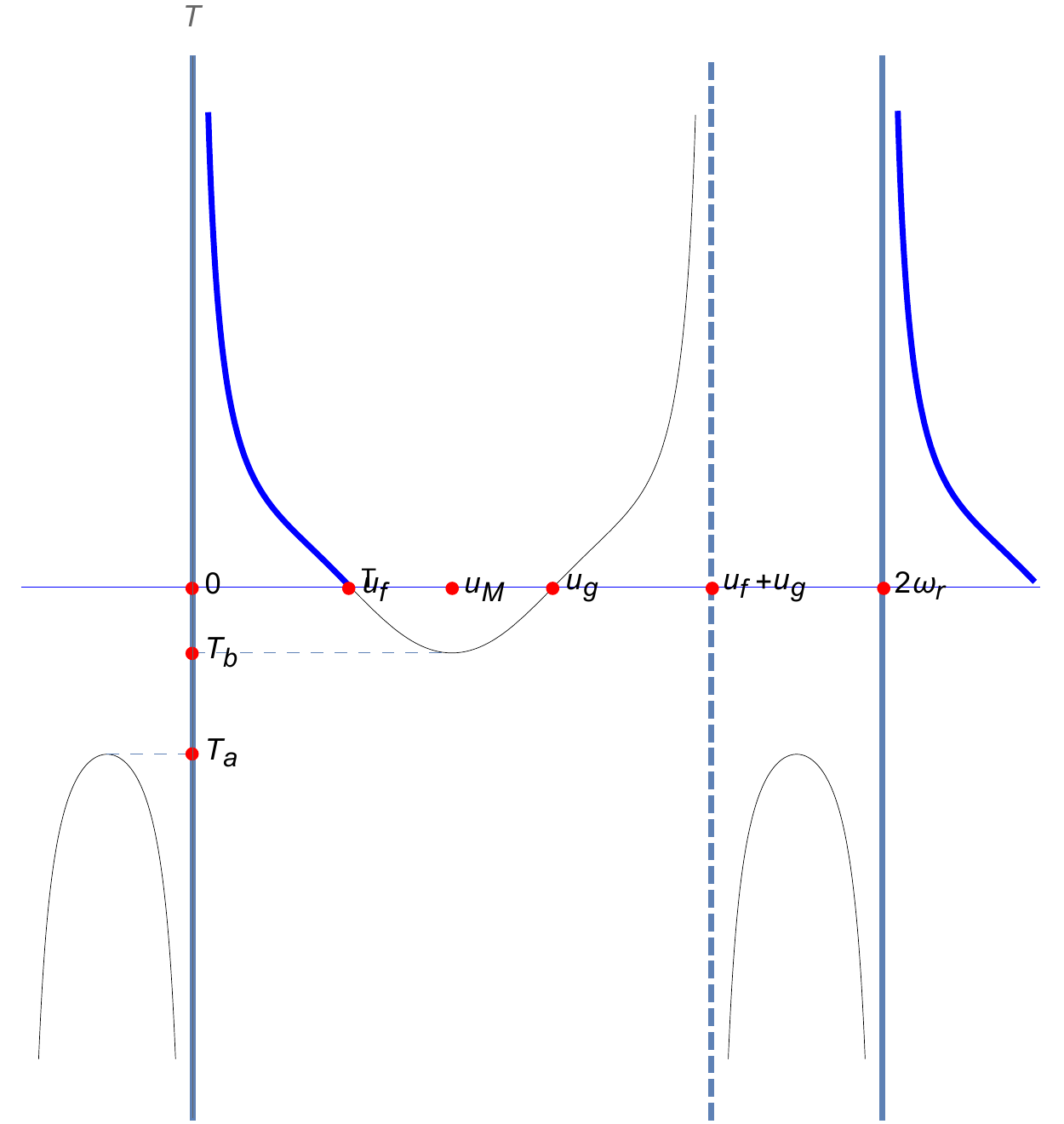} 
\end{center}
\caption{Qualitative behavior of $T(u)$, radiation being taken into account. {\small Special values of the temperature (extrema $T_a, T_b$) and of the conformal time  ($u_f, u_g, u_f+u_g, 2 \omega_r$) are displayed on the figure.
 The width of the connected negative branch (between the two asymptotes) is $2 \delta_c = 2 \omega_r - (u_f+u_g)$, it goes to $0$ when the effect of radiation is neglected, at the same time the connected negative part of the curve is shifted towards $- \infty$ \ie the smallest (negative) extremum $T_a$ becomes very large. Experimentally, the contribution of radiation is small, which means that  $2 \delta_c$ is very small compared to  $2 \omega_r$. The physical branch of the history of the universe is the interval $[0, u_f]$.}}
\label{qualitativeTbis} 
\end{figure}

\subsection{Notations for special values of the conformal time} 
\label{sec:notations}

{\sl The end of  time.} Call $u_f  = \int_0^\infty \, dT/\sqrt{Q(T)}$.  The conformal time reaches $u_f$ when the reduced temperature $T$ becomes equal to $0$. This corresponds to an infinite value of cosmic time. The value $u_f$ is given by the previous integral but one can also determine numerically its value by looking at the first zero of the function $T(u)$ along the real axis.
We call $u_f$ ``the end of time'', but this is only a shortening for ``the value of the conformal time at the end of cosmic time'', since, after that, the conformal time goes on....

{\sl After the end of time: the negative history.} 
If $\alpha = 0$ (no radiation), and for given $\lambda$,  call  $T_M$  the solution of the equation $V(T)=\lambda$. If $\alpha \neq 0$, call $T_a \leq T_b$ the two solutions of the same equation (see graphs \ref{potential} and \ref{potential0}).
In the first case, call $\delta_M = \int_{T_M}^0 \, dT/\sqrt{Q(T)}$.
In the second case, call $\delta_M = \int_{T_b}^0 \, dT/\sqrt{Q(T)}$.
As a function of conformal time, the reduced temperature $T$ has a (negative) minimum $T_M$ (if $\alpha=0$), or $T_b$ (if $\alpha \neq 0$) which is obtained for $u_M = u_f + \delta_M$. This occurs after the end of time.  After the end of time $u_f$, the reduced temperature $T(u)$ becomes negative and continues to decrease until it reaches $T_M$ (or $T_b$) then it starts to increase and vanishes again when $u=u_g$, with $u_g= u_f + 2 \delta_M$.

{\sl  After the negative history: a new beginning.}  When $u > u_g$, the temperature is again positive, it starts from zero and increases to infinity (big crunch), this happens when $u= u_f+u_g= 2 u_M$. 
Although it could be a perfectly allowed region, it is not the branch of the universe in which we live. If the chosen cosmological model uses $\alpha=0$, the value $2 u_M = 2\int_{T_M}^\infty \, dT/\sqrt{Q(T)}$ coincides with  $2 \omega_r$,  the real period. Otherwise (\ie if we take $\alpha \neq 0$),  the curve $T(u)$ develops a negative connected branch for $u$ between $2 u_M$ and the period $2\omega_r$: {\sl the radiation conformal shift\footnote{in ref \cite{CoqueGros} the number $2\delta_c$ was called $u_c$}} $2\delta_c = 2 \omega_r - 2 u_M$, that vanishes if $\alpha=0$, measures the size of this unphysical branch; as we see from graph \ref{potential}, it can be obtained numerically as the integral $\delta_c =  \int_{- \infty}^{T_a} \, dT/\sqrt{Q(T)}$.

{\sl The complex history.} Under the potential, \ie for $T_a < T < T_b$ (using $\alpha \neq 0$), or for $- \infty < T<T_M$ (using $\alpha =0$), the polynomial $Q(T)$ is negative and the conformal time is purely imaginary.  As a function of the complex argument $u$, the function $T(u)$ is doubly periodic in the complex plane.
We have already determined one (real) half-period $\omega_r$, by integrating $dT/\sqrt{Q(T)}$ along the branches $\int_\infty^{T_a} + \int_{T_b}^\infty$, integration between $i T_a$ and $iT_b$ gives us the (complex) half-period $ i \omega_i$. So, in terms of real quantities, we have: $\omega_i = \int_{T_a}^{T_{b}} \, dT/\sqrt{\vert Q(T) \vert}$. If one uses $\alpha = 0$, we just replace $T_a$ by $-\infty$.

Let us summarize the previous discussion (see also fig.~\ref{qualitativeTbis}), by the following :
\begin{equation}
\begin{split}
u_f &= \int_0^\infty  \frac{dT}{\sqrt{Q(T)}}, \; \delta_M =  \int_{T_b}^0 \frac{dT}{\sqrt{Q(T)}},\;  \delta_c =  \int_{-\infty}^{T_a} \frac{dT}{\sqrt{Q(T)}}, \; \omega_i =  \int_{T_a}^{T_b} \frac{dT}{\sqrt{\vert Q(T) \vert} } \\
u_M &= u_f + \delta_M,\;  u_g = u_f+2 \delta_M, \; 2 \omega_r = u_f+ u_g + 2\delta_c = 2 u_M + 2\delta_c. \\
 \text{If one uses}  &\; \alpha =0, \; \text{one sets} \; T_a \rightarrow \infty, \; T_b = T_M, \; \text{and therefore} \; \delta_c = 0, \; \omega_r = u_M.
\end{split}
\label{eq:ufugdelc}
\end{equation}

{\sl The conformal time at the inflection point} $u_I$.  From the associated mechanical system, see figs~\ref{potential} or \ref{potential0},  we see that when the reduced cosmological constant $\lambda$ is such that $\lambda_{-} < \lambda < \lambda_{+}$ (case $\alpha \neq 0$), or is such that $1 < \lambda$ (case $\alpha =0$), the curve $T(u)$ has an inflection point between $0$ and $u_f$, and another one (after the negative history) when $u>u_g$.
This occurs both for models with $k=1$ and $k=-1$, but this point is ``physical'',  \ie it occurs for a positive temperature, only if $k=1$.   Its position $u_I$ can be determined by solving $T^{\prime \prime}(u)=0$.  Rather than solving this equation, one may notice that, assuming $k=+1$, the value $T(u_I)$ is also equal to $T_+$, the positive real value of $T$ for which the potential $V(T)$ is maximal ($T_+$ is $\alpha$-dependent, but $T_+=1$ if $\alpha =  0$), therefore, $u_I$ can be found by solving the equation $T(u_I)=T_+$, if $\alpha \neq 0$, or the equation $T(u_I)=1$, if $\alpha = 0$.
Notice that if $k=0$ the second derivative of $T$ vanishes when $T=0$, \ie when $u=u_f$ (end of time), but does not change sign: there is no inflection point in the physical region.

{\sl The conformal time now} $u_0$. The above special values of the conformal time depend on the parameters used to construct the cosmological model (a universe history). In contradistinction, the conformal time $u_0$ just specifies the date ``today'', \ie where we are on the curve describing a universe history.  This value $u_0$ is defined by the equation $T(u_0) = T_o$ where $T_o$ is in principle taken from experiment (see equation~\ref{eq:UsualTemperature}).\\
Assuming $\alpha = 0$ to simplify, we have obviously $0 < u_f < u_M = \omega_r < u_g < u_g + u_f = 2 \omega_r$. If $T_I > 0$, \ie if $k=1$,  it is of course physically interesting to know if $u_I$ is smaller or larger than $u_0$.
Using the experimental results on the present day values of the densities we will see that $u_I$ is very close to $u_f$ and still in our future,  in other words $u_o < u_I < u_f$.
If $k=0$ this inflection point disappears.

{\sl Periods and Weierstrass invariants}. We saw how to determine two periods $2 \omega_r$ and $2 \omega_i$ by integration (see above). As it is discussed in sections \ref{sec:norad} and \ref{sec:withrad}, as well as in the Appendix, another way to encode the lattice with respect to which the cosmological quantities are elliptic (in particular, doubly periodic) is to introduce the Weierstrass invariants $g_2, g_3$. We shall come back to it, but let us only mention now that 
several mathematical computer packages offer facilities to convert Weierstrass invariants $(g_2,g_3)$ into complex half-periods $(\omega_1, \omega_2)$, and vice versa, but the reader should remember that, given the two invariants, the choice of a base in the corresponding lattice of the complex plane is not unique (cf. Appendix), and the result returned for the periods by a computer package will depend upon the package and may even depend upon the program version!
Calling $\Delta = g_2^3 - 27 g_3^2 = 2^{-4} 3^{-3} \lambda (\lambda - \lambda_+)(\lambda - \lambda_-)$ the modular determinant, and assuming $\Delta < 0$ as it seems to be experimentally the case (see sec.~\ref{closedcase}), we can choose\footnote{At the moment, with Mathematica version 10, \cite{Mathematica} the two half-periods returned by the command $\{\omega_1,\omega_2\} = WeierstrassHalfPeriods[\{g_2,g_3\}]$ are such that $\omega_r + i \omega_i = 2 \omega_2$ and $-i \omega_i = \omega_1$.}
for elementary periodicity cell the rhombus $\{0, \omega_r - i \omega_i,   \omega_r + i \omega_i,  2\omega_r\}$, 
where the numbers $\omega_r$ and $\omega_i$ are real; in particular its diagonal $\{0, 2\omega_r \}$ is real, and as a function of the real variable $u$, the number $2 \omega_r$ is the fundamental period.

\subsection{Analytical solutions}
\subsubsection{The case without radiation ($\alpha=0)$}
\label{sec:norad}
For a solution starting with a Big Bang the contribution of the term $\alpha T^{4}$ to the Friedmann equation is only important when the universe is very young.
As the discussion is in any case easier in that case, we shall first assume $\alpha = 0$. The RHS of eq \ref{eq:Friedmann} is then a cubic polynomial, and setting $T= 6 y + k/2 $ brings this equation to the form 
\begin{equation} 
(dy /d{u})^{2} = {4} y^{3} - g_2 y -  g_3
\label{eq:FriedmannNoradiation} 
\end{equation} 
where the parameters $g_2$ and $g_3$, called the Weierstrass invariants, are given by
\begin{equation} 
 g_2=k^2/12 \quad \text{and}Ê\quad g_3=\dfrac{1}{6^3}(k - 2 \lambda)
 \label{eq: g2g3Noradiation} 
 \end{equation} 
The analytic solution, for the reduced temperature $T$, is immediate: as a function of $u$, it is the (scaled and shifted) Weierstrass elliptic function\footnote{See sec.~\ref{ellipticthings}.}   $y={\mathcal P}$ corresponding to the invariants $g_2$ and $g_3$.
 \begin{equation} 
T({u}) = 6\, {\mathcal P}(u; g_2, g_3)+\dfrac{k}{2}
\label{eq: TasPfunction} 
\end{equation} 
Since ${\mathcal P}(u)  = \dfrac{1}{u^2} + O(u^2)$ as $u \rightarrow 0$, \ie near the Big Bang, we see that $T({u}) \sim \dfrac{6}{u^2} + k/2 \quad (u \rightarrow 0)$ but, on physical grounds, one should not use this approximation of $T(u)$ for small $u$ since one  cannot neglect the effect of radiation (the $\alpha$ term) near the Big Bang.\\
The end of time\footnote{Physically, the cosmic time $t(u)$ develops a logarithmic singularity when $u \rightarrow u_f$, this is indeed ``the end of time''.} occurs at the first zero, $u_f$, of $T(u)$. Calling $u_g$ the next zero, we have $u_f + u_g = 2 \omega_r $ where the RHS is the period along the real axis.

Since $T$ vanishes for $u$ equal to $u_f$ and $u_g=2\omega_r-u_f$, the elliptic function $1/T$ has a pole for these two values, with a behavior dictated by the dominant term $\sqrt{\lambda/3}$ in the RHS of eq~\ref{eq:Friedmann}, so that we can write it immediately\footnote{See the last paragraph of the appendix.} in terms of the Weierstrass function $\zeta(u)$. We obtain in this way an alternative explicit expression for the reduced temperature $T(u)$ (remember that $T(u) = {1}/{(a(u) \sqrt{\Lambda_c})})$: 
 \begin{equation} 
\dfrac{1}{T({u})} = \sqrt{\dfrac{3}{\lambda}} \quad \{ \zeta(u-u_g) - \zeta(u-u_f) +  \zeta(u_g)- \zeta(u_f) \}
\label{eq: TasZetafunction} 
\end{equation} 

The dedicated reader can check that the following provides still another expression for the same function $T(u)$.
 In this formula, $\sigma(u)=\sigma(u;g_2,g_3)$ denotes the Weierstrass $\sigma$ function of the same lattice.
 Rather than using the information that we have about the the poles and principal parts of $1/T$, it uses the fact that we know its poles and zeroes.
 \begin{equation} 
{T({u})} =  T_M \times \dfrac{\sigma^2(\dfrac{u_f+u_g}{2})}{\sigma^2(\dfrac{u_f-u_g}{2})} \times \dfrac{\sigma(u-u_f) \, \sigma(u-u_g)} {\sigma(u) \, \sigma(u-u_f-u_g)} 
\label{eq: TasSigmafunction} 
\end{equation} 
Here $T_M = T_b = T(\omega_r)$, see figs~\ref{potential},\ref{potential0}, is the minimum value of $T(u)$;  it can be determined  from $dT/du = 0$, \ie as $T_M = 6 \, e_2 + \frac{1}{2}$ where $e_2$ is the real cubic root of the polynomial $4 y^3 - g_2 y - g_3$. This is numerically easy to find but it can also be given in closed form: $e_2=a_{+} + a_{-}$, where $a_\pm = -\frac{1}{2} (-g_3 \pm \sqrt{- \Delta/27})^{1/3}$, with $\Delta=g_2^3 - 27 g_3^2$.

Eq~\ref{eq: TasPfunction}  looks simpler than eqs~\ref{eq: TasZetafunction}  or \ref{eq: TasSigmafunction} but the latter is numerically as convenient as the first and, as we shall see, generalize straightforwardly to the $\alpha \neq 0$ case.

\paragraph{The cosmic time.}

Integrating eq~\ref{eq:cosmictime_eq} we obtain
\begin{equation} 
\label{eq:cosmictime_norad} 
\sqrt{\frac{\Lambda}{3}} \; t(u) = ln[\frac{\sigma(u_f+u)}{\sigma(u_f-u)}] - 2 u \, \zeta(u_f)
\end{equation} 
This function is not elliptic, it has a logarithmic singularity at $u=u_f$. In a neighborhood of $u_f$, to the left,  we have $\sqrt{\frac{\Lambda}{3}} \; t(u) \sim -  ln(u_f-u)$.
Between $u_f$ and $u_g$ the cosmic time is not real. 

\subsubsection{The case with radiation ($\alpha \neq 0)$}
\label{sec:withrad}
From the analytic point of view, what happens is that the two poles of $T(u)$ become distinct, each one of them is therefore of first order, since $T(u)$ is still elliptic of order $2$.
The RHS of eq \ref{eq:Friedmann} is now a quartic polynomial  $Q(T) = -V(T)+\lambda/3$, but such a polynomial can be brought to a cubic form by a simple change of variables. 
Let $T_{j}$ be any one of the (possibly complex) roots of the equation $Q(T)=0$, then the fractional linear transformation 
\begin{equation} 
\label{eq:transformationT} 
 y = {Q'(T_{j}) \over 4} {1 \over T-T_{j}} + 
{Q''(T_{j}) \over 24} 
\end{equation} brings eq (\ref{eq:Friedmann}) to the same form as before (eq \ref{eq:FriedmannNoradiation}),  but the invariants $g_{2}$ and $g_{3}$ are now given by
\begin{equation} 
g_{2} = {k^{2} \over 12} + {\alpha \lambda \over 3}  \quad \text{and}Ê\quad g_{3} = {1 \over 6^{3}} (k - 2 \lambda) - {{\alpha \lambda k} \over 18 } 
 \label{eq: g2g3withradiation} 
\end{equation} 
Again, we have $y={\mathcal P}(u; g_2, g_3)$ and one can solve eq \ref{eq:transformationT} for $T$ in terms of $y$ to get an explicit expression for the function $T(u)$.
However, in the present case, it is much simpler to express $T$ in terms of the Weierstrass zeta function $\zeta = \zeta(u; g_2, g_3)$.  The obtained expression is exactly the same as the one (eq~\ref{eq: TasZetafunction}) obtained in the previous section, with the difference that the invariants $g_2$, $g_3$ are now given by eq~\ref{eq: g2g3withradiation},  and that $u_f + u_g \neq 2 \omega_r $ since we have $u_f+u_g+2\delta_c=2\omega_r$ with $2\delta_c = 2 \int_{-\infty}^{T_a} dT/\sqrt{Q(T)}\neq 0$.
With $\alpha \neq 0$, and assuming\footnote{This corresponds to the experimental situation, see our discussion in sec.~\ref{closedcase}.} $\lambda_{-} \leq \lambda \leq \lambda_{+}$, the polynomial $Q(T)$ has two real roots $T_a \leq T_b < 0$ (we had only one, $T_M=T_b$, in the case $\alpha = 0$ and $\lambda > 1$).

\paragraph{The cosmic time.}
Equation (\ref{eq:cosmictime_norad}) is modified as follows:
\begin{equation} 
\label{eq:cosmictime_withrad} 
\sqrt{\frac{\Lambda}{3}} \; t(u) = ln[\frac{\sigma(u_f)\sigma(u-u_g)}{\sigma(u_g)\sigma(u-u_f)}] +  u \,  (\zeta(u_g) -  \zeta(u_f))
\end{equation} 
Compared to the case $\alpha=0$, modifications in the interval $0 \leq u \leq u_f$ occur for very small $u$.

\section{Experimental constraints and evolution of cosmological quantities with conformal time}
\label{experimentalconstraints}

The experimental values given for $\Omega_m^o$ and $\Omega_\Lambda^o$ in table 2 of \cite{Planck2013} assume  $\Omega_K^o = 0$ (the spatially flat ``base $\Lambda$CDM model''). 
This very special value of the curvature density is certainly compatible with present-day experiments but, using observations of the CMB together with the results coming from detection of gravitational lensing, it seems (formula 67b of \cite{Planck2013}) that one can only constrain $\Omega_K^o$ to percent level precision: $100 \, \Omega_K^o = -1.0 ^{+1.8}_{-1.9}$. This small value found for $\Omega_K^o$ is often described in the literature by sentences like ``the universe is spatially flat''.
From the analytical and geometrical points of view, the flat case $k=0$ (that implies $\Omega_K=0$ at all times since densities have constant sign and $k = - sign( \Omega_K)$) is however very special, see section \ref{flatcase}.
For conceptual ---and philosophical--- reasons, some people may  find cases $k=0$ and $k=-1$  a bit unpleasant,  see in particular the discussion in pages 748, 749 of \cite{MTW}, while some other people, also for conceptual ---and philosophical--- reasons, prefer to take $k=0$.
In general we do not assume spatial flatness, and in the next three subsections we shall use values of $\Omega_m^o$ and $\Omega_\Lambda^o$ that allow for a non-zero curvature density  $\Omega_K^o$ within the experimental bounds. For definiteness, and since the latter is only constrained to be smaller than a few percents (not such a small number), we shall take $\Omega_K^o=-0.01$, so $k=+1$ (closed case).
Subsection (\ref{flatcase}) will be devoted to the case $k=0$.

\subsection{Experimental present values for density parameters, CMB temperature, and Hubble function}
\label{closedcase}

{\sl Newton constant and Stefan-Boltzmann constants}. With $\hbar = c = 1$, the value of the Stefan-Boltzmann constant is  $\sigma = {\pi^2 k_B^4 / 60} = 59.8 \, cm^{-4}$,
where $k_B$ is the Boltzmann constant and the Newton constant (sometimes called ``Planck area'' in those units) is $G = 2.61  \;  10^{-66} cm^2$. 

{\sl CMB temperature}. The present day value of the CMB temperature given by \cite{Planck2013} is  $\widetilde T_o = 2.7255 \; K$.

%{\sl Hubble parameter}. The present value of the Hubble ``constant'' (\ie now), given by  \cite{Planck2013} is $H_o = 100 \, {\sl h} \,  km \, sec^{-1} \, Mpc^{-1}$, with ${\sl h} \simeq 0.673 \pm 0.012 $.
%Its mean experimental value, in natural units, is therefore $H_o=   (7.27  \pm 0.13) \, 10^{-29} \, cm^{-1}$, therefore $(H_o)^{-1}=(13.75 \pm 0.25) \, 10^{27} \, cm$. 

{\sl Hubble parameter}. The present value of the Hubble ``constant'' (\ie now), given by  \cite{Planck2013} is $H_o = 100 \, {\mathit h} \,  km \, sec^{-1} \, Mpc^{-1}$, with ${\mathit h} \simeq 0.688 \pm 0.008 $.
Its mean experimental value, in natural units, is therefore $H_o=   (7.437  \pm 0.13) \, 10^{-29} \, cm^{-1}$. In the following we shall take $h$ (later called $h_o$) equal to $0.688$.

{\sl Radiation density} $\Omega_r$.
From eq.~\ref{eq:UsualTemperature} and using the present day experimental value of the CMB temperature $\widetilde T_o$, one finds $\alpha \,  \Lambda_c \,  T_o^4 = 5.23 \, 10^{-63} \, cm^{-2}$.
From eq.~\ref{eq:TheOmegas} for $\Omega_r$, and using the experimental value for $H_o$, one finds  $\Omega_r^o = (5.45 \pm 0.20) \, 10^{-5}$.  
This value is therefore obtained from the measurements of $\widetilde T_o$ and $H_o$. Taking into account the effect of all relativistic particles (massless neutrino) is not expected to change this value much. The term $\Omega_r^o$,  which is well determined but  very small compared to the other terms, is often dropped from eq~\ref{eq:CosmicTriangle} written $1=\Omega_r^o + \Omega_K^o + \Omega_\Lambda^o+ \Omega_m^o$. It is certainly legitimate to perform this approximation at present times, but of course not at all times and certainly not at the beginning of the expansion.\\

{\sl Curvature density} (\cf discussion at the beginning of this section).  In contradistinction to the latter, the present day curvature density $\Omega_K^o$ is not well determined, but it seems to be also quite small. For this reason it may look legitimate to drop the term $\Omega_K^o$ from equation eq.~\ref{eq:CosmicTriangle}. Doing so is certainly valid at present times, but setting $k=0$ (which implies $\Omega_K =0$ at all times) is a strong and disputable hypothesis on the topology and the dynamics of the mathematical model chosen for our universe. In any case we want to study and display the {\sl evolution} of $\Omega_K$ as a function of $u$. We take  $\Omega_K^o=-0.01$, a value compatible with the experimental bounds. This implies\footnote{The spatial universe is then, topologically, a sphere $S^3 \sim SU(2)$ or  a quotient of the latter by a (discrete) binary polyhedral subgroup of $SU(2)$.} $k=+1$.

{\sl Matter density}. 
According to the recent measurements (see in particular \cite{Planck2013}), $\Omega_m^o$ is about $0.3$.  The value quoted by  \cite{reviewcosmo2014} for CMB $+$ WMAP $+$ BAO is  $\Omega_m^o = 0.293 \pm 0.010$.
For definiteness, we shall take $\Omega_m^o = 0.293$ in the following numerical experiments. One could  for instance assume a contribution of  0.047 from baryons and 0.246 from dark matter but
 the fact that matter density seems to be dominated by its so-called ``dark matter'' component (compared to its baryonic component) is irrelevant for the present analysis since both contribute in the same way to Friedmann equations.

{\sl Vacuum density (often called ``dark energy density'')}.  The experimental constraints --- see those given by \cite{Planck2013}, in particular fig.~25 of \cite{Planck2013} --- are $\Omega_{\Lambda}^o = 0.707 \pm 0.010$.
We shall take $\Omega_{\Lambda}^o = 0.717$. 

From equation \ref{eq:lambdaalphacube} and using the experimental values of $\Omega_r^o$, $\Omega_m^o$ and $\Omega_\Lambda^o$, one finds $\lambda \alpha^3 \simeq 8.20 \; 10^{-12}$.
This value is independent of the hypothesis made about the curvature density (in particular it also holds if $k=0$).

If we assume $k=\pm 1$, the equations \ref{alphaandlambdaasfunctionofomega} make sense and give individual values --- or constraints --- for the parameters $\alpha$ and $\lambda$. The sum of the four densities should be $1$ at all times, but since they enter multiplicatively in the expressions of $\alpha$ and $\lambda$, one  obtains reasonable bounds by keeping $\Omega_K^o$ as a free variable in eqs \ref{alphaandlambdaasfunctionofomega}. Doing so leads to $\alpha \simeq 2.70 \, 10^{-4} \; \vert \Omega_K^o \vert$ and $\lambda \simeq 0.415/ \vert \Omega_K^o \vert^3$. If we assume that the present day value of the curvature density is bounded, in norm, by $10^{-2}$, we find $\alpha < 2.7 \, 10^{-6}$ and $\lambda > 4.1 \; 10^5$. As already written, we shall use  $\vert \Omega_K^o \vert = 10^{-2}$ in the following numerical experiments.

Warning: if we assume $k=0$, so that $\Omega_K$ is strictly zero at all times, we cannot use eqs \ref{alphaandlambdaasfunctionofomega} to provide individual values (or bounds) for $\alpha$ or $\lambda$. The only constraint that we have is on the value of $\lambda \alpha^3$.

%%%%%%
We called $\lambda_{\pm}/3$  the ordinates of the left and right  maxima of the potential $V_{\alpha, k}(T)$ of the associated mechanical system (see fig.~\ref{potential} in sec.~\ref{sec:qualitativebehavior}). \\
If $k=0$, one has $\lambda_{+}=\tfrac{1}{16\, \alpha^3}$ and $\lambda_{-}=0$. The numerical value obtained previously for the product $\lambda \alpha^3$ is clearly much smaller than $\tfrac{1}{16}$.
Therefore $\lambda << \lambda_{+}$. On the other hand, experimentally, ${\Lambda}$ and therefore $\lambda$, are positive. So we have the constraint $\lambda_{-}=0 < \lambda << \lambda_{+}$. \\
If  $k=+1$, from the experimental bounds on  $\Omega_K^o$, we conclude, as before,  that the value of $\alpha$ is itself very small, so that we can use for $\lambda_{\pm}$ the approximations $\lambda_{-}\simeq (1-3 \alpha )$ and 
$\lambda_{+}\simeq {\tfrac{1}{ 16\alpha^{3}}}(1 + 12 \alpha)$ given in \ref{sec:qualitativebehavior}, or even $\lambda_{-}\simeq 1$ and $16 \lambda_{+} \alpha^{3} \simeq 1$, but $\lambda$ is at the same time much larger than $1$ and such that  $\lambda \alpha^3  \simeq 8.20 \; 10^{-12}$. As a result: $\lambda_{-} < \lambda << \lambda_{+}$. \\
The case $k=-1$ can  be discussed similarly, but here $\lambda_{-}$ is negative and corresponds to a minimum of the potential; it is  also associated with an inflection point of the curve $T(u)$ but  this occurs when $T$ is negative, \ie in the unphysical region (after the end of time).

In all cases, the constraint $\lambda_{-} < \lambda < \lambda_{+}$ tells us (see fig.~\ref{potential}) that, as a function of  conformal time $u$,  the temperature will decrease from plus infinity to $0$, for some value $u_f$ (the end of time), it will then become negative until it reaches a minimum (the associated mobile bumps against the potential since  $\lambda \ll \lambda_{+}$), and goes back to (plus) infinity. This branch is followed, since $\alpha \neq 0$, by a round trip of the mobile in the non-physical region (a negative branch of the curve $T(u)$). In the case $k=1$ the curve $T(u)$ has an inflection point in the physical region; this inflection point disappears if $k=0$, and moves to the unphysical region if $k=-1$, but in the three cases, given the experimental constraints on the densities and the corresponding bounds on $\lambda$, the overall features of the temperature, as a function of the conformal time, are the same.

\subsection{Evolution of the cosmological quantities as functions of conformal time}
\label{sec:evolutionofallquantities}

In the cases $k = \pm 1$, from the independent measurements of $H_o$, $\Omega_m^o$, $\Omega_\Lambda^o$ and ${\widetilde T}_o$,  one can determine the evolution of all cosmological quantities in terms of the conformal time $u$.
Here we take $k=+1$.  The various quantities are obtained, in turn, as follows.
Eq.~\ref{eq:UsualTemperature}, using $\widetilde T_o$, give the product $\alpha \Lambda_c T_o^4$.
Then eq.~\ref{eq:TheOmegas}, with $H_o$, gives $\Omega_r^o$.
The densities $\Omega_m^o$ and $\Omega_K^o$, together with eq.~\ref{eq:CosmicTriangle} gives $\Omega_K^o$.
The parameters $\alpha$ and $\lambda$ are then obtained from eq.~\ref{alphaandlambdaasfunctionofomega}.
$\Lambda$ comes from $\Omega_\Lambda^o$ and $H_o$, using eq.~\ref{eq:TheOmegas}.
This gives $\Lambda_c$ since $\lambda$ is known, and also $T_o$, since $\alpha \Lambda_c T_o^4$ is known.
Eqs.~\ref{eq: g2g3withradiation} give the two invariants $g_2$ and $g_3$.
The Weierstrass functions are then perfectly determined.
The different cosmological quantities of interest, as functions of $u$, are finally given by eqs.~\ref{eq:HubbleDef}, \ref{eq: TasZetafunction}, together with  \ref{eq:Qpolynomial}, \ref{eq:ufugdelc}
 (or \ref{eq: TasPfunction} if radiation is negelected), and \ref{eq:TheOmegas}.

\subsubsection*{Temperature  $\widetilde{T}$}

\begin{figure} 
 \begin{center}
\includegraphics*[scale=0.35]{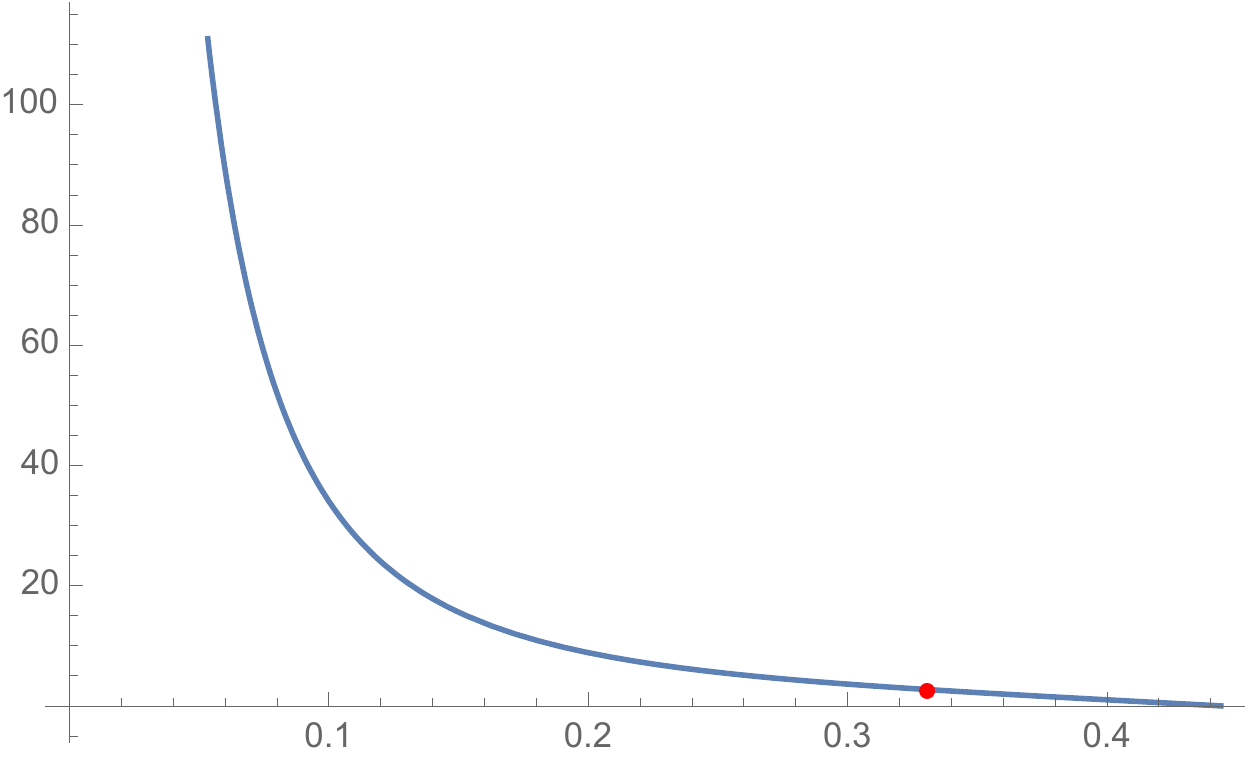} 
 \hspace{1.0cm}
\includegraphics*[scale=0.35]{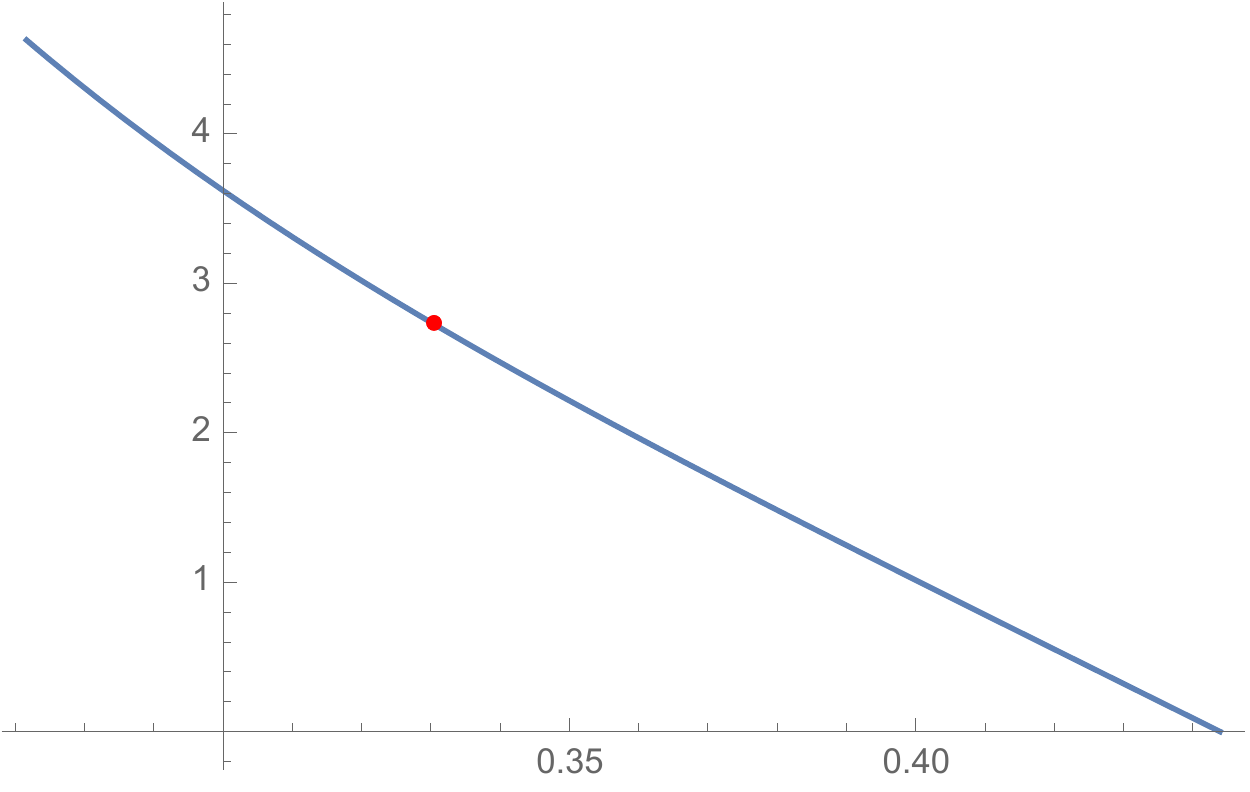} 
 \hspace{1.0cm}
\includegraphics*[scale=0.35]{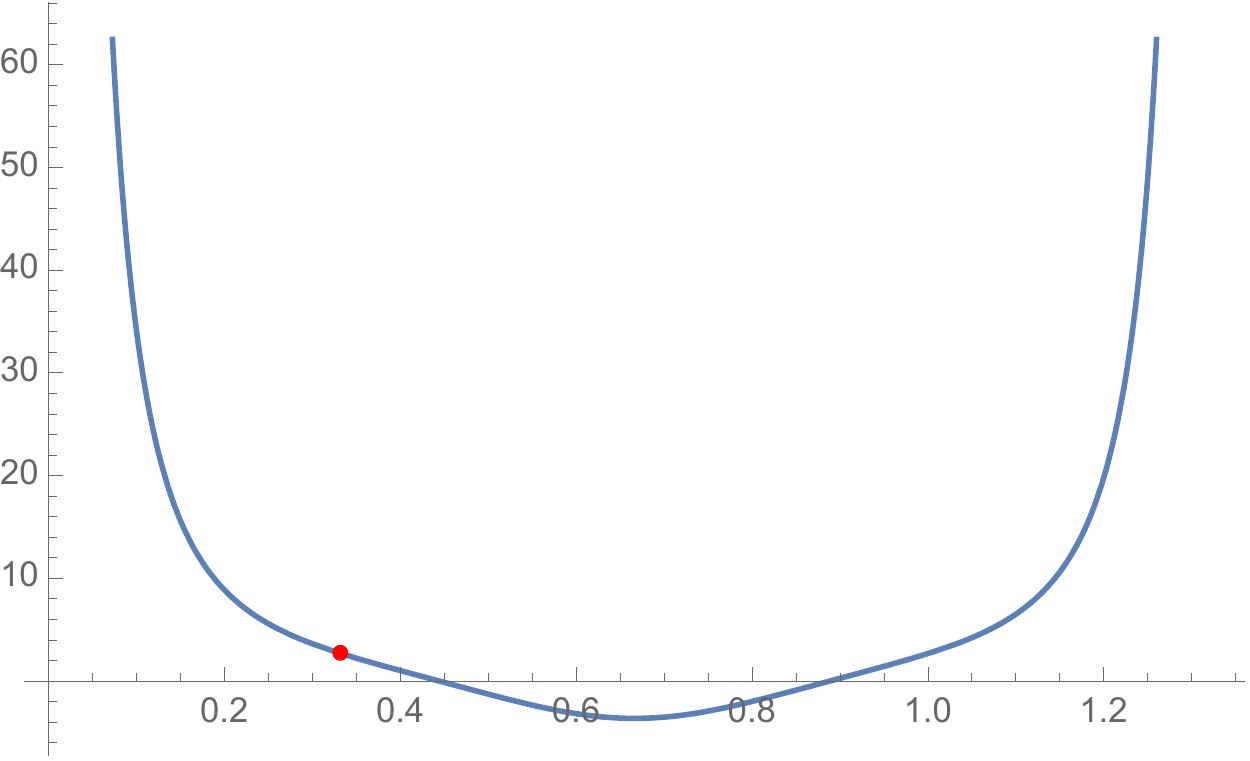} 
\end{center}
\caption{Evolution of the universe temperature $\widetilde{T}$ as a function of conformal time. 
{\small Cases: (i) In the interval $[0, u_f]$, (ii) Around the present value $u_o$, (iii) For a real period $[0, u_f+u_g \lesssim  2\omega_r]$. 
The physical branch is the interval $[0, u_f]$ where the curve intersects the real axis (first zero): $u_f$ is the end of time.
The curve possesses a negative branch in the interval $[u_f+u_g, 2 \omega_r]$ which is not displayed (see figs \ref{qualitativeT}, \ref{qualitativeTbis} and \ref{fig:smallbranchTandH}).}}
\label{fig:evolutionoftemperature} 
\end{figure}

As a function of conformal time, the evolution of the usual temperature $\widetilde{T}$ is displayed in fig.~\ref{fig:evolutionoftemperature}.
Remember that it is obtained by multiplying $T$ by the known (and constant) scaling factor given by eq.~\ref{eq:UsualTemperature}.
As already mentioned several times, this function is periodic in the complex plane, and in particular along the real axis, but the physical branch (\ie the history of our universe) is only given by the interval $[0, u_f]$, the value $u_f$ being the ``end of  time'', which corresponds to $t \rightarrow \infty$ in terms of cosmic time. 
With $k=1$, \ie in a universe with the topology of $S^3$ (or a discrete quotient of the latter), the fact that $u_f \leq 2 \pi$ means that an observer looking forward will never be able to see the back of his head, even if he waits for a very long time.

\subsubsection*{Radiation, matter, vacuum, curvature densities $\Omega_r$, $\Omega_m$,  $\Omega_\Lambda$, $\Omega_K$ and $\Omega = \Omega_m + \Omega_\Lambda + \Omega_r $}

\begin{figure} 
 \begin{center}
 \includegraphics*[scale=0.35]{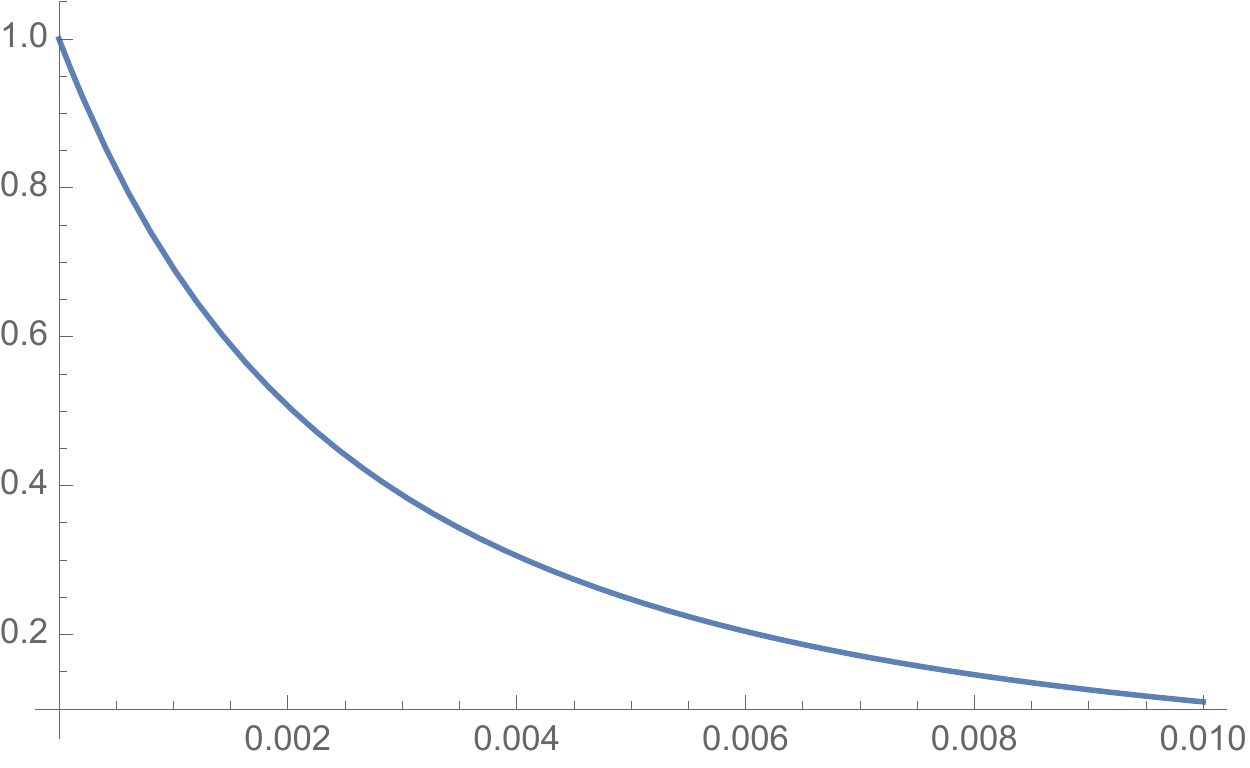} 
 \hspace{1.0cm}
\includegraphics*[scale=0.35]{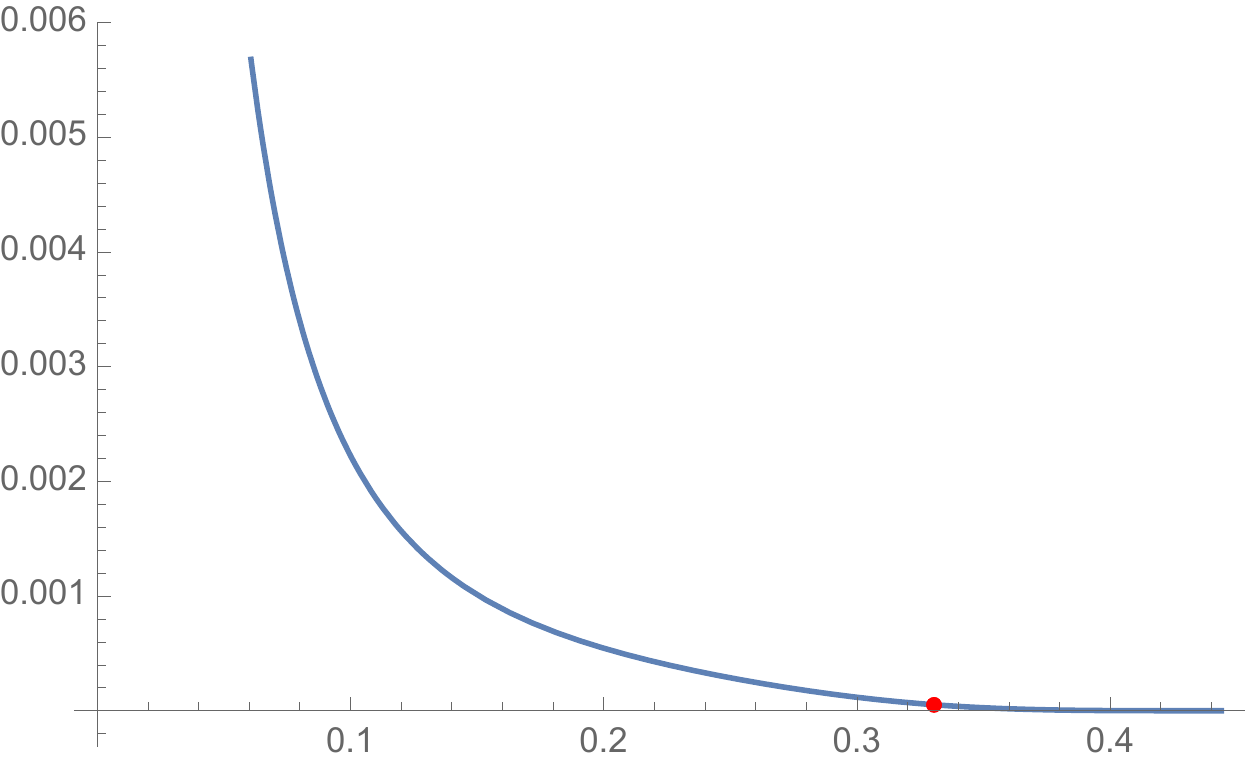} 
 \hspace{1.0cm}
\includegraphics*[scale=0.35]{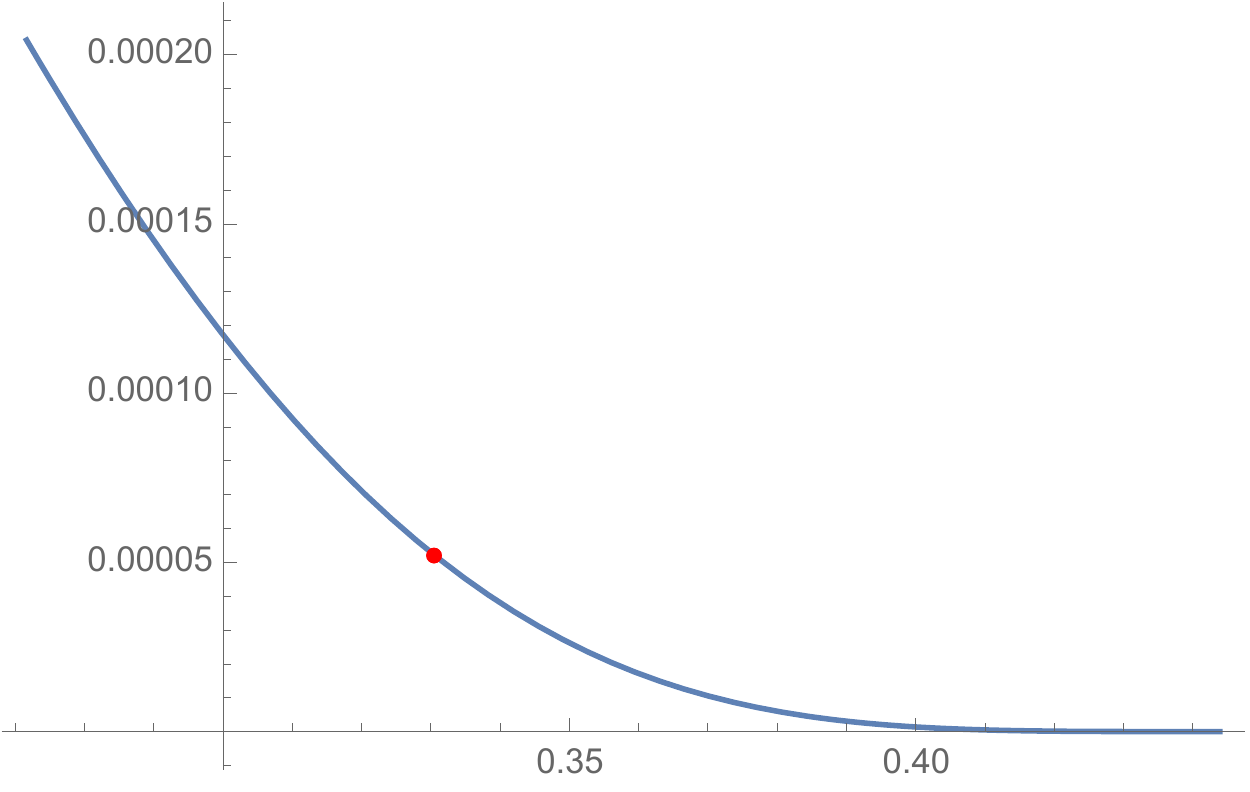} 
 \hspace{1.0cm}
\includegraphics*[scale=0.35]{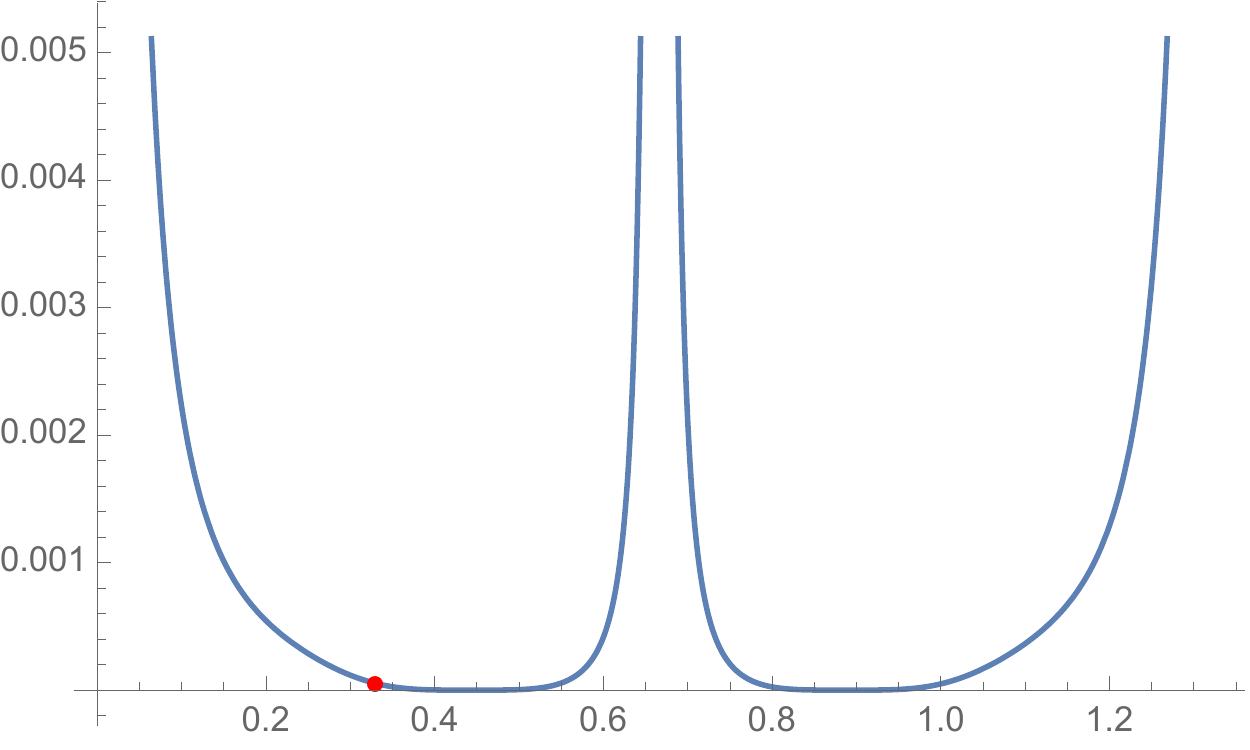} 
\end{center}
\caption{Evolution of the radiation density $\Omega_r(u)$ as a function of conformal time. 
{\small Cases: (i) Near the Big Bang. (ii) In the interval $[0, u_f]$, the physical branch. (iii) Around the present value $u_o$. (iv) For a real period $[0, u_f+u_g \lesssim  2\omega_r]$. 
 In the tiny interval $[u_f+u_g, 2 \omega_r]$ the curve has two infinite positive branches that are not displayed. }}
\label{fig:evolutionofradiation} 
\end{figure}

\begin{figure} 
 \begin{center}
 \includegraphics*[scale=0.35]{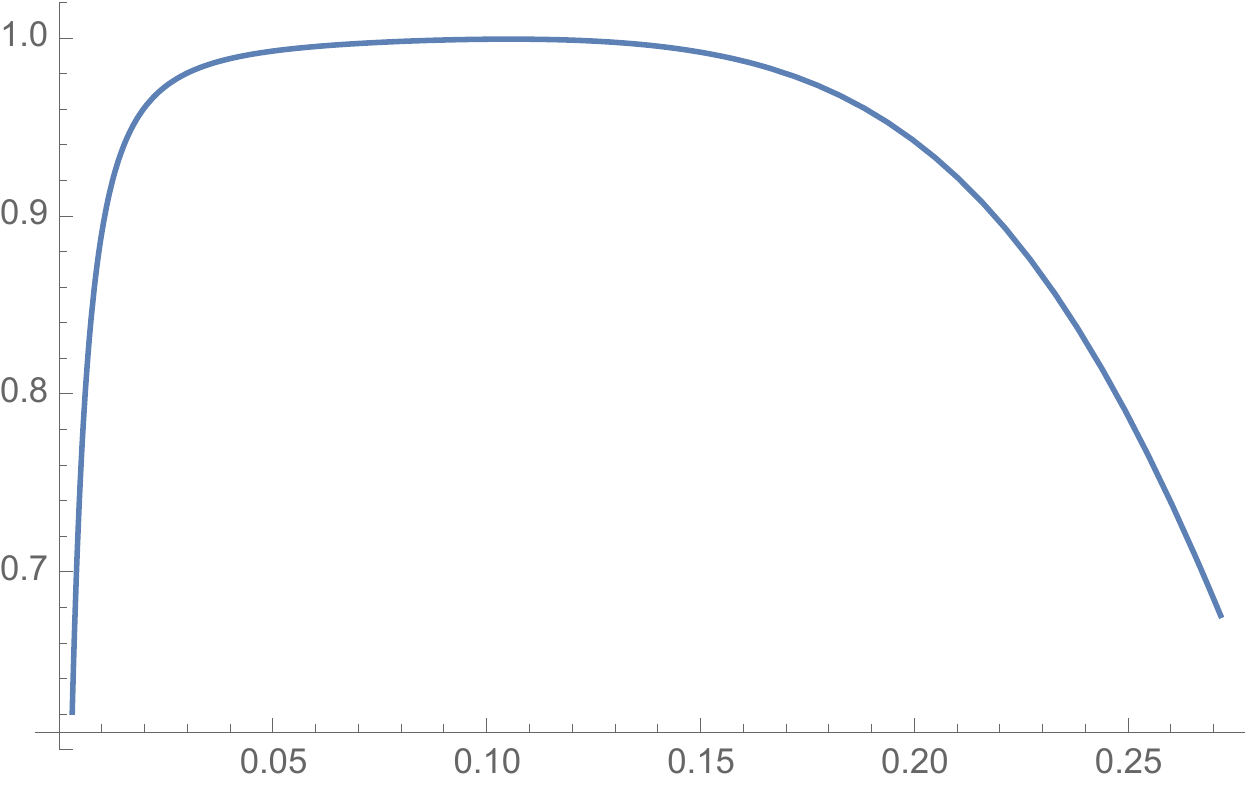} 
 \hspace{1.0cm}
\includegraphics*[scale=0.35]{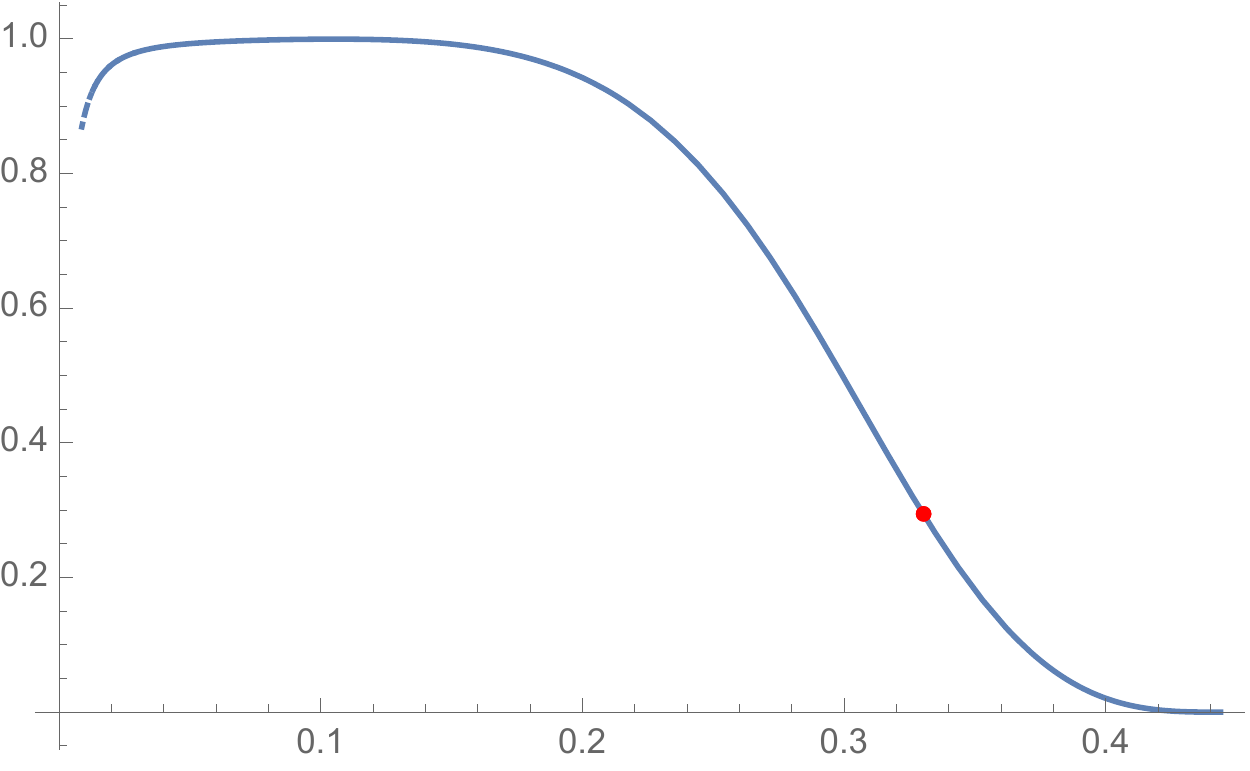} 
 \hspace{1.0cm}
\includegraphics*[scale=0.35]{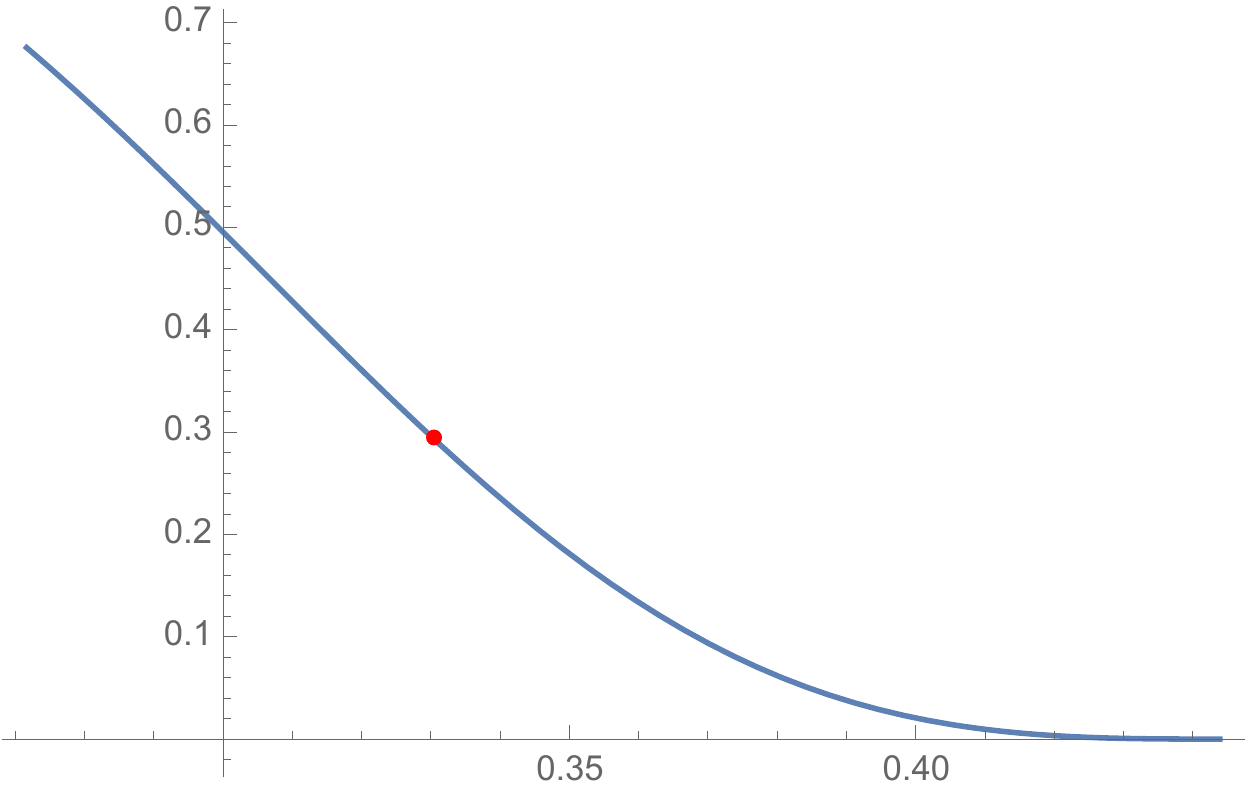} 
 \hspace{1.0cm}
\includegraphics*[scale=0.35]{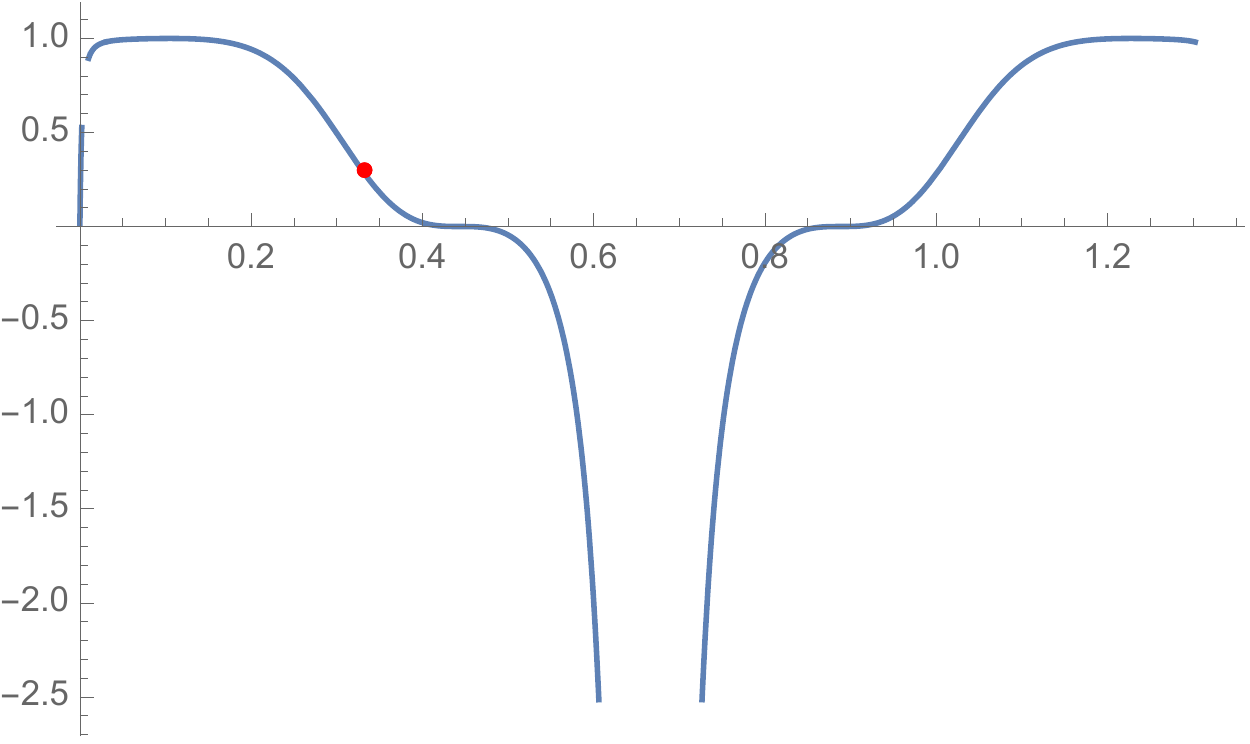} 
\end{center}
\caption{Evolution of the matter density $\Omega_m(u)$ as a function of conformal time. 
{\small Cases: (i) Near the Big Bang. (ii) In the interval $[0, u_f]$, the physical branch. (iii) Around the present value $u_o$. (iv) For a real period $[0, u_f+u_g \lesssim  2\omega_r]$. 
 In the tiny interval $[u_f+u_g, 2 \omega_r]$ the curve has two infinite negative branches that are not displayed.}}
\label{fig:evolutionofmatter} 
\end{figure}

\begin{figure} 
 \begin{center}
\includegraphics*[scale=0.35]{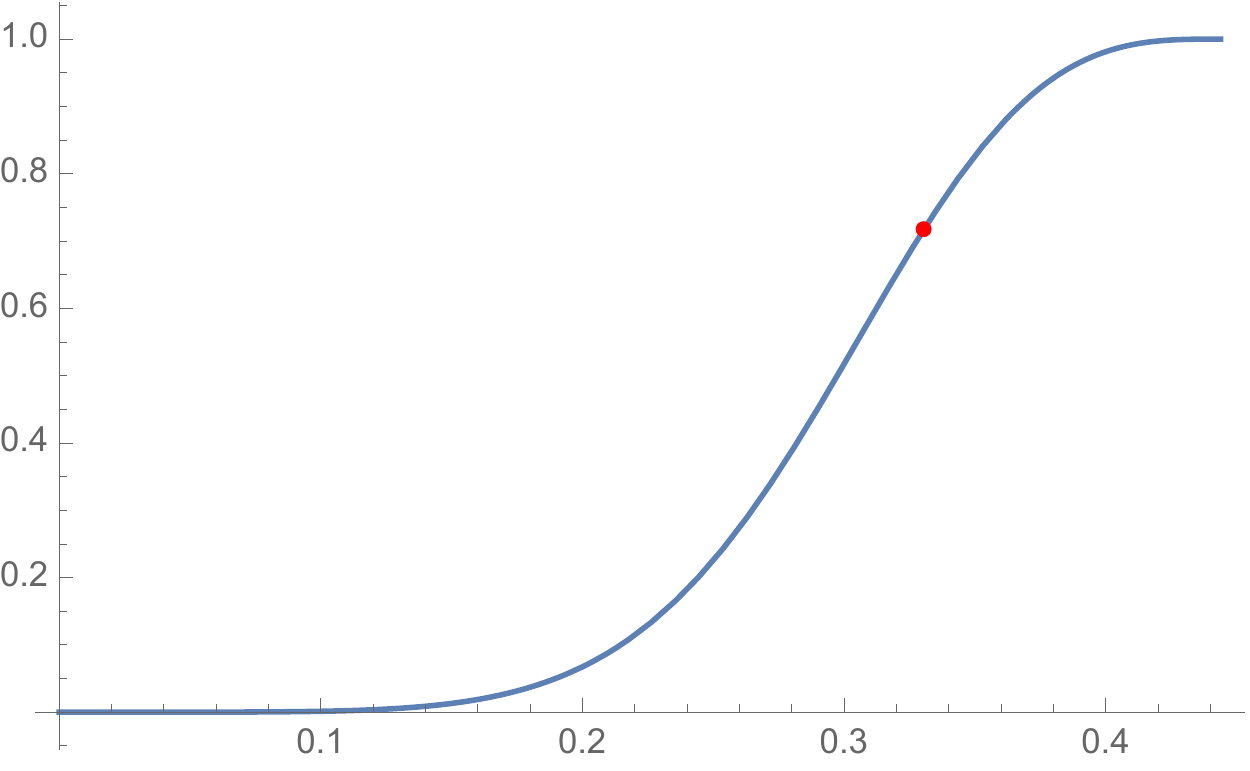} 
 \hspace{1.0cm}
\includegraphics*[scale=0.35]{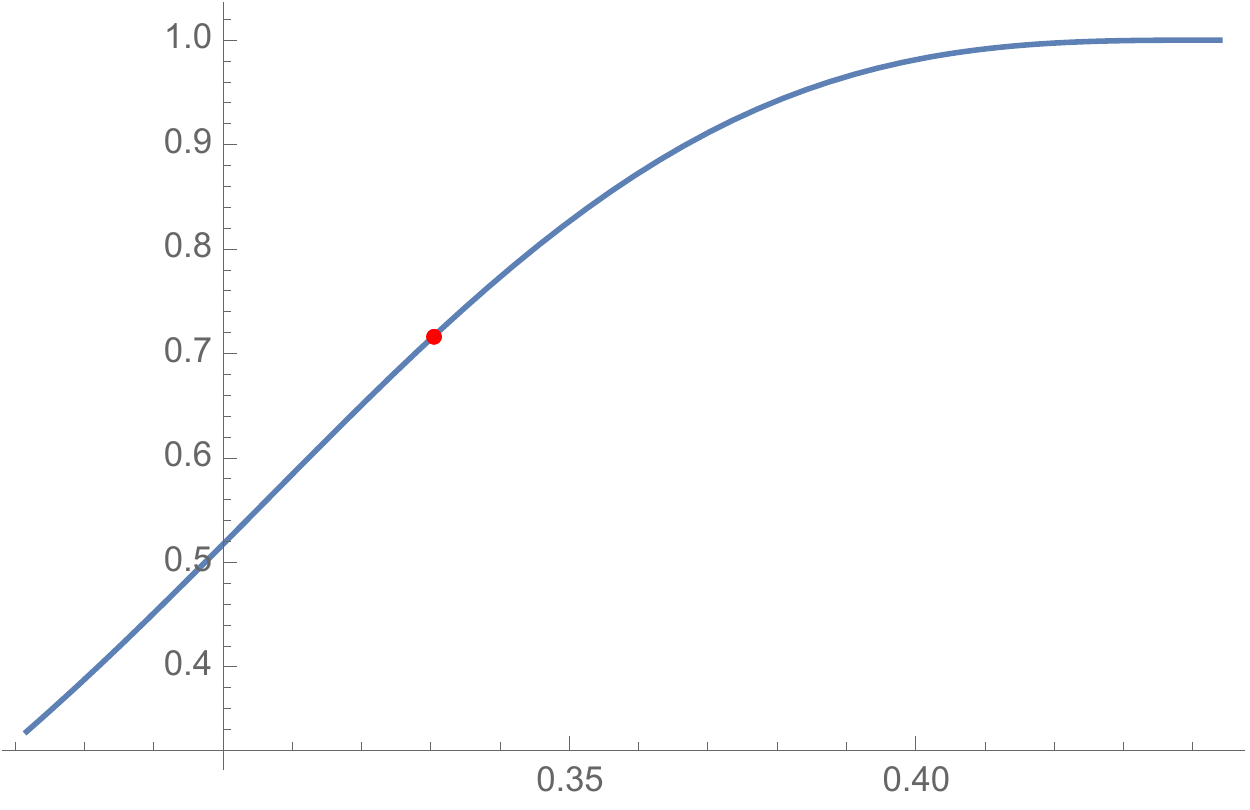} 
 \hspace{1.0cm}
\includegraphics*[scale=0.35]{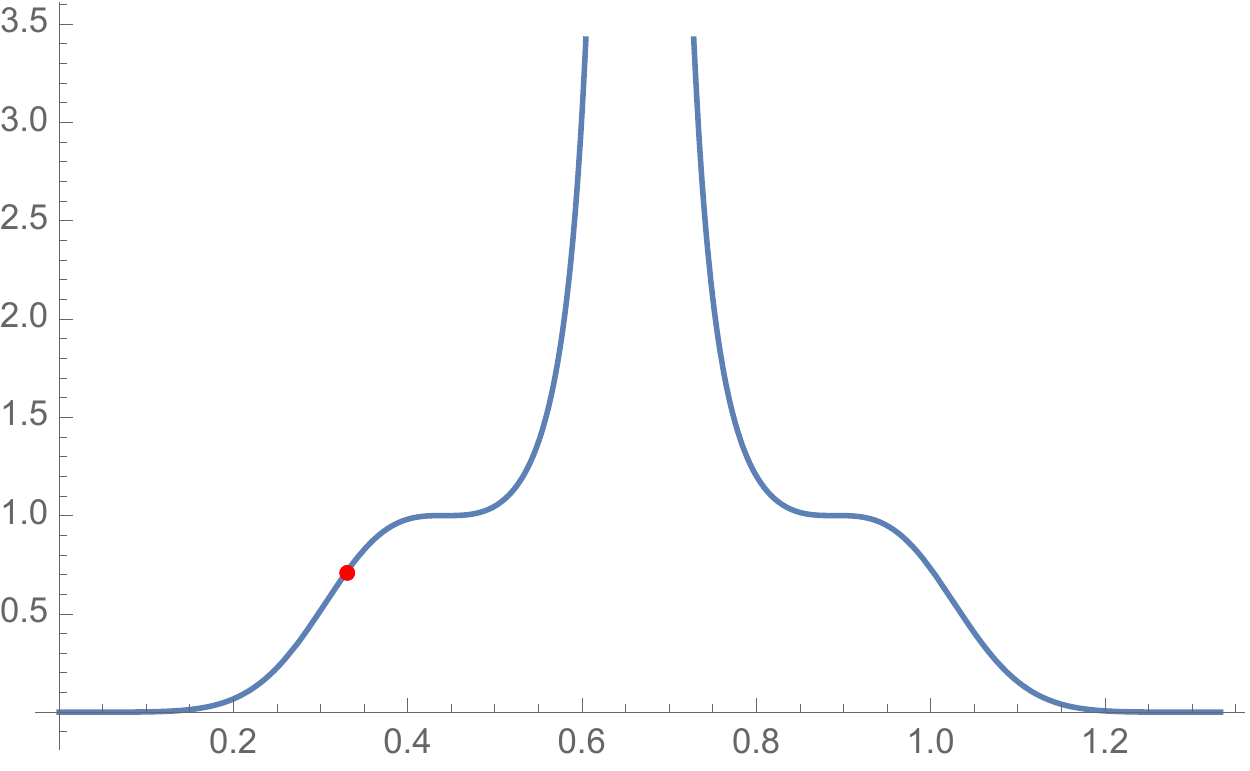} 
\end{center}
\caption{Evolution of the vacuum density $\Omega_\Lambda(u)$ as a function of conformal time. 
{\small Cases: (i) In the interval $[0, u_f]$, the physical branch. (ii) Around the present value $u_o$. (iii) For a real period $[0, u_f+u_g \lesssim  2\omega_r]$. 
 In the tiny interval $[u_f+u_g, 2 \omega_r]$ the curve has two infinite positive branches that are not displayed.}}
\label{fig:evolutionofvacuum} 
\end{figure}

\begin{figure} 
 \begin{center}
\includegraphics*[scale=0.35]{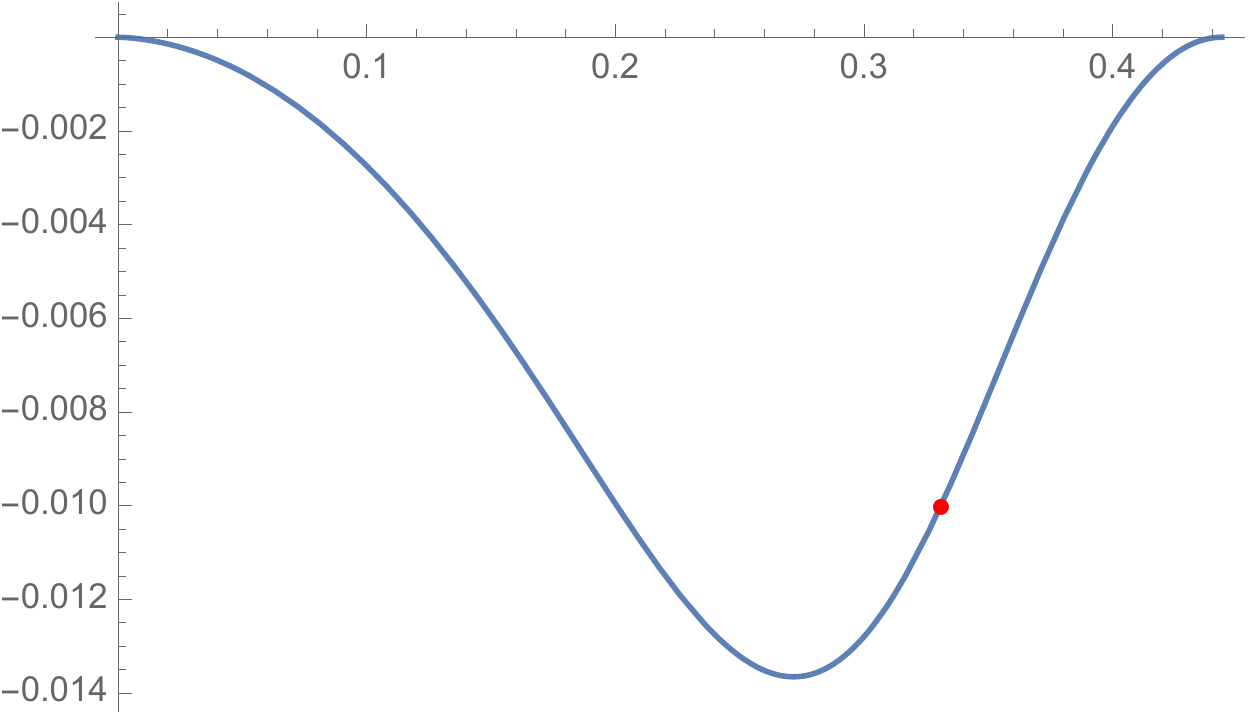} 
 \hspace{1.0cm}
\includegraphics*[scale=0.35]{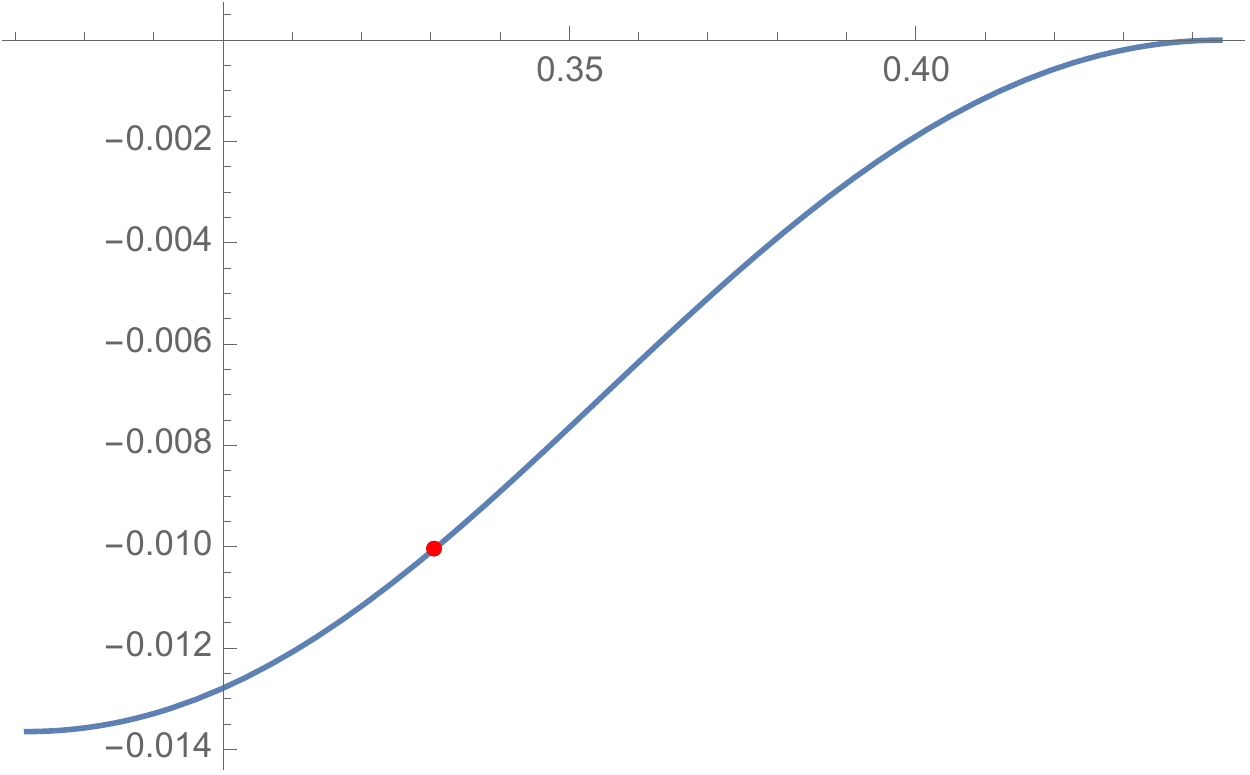} 
 \hspace{1.0cm}
\includegraphics*[scale=0.35]{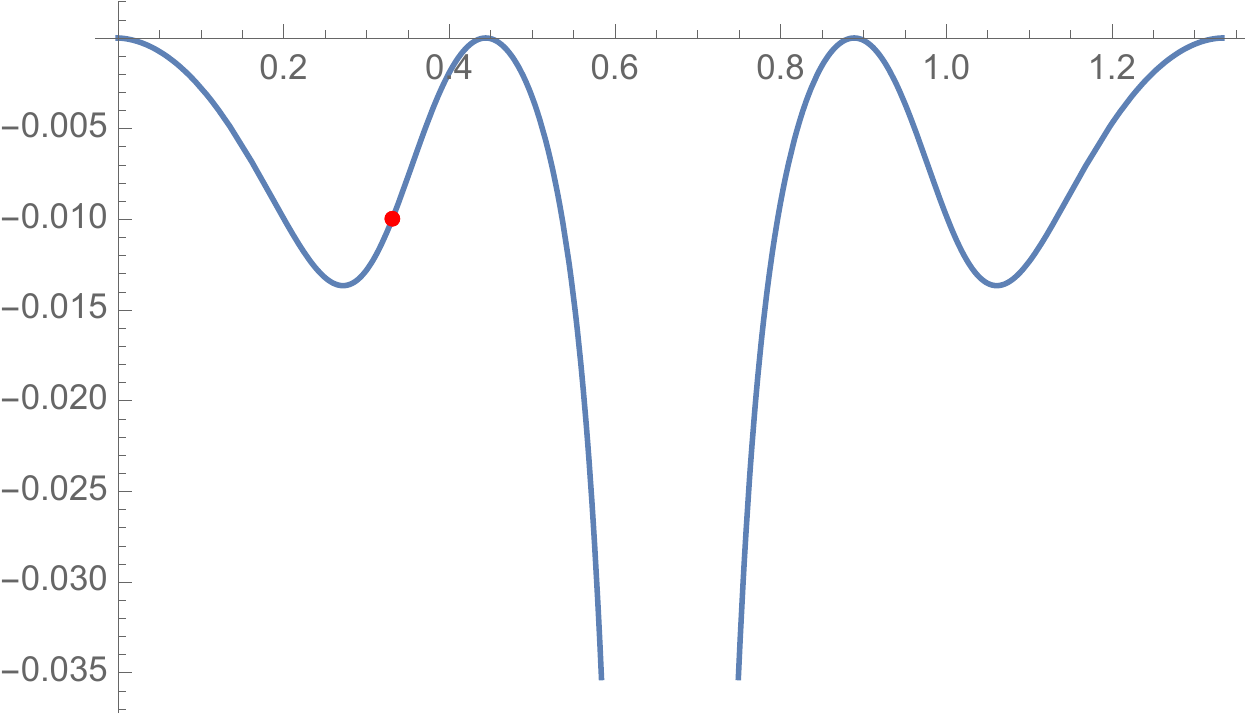} 
\end{center}
\caption{Evolution of the curvature density $\Omega_K(u)$ as a function of conformal time. 
{\small Cases: (i) In the interval $[0, u_f]$, the physical branch. (ii) Around the present value $u_o$. (iii) For a real period $[0, u_f+u_g \lesssim  2\omega_r]$. 
In the tiny interval $[u_f+u_g, 2 \omega_r]$ the curve has two infinite negative branches that are not displayed.}}
\label{fig:evolutionofcurvature} 
\end{figure}

\begin{figure} 
 \begin{center}
\includegraphics*[scale=0.35]{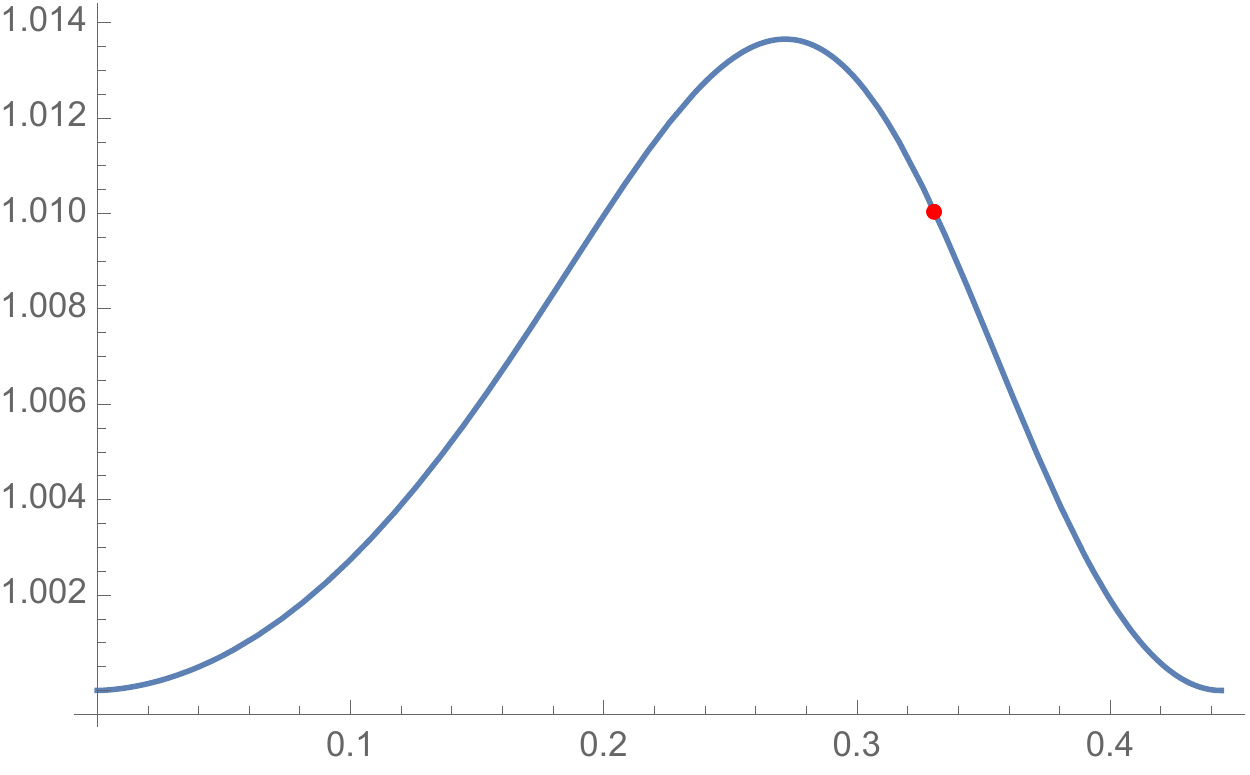} 
 \hspace{1.0cm}
 \includegraphics*[scale=0.35]{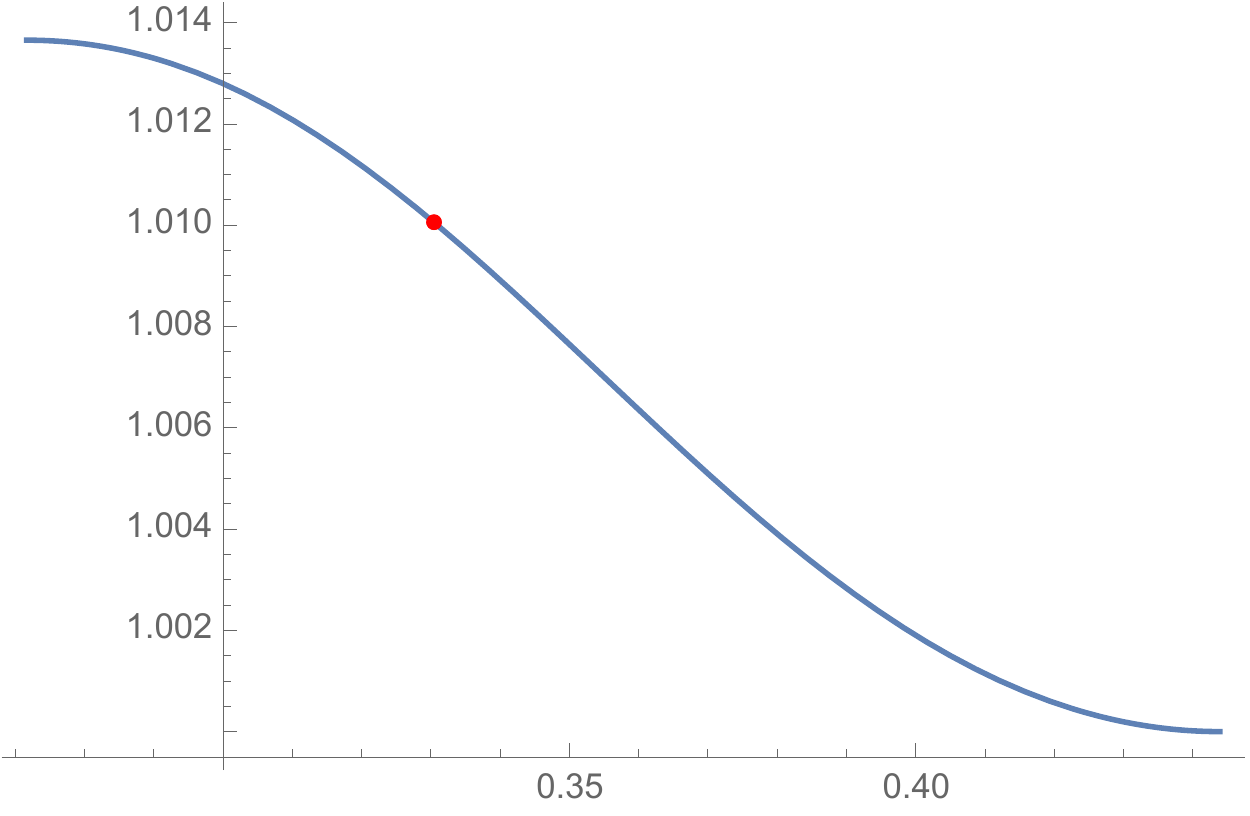} 
 \hspace{1.0cm}
 \includegraphics*[scale=0.35]{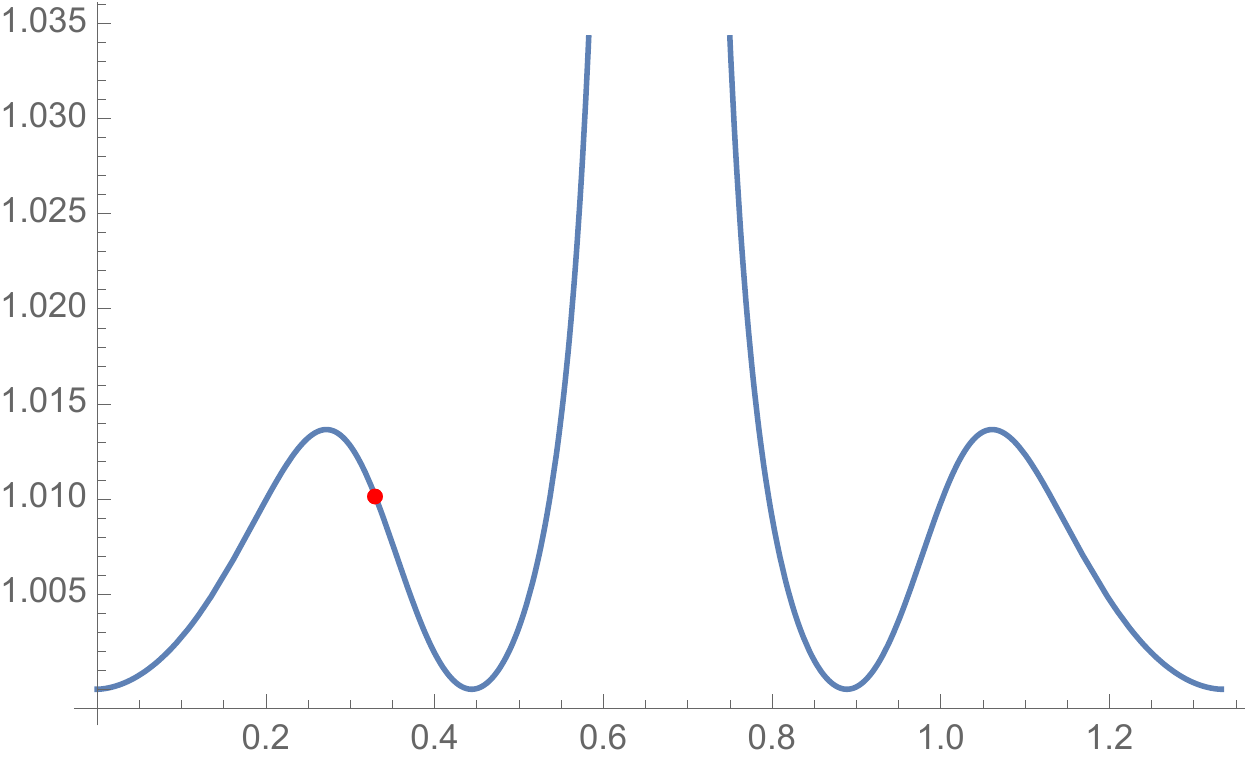} 
 \hspace{1.0cm}
\end{center}
\caption{Evolution of the sum $\Omega(u)=\Omega_m(u)+\Omega_\Lambda(u)+\Omega_r(u)$ as a function of conformal time. 
{\small Cases: (i) In the interval $[0, u_f]$, the physical branch. (ii) Around the present value $u_o$. (iii) For a real period $[0, u_f+u_g \lesssim  2\omega_r]$. 
In the tiny interval $[u_f+u_g, 2 \omega_r]$ the curve is not displayed.}}
\label{fig:evolutiontotalomega}.
\end{figure}

The curves displaying the behavior of those densities, as functions of conformal time, are respectively given in figures \ref{fig:evolutionofradiation}, \ref{fig:evolutionofmatter}, \ref{fig:evolutionofvacuum}, \ref{fig:evolutionofcurvature}, \ref{fig:evolutiontotalomega}.
We already commented about the fact that we choose (arbitrarily) the value $\Omega_K^o = - 0.01$ (so $k=+1$), which is non-zero, but compatible with the experimental bounds, to perform the calculations leading to the following plots.
The values obtained for the conformal time $u$ of the various points of interest (extrema, inflection point, present day value, end of time, etc.), as well as the values obtained for the modular parameters (Weierstrass invariants, periods, j-function, etc),  are quite sensitive to the choice of  $\Omega_K^o$; however, the overall features of the curves are rather stable, given the present values of the other densities.

\subsubsection*{Hubble function $H$ and deceleration function $q$}

\begin{figure} 
 \begin{center}
\includegraphics*[scale=0.35]{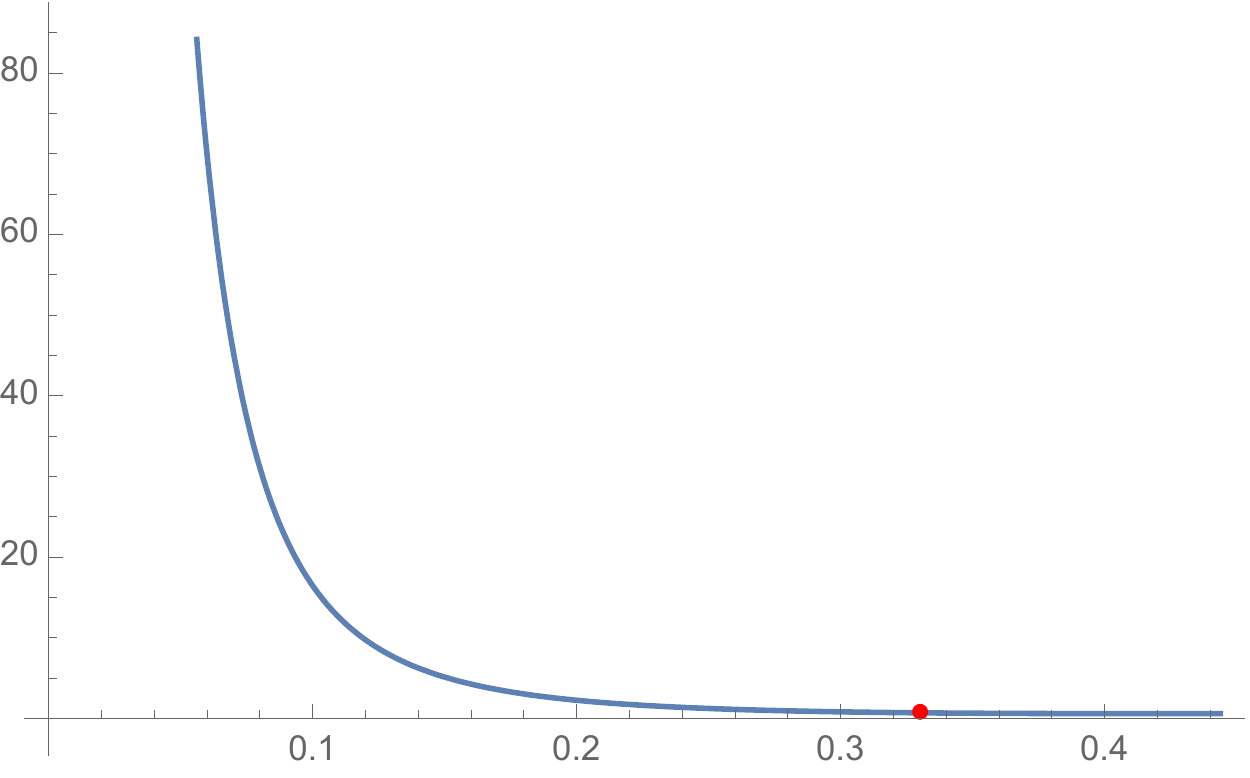} 
 \hspace{1.0cm}
\includegraphics*[scale=0.35]{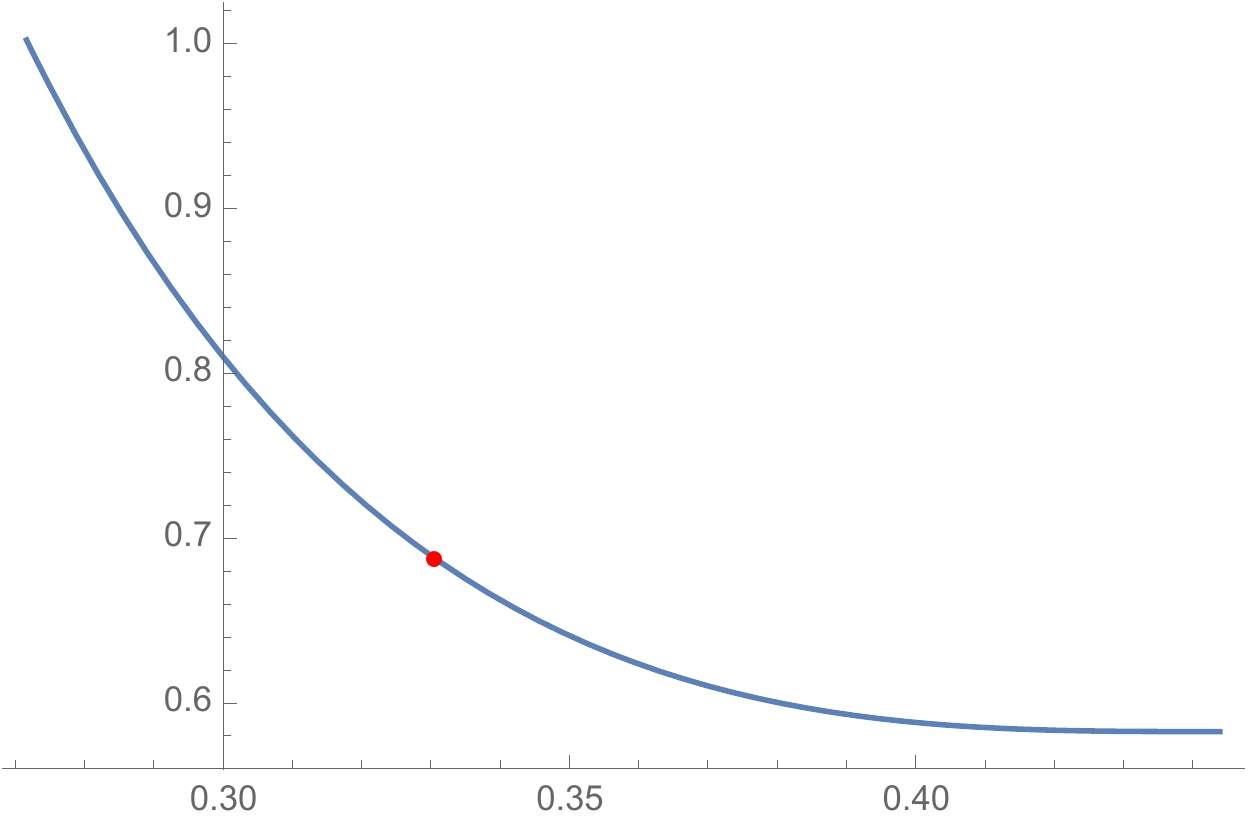} 
 \hspace{1.0cm}
\includegraphics*[scale=0.35]{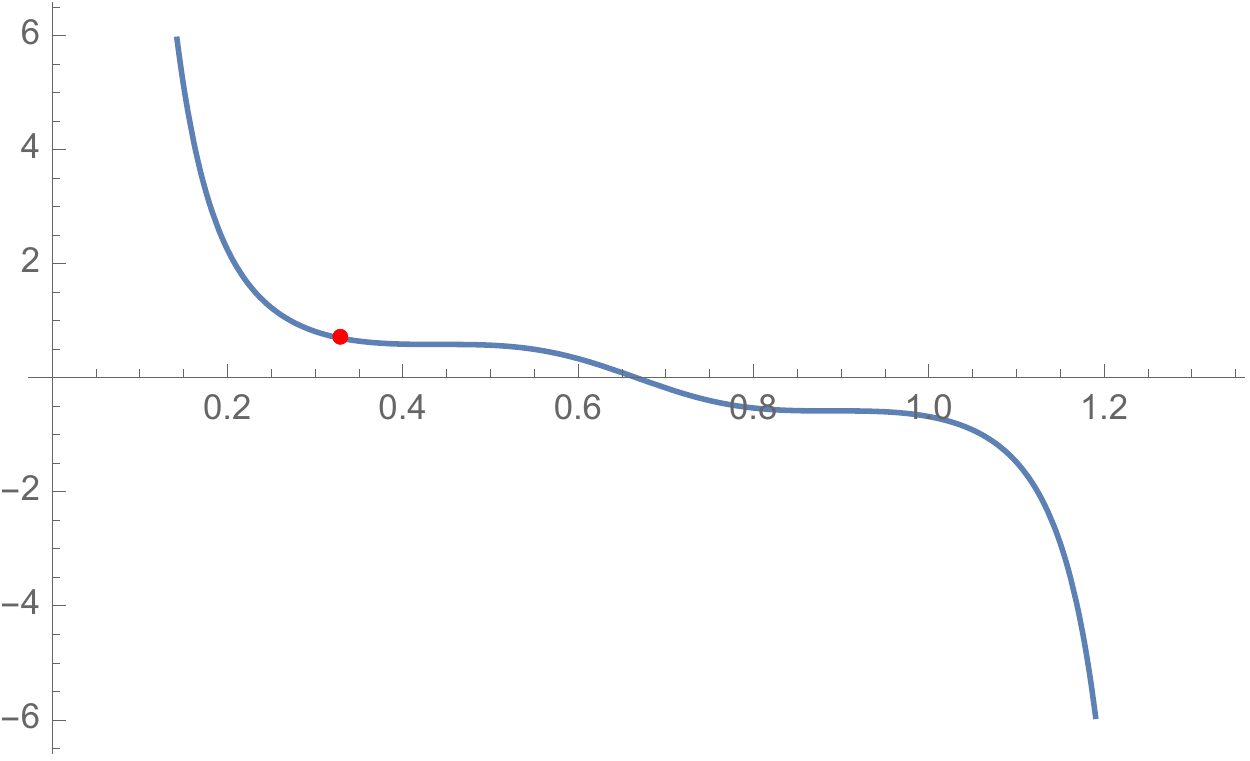} 
\end{center}
\caption{Evolution of the Hubble ``constant'' $H(u)$ as a function of conformal time.  {\small We plot $h(u)$, with $H(u)= 100\;  h(u)\;  km \; s^{-1} \; Mpc^{-1}$.
{\small Cases: (i) In the interval $[0, u_f]$, the physical branch. (ii) Around the present value $u_o$. (iii) For a real period $[0, u_f+u_g \lesssim  2\omega_r]$. 
The branch of the curve in the tiny interval $[u_f+u_g, 2 \omega_r]$  is not displayed.}}}
\label{fig:evolutionofhubble} 
\end{figure}

Rather than plotting the Hubble function $H(u) \times cm$, we give in fig.~\ref{fig:evolutionofhubble} the evolution of the $h$ coefficient as a function of conformal time.
As usual, $H = (100\,  h \, 10^5 \, cm)/(s \, Mpc )$, so  $H \times cm  = 1.0810076 \, 10^{-28} \,  h$, with $h_o = 0.688$, the present value of $h$. 

\begin{figure} 
 \begin{center}
 \includegraphics*[scale=0.35]{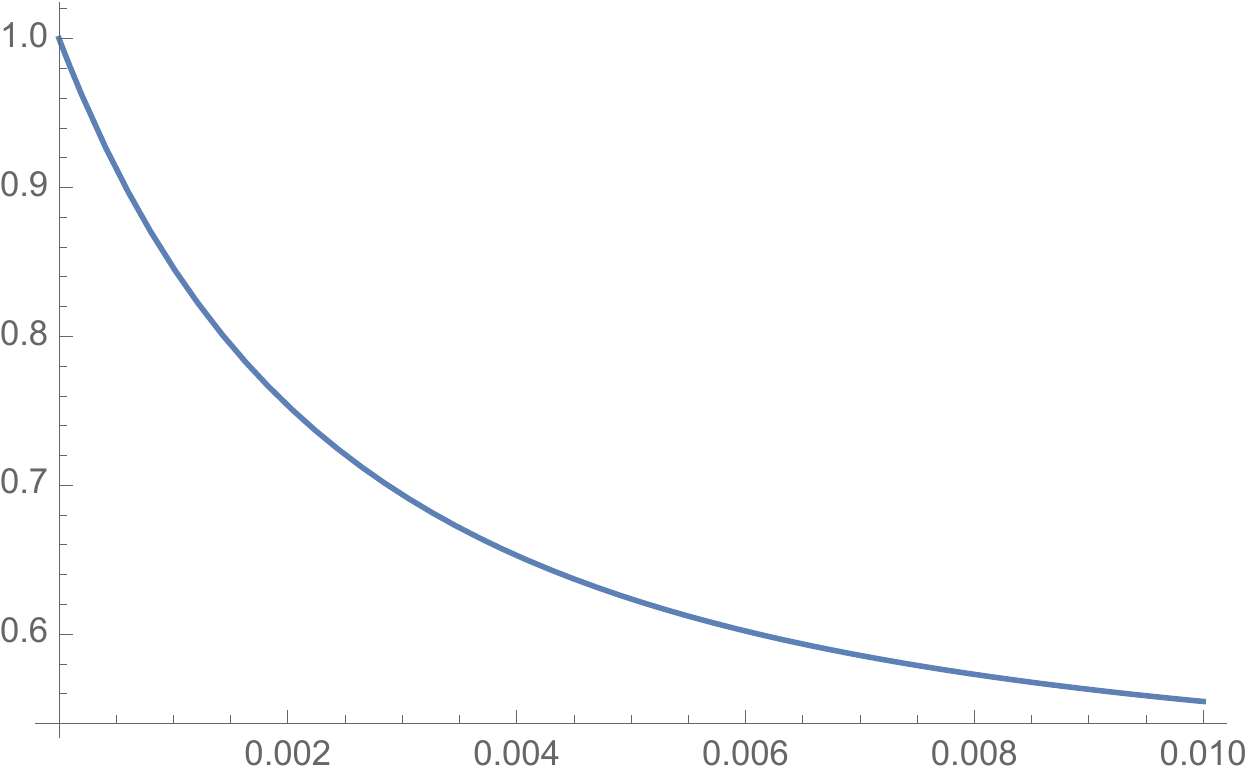} 
 \hspace{1.0cm}
\includegraphics*[scale=0.35]{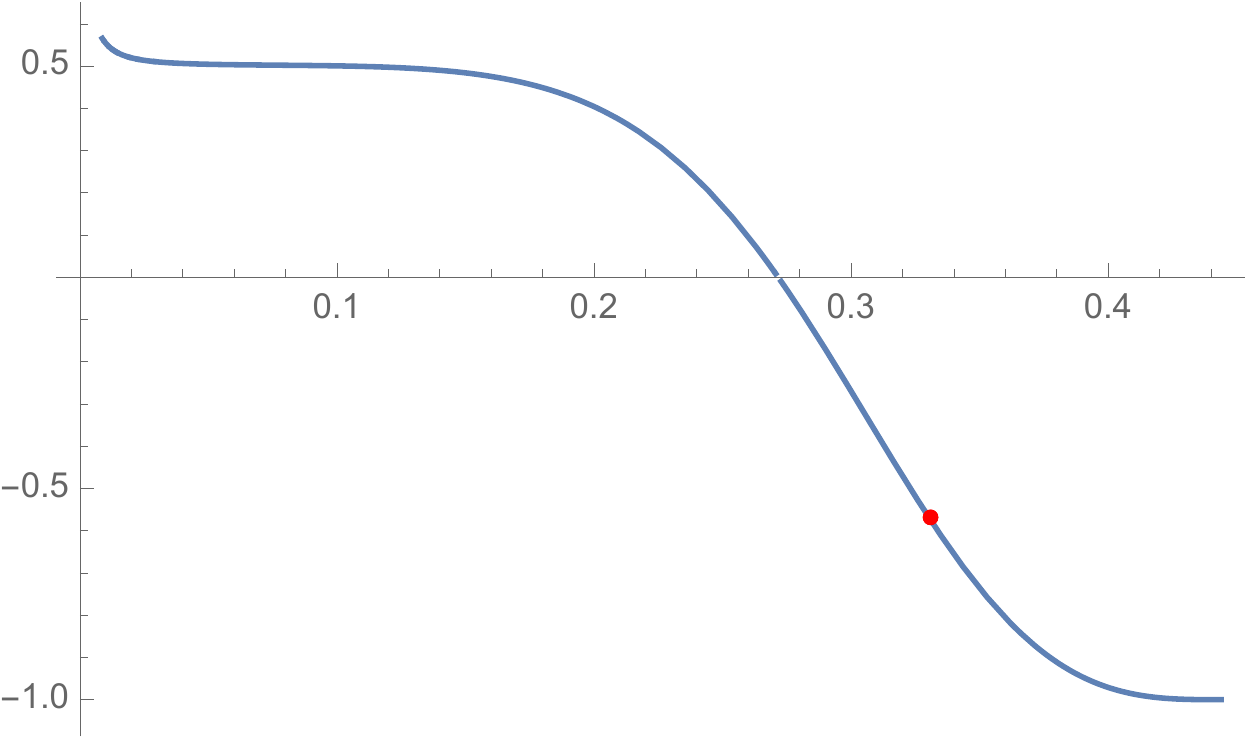} 
 \hspace{1.0cm}
\includegraphics*[scale=0.35]{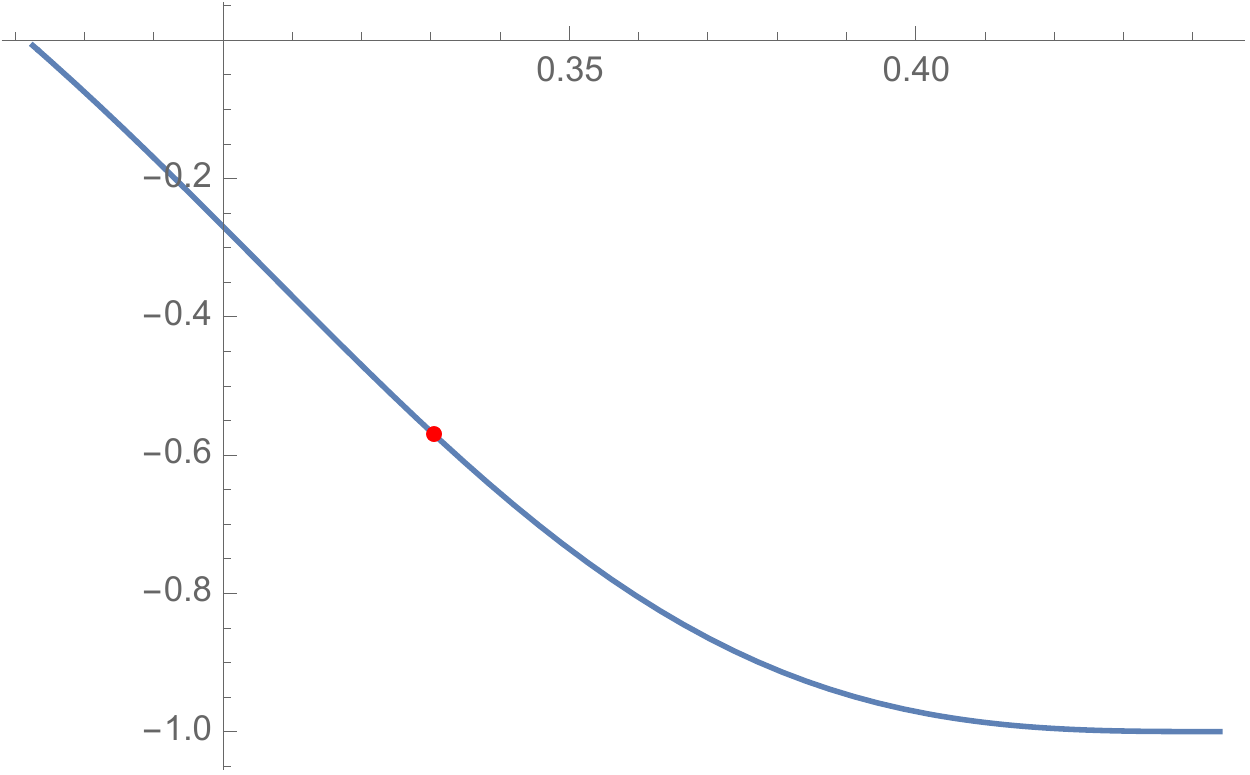} 
 \hspace{1.0cm}
\includegraphics*[scale=0.35]{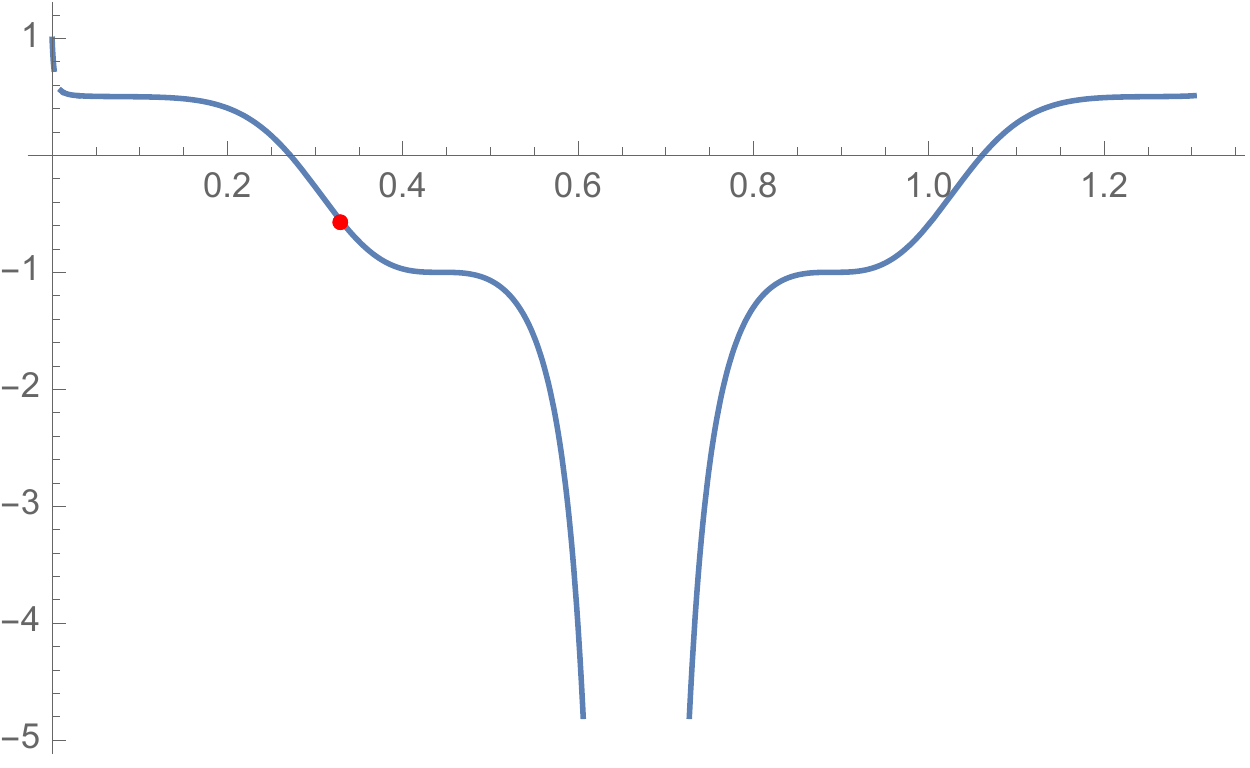} 
\end{center}
\caption{Evolution of the deceleration function $q(u)$ as a function of conformal time. 
{\small Cases: (i) Near the Big Bang. (ii) In the interval $[0, u_f]$, the physical branch.  (iii) Around the present value $u_o$. (iv) For a real period $[0, u_f+u_g \lesssim  2\omega_r]$. 
In the tiny interval $[u_f+u_g, 2 \omega_r]$ the curve has two infinite positive branches that are not displayed.}}
\label{fig:evolutionofdecceleration} 
\end{figure}

Another useful quantity is the {\sl deceleration function} --- a misnomer\footnote{In the old books it is not uncommon to find $q$ defined as $\Omega_m/2$ since it was then fashioned to set to zero both the vacuum and the radiative contributions.} since we seem to live (now) in an accelerating expanding universe: 
$ q = - a(da/dt)^{-2} (d^{2}a /dt^{2}) = \Omega_m/2 - \Omega_\Lambda + \Omega_r$. It can  be expressed in terms of $T({u})$ as follows: 
 $$ 
q = {- {\lambda \over 3} + {1 \over 3} T^{3} + \alpha T^{4} \over 
 {\lambda \over 3} - k T^{2} + {2 \over 3} T^{3} + \alpha T^{4}} 
$$ 
The behavior of this function $q(u)$ is given in fig.~\ref{fig:evolutionofdecceleration}.

\subsubsection*{The behavior of $\widetilde T$ and $h$ in the small interval $[u_f+u_g, 2 \omega_r]$}

\begin{figure} 
 \begin{center}
 \includegraphics*[scale=0.35]{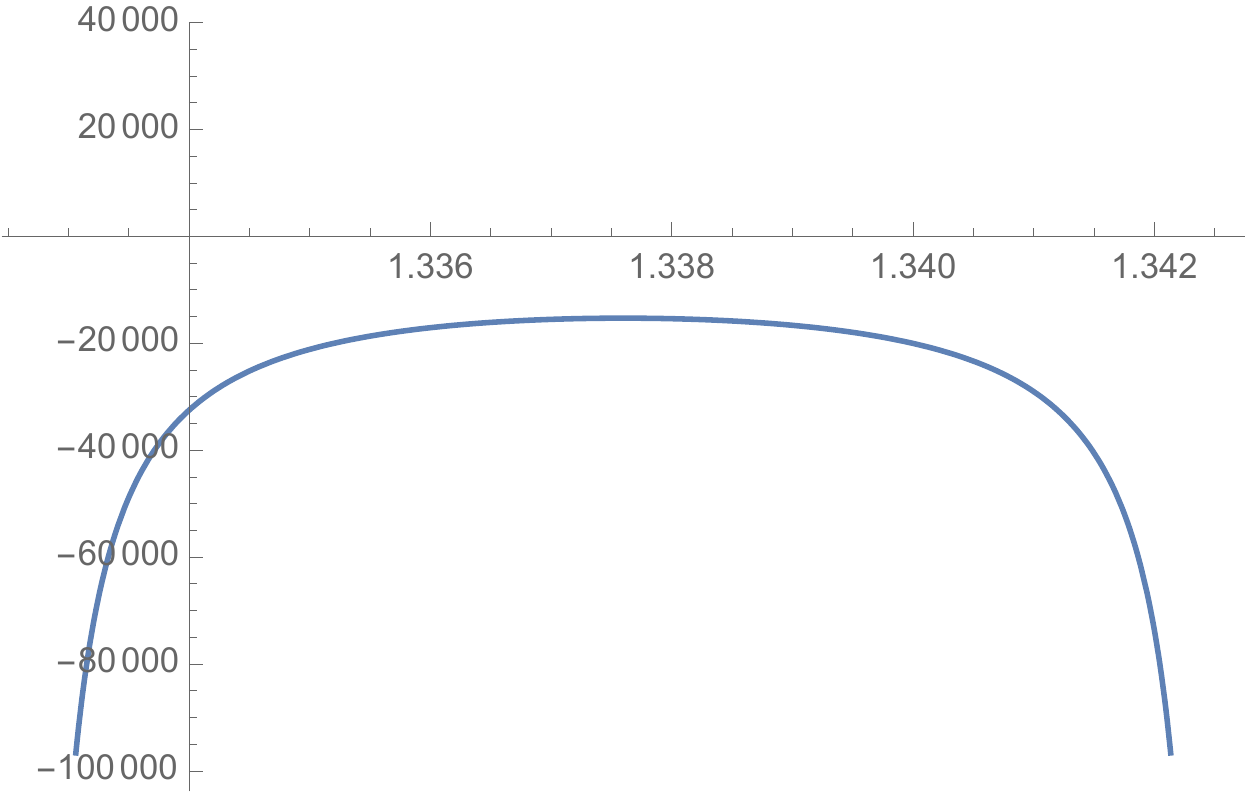} 
 \hspace{1.0cm}
\includegraphics*[scale=0.35]{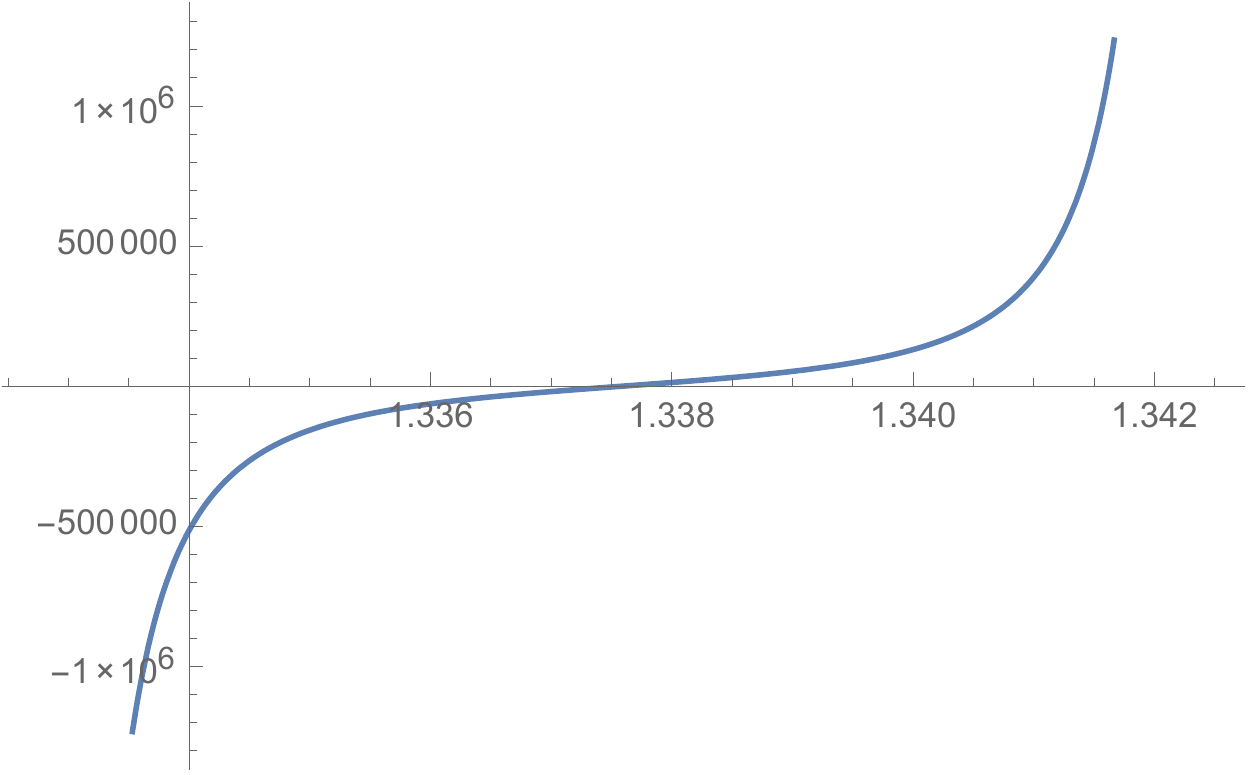} 
\end{center}
\caption{Evolution of $\widetilde{T}(u)$ and $h(u)$ in the (unphysical) tiny interval $[u_f+u_g, 2 \omega_r]$}
\label{fig:smallbranchTandH} 
\end{figure}

In the plots \ref{fig:evolutionofradiation} -- \ref{fig:evolutionofdecceleration} , we did not display the branches of the curves in the tiny interval $[u_f+u_g, 2 \omega_r]$, located just before the period $2 \omega_r$ and far above the physical region $[0, u_f]$.
Nevertheless, for completeness sake, we display in fig.~\ref{fig:smallbranchTandH} the behavior of the functions $\widetilde{T}(u)$ and $h(u)$ ---the latter being essentially the derivative of the former--- in this interval.
We remind the reader that this interval shrinks to zero when the radiation term (in $\alpha T^4$) is set to zero in the Friedmann  equations.

\subsubsection*{The behavior of $a$ and $t$}

\begin{figure} 
 \begin{center}
\includegraphics*[scale=0.35]{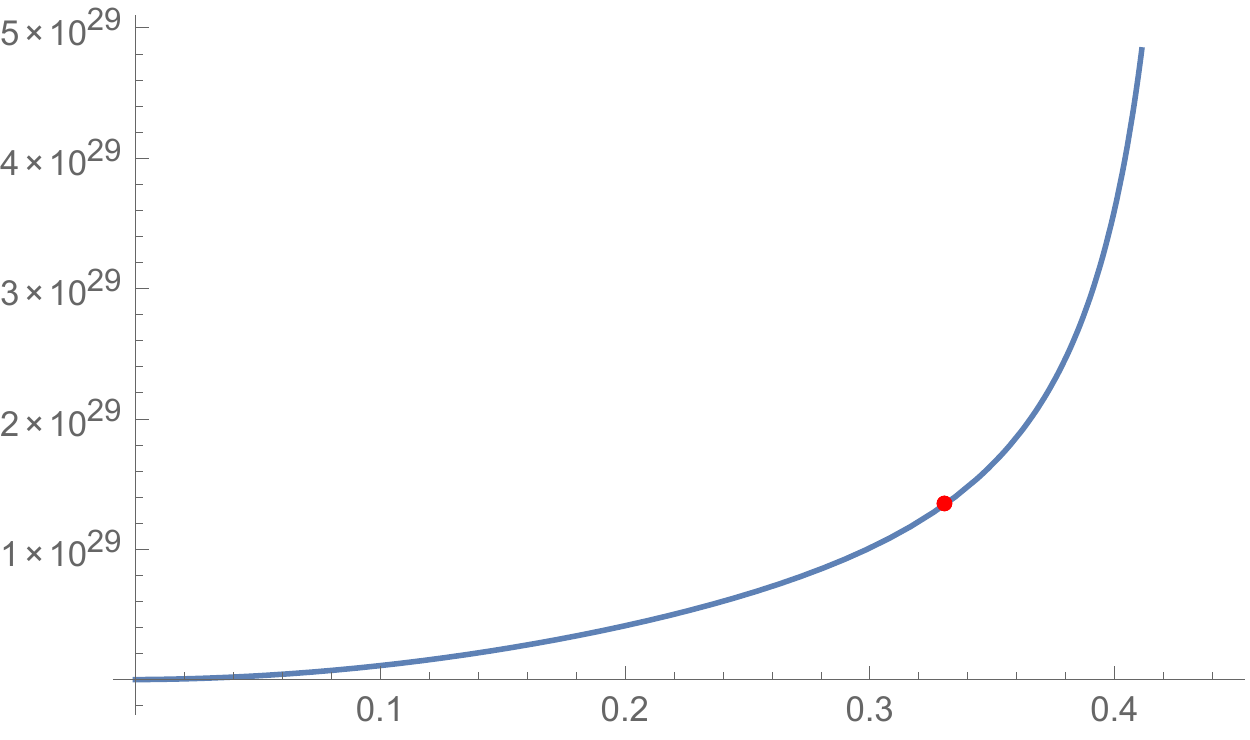} 
 \hspace{1.0cm}
\includegraphics*[scale=0.35]{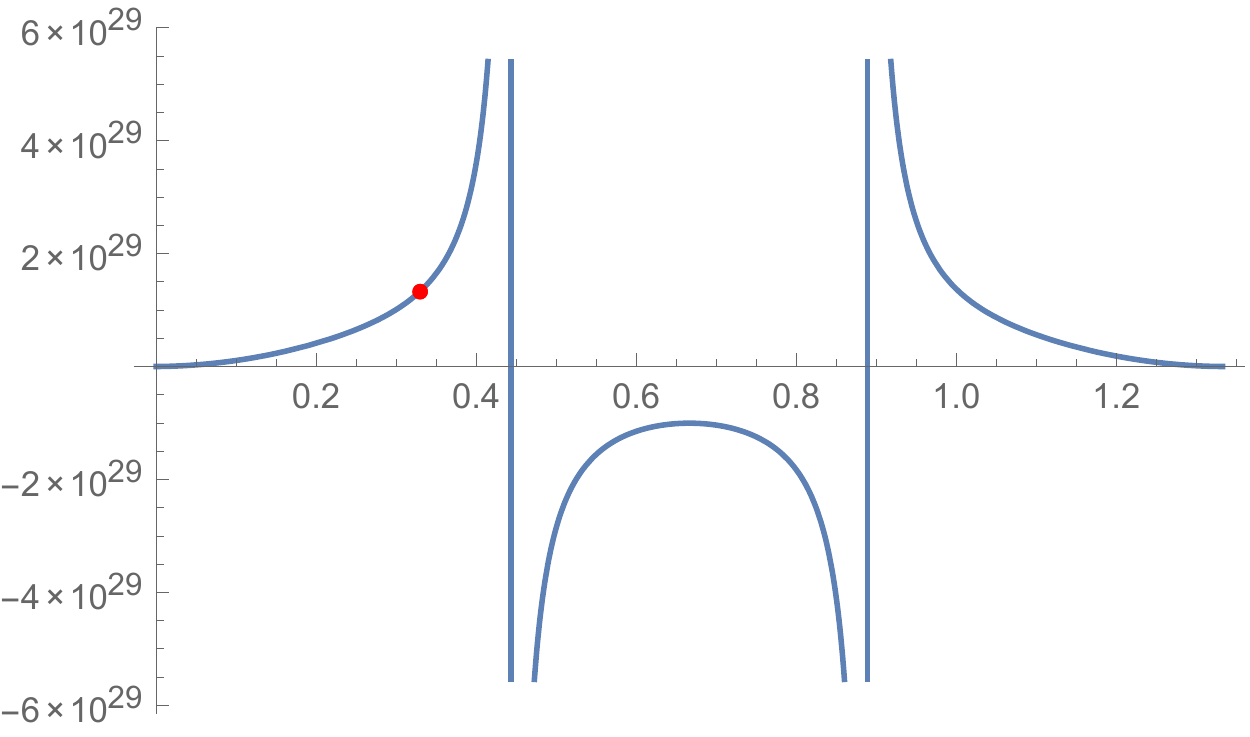} 
\hspace{1.0cm}
\includegraphics*[scale=0.35]{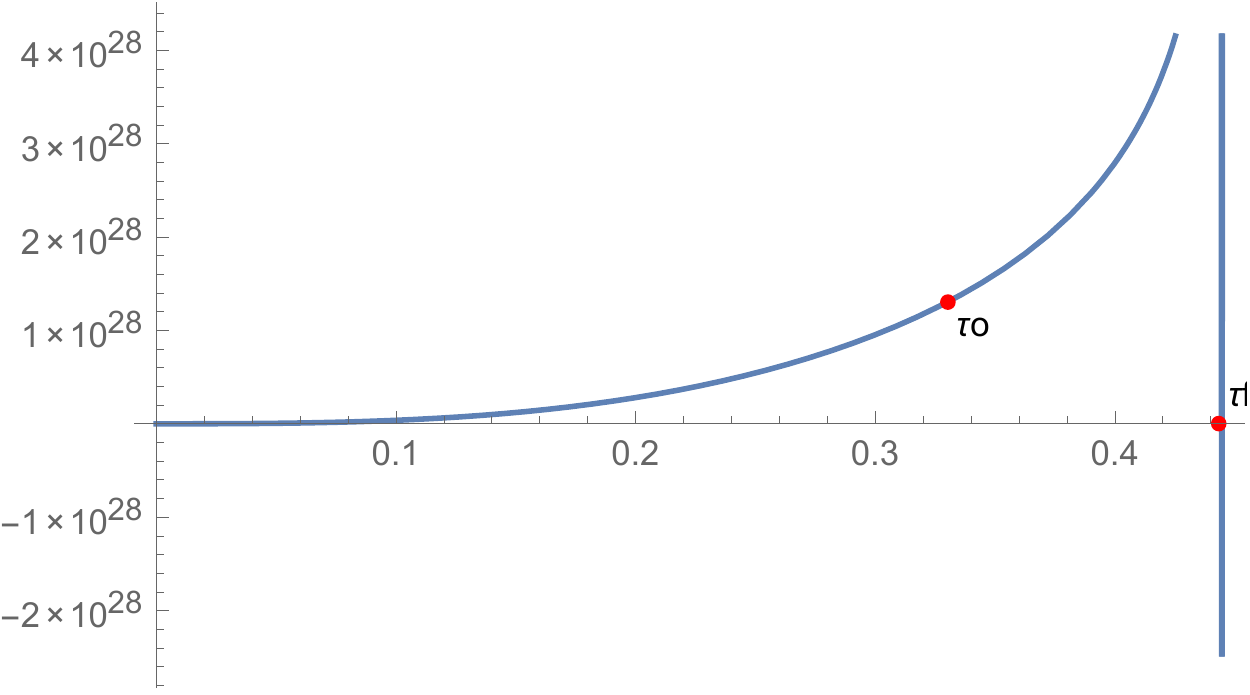} 
\end{center}
\caption{
\small{
 Evolution of the spatial scale factor $a(u)/cm$ as a function of conformal time,
(i) in the interval $[0, u_f]$, the physical branch.
(ii) For a real period $[0, u_f+u_g \lesssim  2\omega_r]$. In the tiny interval $[u_f+u_g, 2 \omega_r]$ the curve is not displayed.
(iii) Evolution of the cosmic time $t(u)/cm$ as a function of conformal time.}}
\label{fig:evolutionofaovercmandtovercm} 
\end{figure}

Finally, we display in fig.~\ref{fig:evolutionofaovercmandtovercm} the evolution of $a(u)$, the spatial scale factor (essentially the inverse of $T(u)$), and of the cosmic time $t(u)$, as functions of the conformal time $u$.\\
The function $t(u)$ is not elliptic and has a logarithmic singularity when $u\rightarrow u_f$.  In this neighborhood,  {\small $\sqrt{\frac{\Lambda}{3}} \; t(u) \sim - ln(u_f-u)$, 
$\sqrt{\frac{\Lambda}{3}} \; a(u) \sim  \frac{1}{u_f - u}$, \ie $a(t) \sim \sqrt{\frac{3}{\Lambda}} \; e^{t \; \sqrt{\Lambda/3}}$} and the universe approaches the empty de Sitter space-time.
For $u=u_o$ (now), one finds $\mathbf{a_o =  1.42 \; 10^{11} \; yr }$ and $\mathbf{t_o=1.38\;10^{10} \; yr}$.

Near the Big Bang, we have 
{\small $\sqrt{\Lambda /3} \; a(u) = ({\mathcal P}(u_f) - {\mathcal P}(u_g)) \, u + \tfrac{1}{2} \, ({\mathcal P}^\prime (u_g) - {\mathcal P}^\prime(u_f)) \, u^2 + O(u)^3$}
and 
{\small $\sqrt{\Lambda /3} \; t(u) = \tfrac{1}{2} ({\mathcal P}(u_f) - {\mathcal P}(u_g)) \, u^2 +   \tfrac{1}{6}   ({\mathcal P}^\prime (u_g) - {\mathcal P}^\prime(u_f)) \, u^3 + O(u)^4$}
so,
{\small $a(t) \sim \sqrt{2}Ê\; (\tfrac{\Lambda}{3})^{-1/4} ({\mathcal P}(u_f) - {\mathcal P}(u_g))^{1/2} \; t^{1/2}$.}\\
If radiation is neglected, the first term in the Taylor expansions of $a(u)$ and $t(u)$ vanish, since in that case {\small ${\mathcal P}(u_f) = {\mathcal P}(u_g) = -k/12$}, and we have instead
{\small $a(t) \sim  (\tfrac{\Lambda}{3})^{-1/6}\;  3^{1/3} \;   (\tfrac{3}{2})^{1/3} \;  ({\mathcal P}^\prime (u_g)-{\mathcal P}^\prime (u_f))^{1/3} \; t^{2/3}$.}\\
We recover, in both cases, the known results for the power-law behavior of $a(t)$ which can also be derived directly by keeping only the leading term in the differential equation for the function $a(t)$.

\subsection{Tables}
\label{sec:tables}
The experimental inputs are in boldface. $\Omega_r(u_o)$ is exactly deduced from the measurements of the temperature $\widetilde T[ u_o]$ and the Hubble constant $H_o$.
Remember that we choose for $ \Omega _K{[ u_o]}$ an arbitrary non-zero negative value compatible with the experimental bounds (so $k=+1$).
We recall the notations used for special values of the conformal time $u$: $0$ (Big Bang), $u_o$ (present value), $u_f$ (first zero of $T$,  \ie end of cosmic time: $t\rightarrow \infty$),  $u_g$ (second zero of $T$), $2 \omega_r$ (real period of $T$),  $u_I$ (inflection point for $T$),  $u_k$ (value of $u$ for which $\Omega_K$ as an extremum in the interval $[0, u_f]$). 
The other quantities are either standard in cosmology (their definitions have been recalled in the text), or they are related to modular considerations: the Weierstrass invariants $g_2, g_3$ have been defined in sections \ref{sec:norad},  \ref{sec:withrad}, see also the end of sec.~\ref{sec:notations} for the half-periods $(\omega_1$, $\omega_2)$,
and we refer to the appendix for the other quantities (modular discriminant $\Delta$,  modular parameter $\tau$ and Klein invariant $j$).
%\red why is the modular lambda function (called $m$ here) equal to the opposite of the modular parameter $\tau$ ? \redend

{\small
$$
\begin{array}{cccccc}
 {H_o * cm = }  \bf{{7.44 \,   {10}^{-29}}} & {a_o / cm = }  1.31\, 10^{28} & {t_o / cm = } 1.31\, 10^{28} & {H_o * t_o = } 0.973 &  \\
 { \Lambda *  cm^2 =}  {1.19 \, {10}^{-56}} & { \Lambda_c * cm^2 =} {2.90 \,  {10}^{-62}} & { \lambda   = } 409049. & { \alpha   = } {2.72  \,  {10}^{-6}} &  \\
 { \rho_m[ u_o]*cm^4 =}  7.41\,  10^7 & { \rho_{vac}* cm^4 =} 1.81\,  10^8 & {C_m/cm = } 3.91\,  10^{30} & {q_o = } -0.57 & \\
 \end{array}
$$
%%%%%%
%\begin{alphafootnotes}
\renewcommand{\thefootnote}{\alph{footnote}}
$$
\begin{array}{cccccccccccc}
 { u  [0] = }  0 & { u_k =} 0.272 & { u_o =} 0.330 & { u_I =} 0.441 & { u_f =}  0.444 & & & &\\
  {\widetilde T[0] = } \infty&  {\widetilde T[ u_k] = } 4.626 &  {\widetilde T[ u_o] = } \bf{2.725} &  {\widetilde T[ u_I] = } 0.062 &  {\widetilde T[ u_f] = }  0 & {} & {} & {} & {} \\
 {T[0] = } \infty& {T[ u_k] = } 74.22 & {T[ u_o] = } 42.72 & {T[ u_I] = } 1 & {T[ u_f] = }  0 & {} & {} & {} & {} \\
  & & & & & & & & & & &\\
 \Omega _m{[0] = }  0\footnotemark[1] & \Omega _m{[ u_k] = } 0.675 & \Omega _m{[ u_o] = } \bf{0.293} & \Omega _m{[ u_I] = } {4.89 \,  {10}^{-6}} & \Omega _m{[ u_f] = }  0 & {} & {} & {} & {} \\
 \Omega _{\Lambda }{[0] = }  0 & \Omega _{\Lambda }{[ u_k] = }  0.338 & \Omega _{\Lambda }{[ u_o] = } \bf{0.717} & \Omega _{\Lambda }{[ u_I] = } 1. & \Omega _{\Lambda }{[ u_f] = }  1. & {} & {} & {} & {} \\
 \Omega _K{[0] = }  0 & \Omega _K{[ u_k] = } -0.014 & \Omega _K{[ u_o] = } \bf{-0.01} & \Omega _K{[ u_I] = } -{7.33 \, {10}^{-6}} & \Omega _K{[ u_f] = }  0 & {} & {} & {} & {} \\
  \Omega _r{[0] = }  1\footnotemark[2] & \Omega _r{[ u_k] = } 2.043\, 10^{-4} & \Omega _r{[ u_o] = } \bf{5.21992 \,10^{-5}} & \Omega _r{[ u_I] = } {1.992 \, {10}^{-11}} & \Omega _r{[ u_f] = }  0 & {} & {} & {} & {} \\
   & & & & & & & & & & &\\
 {a[0] = }  0 & {a[ u_k]/cm = } 7.90\,  10^{28} & {a[ u_o]/cm = } 1.34\,  10^{29} & {a[ u_I]/cm = } 5.86\,  10^{30} & {a[ u_f]/cm = } & \infty  & {} & {} & {}  \\
 {t[0] = }  0 & {t[ u_k]/cm = } 7.00\,  10^{27} & {t[ u_o]/cm = } 1.31\,  10^{28} & {t[ u_I]/cm = } 7.22\,  10^{28} & {t[ u_f]/cm = } & \infty  & {} & {} & {} 
  \end{array}
 $$
 \footnotetext[1]{{If one does not include the radiation density  in the Friedmann equations, \ie if one sets $\Omega _r=0$ at all times, one finds $\Omega _m[0]=1$ instead of $0$, in agreement with the fact that, in that case,  $\Omega _m[0]+\Omega _\Lambda[0]+\Omega _K[0]=1.$}}
 \footnotetext[2]{This is in agreement with the fact that $\Omega _m[0]+\Omega _\Lambda[0]+\Omega _K[0]+\Omega _r[0]=1$, see previous footnote.}
%\end{alphafootnotes}
\renewcommand{\thefootnote}{\arabic{footnote}}
%%%%%
 $$
\begin{array}{cccc}
&  g_2{ = }  0.454& g_3{ = } -3787. &   \\
 { \omega_1   = } -0.387\, i & { \omega_2= } 0.336\, + 0.194 \, i &  { \omega_r =}  0.671 & { u_g =}  0.889 \\
 {\tau = } - 0.50\, +0.86 i & {j  =} -{2.412  \,  {10}^{-10}} & { \Delta   = } -3.87 \;  10^8 & {}Ê
\end{array}
% {\ell = } 0.50-0.866 i 
$$
}

\subsection{Comments}

Very quickly after the Big Bang, taking into account the radiation term in  Friedmann equations, \ie using eq. \ref{eq: TasZetafunction} together with eq. \ref{eq:ufugdelc} rather than using only  eq. \ref{eq: TasPfunction}, only changes, in the previous table, the third decimal digit in the values of all quantities of interest, at least in the ``physical branch'' of the history ($u < u_f$). However the influence of this term is very important when $u \rightarrow 0$. See in particular the values of the densities at the point $u=0$. Analytically the curve $T(u)$ also develops a connected negative branch, between $u_f+u_g$ and $2 \omega_r$, but this is physically irrelevant since it occurs long after the end of time $u_f$. If radiation is neglected ($\alpha =0$), the quartic term in Friedmann equations disappears, and $2 \omega_r = u_f+u_g$.

The (negative) curvature density $\Omega_K$ has a minimum for a value $u_k$ in the physical branch, see fig.~\ref{fig:evolutionofcurvature} and the table in sec.~\ref{sec:tables}, showing that $0 < u_k < u_o$, so this happened in our past.
It is incorrect to think or claim that the experimental smallness of $\Omega_K^o$ implies that it was even smaller in the past:  indeed, although $\Omega_K$ goes to $0$ at the Big Bang, it has -- actually it had -- an extremum for $u=u_k$.

If one takes radiation into account, as one should, one finds that the matter density $\Omega_m$ is zero at the Big Bang, and quickly increases --- see fig.~\ref{fig:evolutionofmatter} --- to reach a maximum almost equal to $1$ because the other densities are negligible in this region. Using the same experimental values as before, the maximum is reached for a value of conformal time $u_m \simeq 0.10526$.  After that, $\Omega_m$ stabilizes for a while and starts to slowly decrease ($\Omega_m^o \simeq 0.293$ nowadays) to reach $0$ at the end of time. The sharp increase of $\Omega_m$ at the Big Bang is already encoded in the Friedmann equations (with radiation, of course). Notice that we do not introduce any inflation mechanism in this paper.
Near $u=0_+$, this density behaves as $\tfrac{2 u}{3 \sqrt{\alpha}}$. If one incorrectly forgets the radiative term contribution, the matter density behaves instead as  $1 + \tfrac{u^2}{4}$ when $u\rightarrow 0$; notice that the former expression does not go continuously to the latter when $\alpha \rightarrow 0$. 
%  T  sim 1 / (sqrt(alpha) u) ,  t sim (1/2) sqrt{alpha/Lambdac) u^2

With the given experimental values, one finds $0 < u_m < u_k < u_o < u_I < u_f$. The inflection point $u_I$ of $T(u)$ is therefore still in our future, but it is located very near the end of time $u_f$,  and is hard to see on the figures. 
 
The table gives $u_f=0.444$,  therefore, looking forward, you cannot hope to see the back of your head (assuming $k=1$) since $u_f < 2 \pi$, even if you wait a very long time. However, as already mentioned, this value is quite sensitive to the (arbitrary) chosen input for $\Omega_K^o$, so, there is still hope!

In contradistinction with other quantities (like $g_2, g_3, \lambda, \alpha$, \etc), the numerical values found for $t_o$, the cosmic time, now, and $a_o$, the cosmic scale factor, now, are quite stable with respect to the choice of $\Omega_K^o$. In terms of cosmic time, here are a few important dates in the history of the universe:  $t(0)=0$, $t(u_m)= 4.58\;10^8 \; yr$, $t(u_k)=7.40\;10^9 \; yr$, $t(u_o)=t_o=1.38\;10^{10} \; yr$, $t(u_I)=7.63\;10^{10} \; yr$, $t(u_f)=\infty$.

One should remember that the function $T$, introduced in the first section, differs from the usual temperature $\widetilde T$ by a multiplicative dimensionless factor  $({3 \over 8 \pi G} {\alpha \Lambda_{c} \over 4 \sigma})^{1/4}$, see eq.~\ref{eq:UsualTemperature}.
Using the values of the previous table, this factor is about $6.23 \, 10^{-2}$. It is quite sensitive to the input chosen for $\Omega_K$.

Analytically, the main qualitative feature that distinguishes the case $\alpha \neq 0$ from the  case $\alpha=0$ is that, in the former, the two poles of $T(u)$ are distinct, consequently, each of them is of first order.

Because of the smallness of the $j$-invariant, the modular parameter $\tau$ appears to be numerically very close to $exp(2 i \pi/3)$, but it cannot be equal to this special value since $k=1$ and $\alpha \lambda > 0$ imply $g_2 \neq 0$ (see eq.~\ref{eq: g2g3withradiation}), so $j$, given by eq.~\ref{eq:defKleinInvariantJ}, is strictly non-zero.  We shall return to this in sec.~\ref{flatnoradiation}.

\section{Some specific features of the flat case $k=0$}
 \label{flatcase}
 
 If one assumes $k=0$ from the very beginning, the Friedmann equation simplifies (the quadratic term vanishes), 
  and becomes $ (\tfrac{dT}{du})^2 = \alpha T^4 + \tfrac{2}{3} T^3 + \tfrac{\lambda}{3}$, with, as usual, $\lambda = \tfrac{\Lambda}{\Lambda_c}$, $\sqrt{\Lambda_c} = \tfrac{2}{3 C_m}$ and $\alpha=C_r \Lambda_c$. We have immediately $\tfrac{\Omega_m}{\Omega_\Lambda} = \tfrac{2}{\lambda} T^3$,  $\tfrac{\Omega_r}{\Omega_m} = \tfrac{3}{2\alpha} T$ but this information, taken at conformal time $u_o$, is insufficient to obtain separately $T_o$, $\alpha$ or $\lambda$. In other words,  given the values of $H_o$, ${\widetilde T}_o$, $\Omega_m^o$,  $\Omega_\Lambda^o$, and $\Omega_r^o$ (the sum of the last three being set to $1$), one cannot obtain explicit values for all the cosmological quantities appearing in the table given in section \ref{sec:tables}, for instance one cannot obtain the values of $\lambda$, of $\alpha$, or of the conformal time ``now'' \ie $u_o$: one extra piece of  data, that could be for instance $\Lambda_c$, is missing.
However, as we shall see below, the experimental data singles out a well-defined isomorphism class of elliptic curves.

\subsection{General study}

If we don't neglect radiation, the invariants are given by equations  \ref{eq: g2g3withradiation} where we set $k=0$. So  
\begin{equation}
g_2 = \alpha \, \lambda /3, \quad \text{and} \quad  g_3 = - 2\lambda/6^3
\end{equation}
From equations \ref{eq:TheOmegas}, one immediately obtains:
\begin{equation}
\frac{27}{16} \;   \lambda \, \alpha^3 = \frac {\Omega_r^3 \, \Omega_\Lambda}{\Omega_m^4}
\end{equation}
As $\lambda \, \alpha^3$ does not depend upon the value of the conformal time, it can be evaluated by replacing the densities by their present-day values. Notice that this quantity is positive, although rather small; indeed, using the same experimental values as in the previous section, one finds $\lambda \, \alpha^3 \simeq 8.20 \; 10^{-12}$.

We cannot, in the case $k=0$, obtain separately the values of $\lambda$ or $\alpha$ from the measurement of densities.  
However, since the product $\lambda \, \alpha^3$ is fixed, we can investigate what happens under scaling. 
The value of the cosmological constant $\Lambda$ being determined, changing the parameter $\Lambda_c$ (or, equivalently the ``mass'' of the universe) amounts to change the value of their ratio, $\lambda$.
We therefore set $\alpha^\prime = \alpha \; s^2 $ and $\lambda^\prime = \lambda/s^6$, where $s$ is an arbitrary parameter;  in this way $\lambda^\prime \, {\alpha^\prime}^3 = \lambda \, \alpha^3$ is kept fixed.
From the expressions of the Weierstrass invariants $g_2$ and $g_3$, we find immediately $g_2^\prime = g_2/s^4$ and  $g_3^\prime = g_3/s^6$.
Setting $\Delta = {g_2}^3 - 27 {g_3}^2$, $j = {g_2}^3/\Delta$ and using the same definitions for $\Delta^\prime$ and $j^\prime$ in terms of the prime invariants $g_2^\prime, g_3^\prime$, we notice that
$\Delta^\prime = {g_2^\prime}^3 - 27 {g_3^\prime}^2 =  ({g_2}^3 - 27 {g_3}^2)/s^{12} = \Delta/s^{12}$ so that  $j^\prime=j$.
The quantity $\Delta$ is known as the modular discriminant, and $j$ as the Klein invariant (see the Appendix for details).
In other words, if one assumes $k=0$, the measurement of $\Omega_m^o$,  $\Omega_\Lambda^o$, and $\Omega_r^o$ (the sum of the three being set to $1$), $H_o$ and ${\widetilde T}_o$ does not specify a single lattice in the complex plane, or the corresponding field of elliptic functions, but a class of homothetic lattices. Equivalently, this experimental data does not single out an elliptic curve but an isomorphism class of elliptic curves specified by the $j$-invariant.
If one introduces rescaled conformal time parameters $u^\prime = s\; u$ (with the same rescaling for periods, of course),  the invariants $g_2, g_3$, and the discriminant $\Delta$, are rescaled as previously, in agreement with the general results recalled in the Appendix.

 In terms of cosmological parameters the modular discriminant $\Delta$ is readily evaluated, one finds 
%%%  $\Delta=\tfrac{\lambda^2}{3^2} (\tfrac{\alpha^3 \lambda}{3} - 27)$,  ???
 $\Delta=\tfrac{\lambda^2}{2\times 6^3} (-1 + 16  \lambda \alpha^3)$, 
 so that $\Delta < 0 \Leftrightarrow \lambda \, \alpha^3  < 1/16$, a condition that obviously holds experimentally, since $\lambda \, \alpha^3 \simeq 8.20 \; 10^{-12}$.
 % so that $\Delta \leq 0 \Leftrightarrow \lambda \, \alpha^3 > 108$, a condition that is experimentally ruled out. ???
  $\Delta$ is therefore negative. From the expression of Weierstrass invariants $g_2$, $g_3$, one obtains the Klein invariant  
  % $j = 1728 \,  (1 - \tfrac{81}{\lambda \, \alpha^3})^{-1}$.  ???
  \begin{equation}
  j = 1 + \frac{1}{-1 + 16 \lambda \alpha^3} = 1 + \frac{1}{-1 + \frac{256}{27} \frac{\Omega_\Lambda \, \Omega_r^3}{\Omega_m^4}}
  \label{flatKleinInvariant}
  \end{equation} 
% In terms of densities, it reads: 
%\begin{equation}
%{j}/{1728} = {1}/({1-\frac{2187 \; \Omega_m^4 }{16 \; \Omega_r^3 \, \Omega_\Lambda}})
%\end{equation}
Using the same experimental values as in the previous section, namely 
$\Omega_m^0 = 0.293$, $\Omega_\Lambda^0= 0.717$, and $\Omega_r^0 = 5.20 \, 10^{-5}$ one finds 
%$j \simeq - 6.42 \, 10^{-2}$. ??
$j \simeq - 7.50 \, 10^{-14}$.
\\
A value of $j$ being given, there are several methods to find a modular parameter $\tau$, in the upper half-plane , such that $j(\tau)=j$.  One method to determine this parameter (defined only up to a modular transformation) is recalled in the Appendix, but in our case, the numerical precision on the -- very small -- value of $j$ is insufficient to distinguish $\tau$ from the boundary point $exp(2 i \pi/3)$ where $j$ is exactly $0$, the latter situation occurring when the radiation contribution to Friedmann equation is neglected (next subsection).

\smallskip

The expressions giving the function $T(u)$, the densities, etc, in terms of Weierstrass functions, are the same as in the previous section, but for the fact that the invariants $g_2, g_3$ are slightly simpler since k=0.
The overall features of the curves showing the evolution of cosmological quantities of interest are also similar, but for the fact $\Omega_K$ is now strictly zero, and that there is no inflection point in the physical region for the curve $T(u)$, as it is a priori clear by looking at the behavior of the associated mechanical system for $k=0$, in fig. \ref{potential} and \ref{potential0}.

\subsection{Neglecting radiation}
\label{flatnoradiation}
$k$ being set to $0$, let us assume moreover that we neglect the radiative contribution (\ie we also set $\alpha = 0$).
This is of course only an approximation, as the density $\Omega_r$, as measured today, and although small, is clearly non-zero. Nevertheless, this limit case allows one to obtain several interesting exact results.
  Setting $T = 6\, y$ in the Friedmann equation gives:  $\tfrac{dy}{du}^2 = 4 y^3 + \tfrac{\lambda}{108}$. The solution is immediate: $y$ is the Weierstrass elliptic function, with invariants $g_2=0$ and $g_3= - \lambda/108$.
 $$ T(u) = 6 \, {\mathcal P}(u; g_2=0, g_3=  - \frac{\lambda}{108})$$
The end of  time $u_f$, \ie the first (real) positive zero of $T$ will exist only if $g_3 < 0$, \ie iff $\lambda >0$, otherwise it is complex. Since, experimentally (nowadays), the cosmological constant $\Lambda$ is positive, we have also $\lambda >0$ and $u_f$ exists.
The determination of the zeros of the Weierstrass ${\mathcal P}$ function, in one periodicity cell, is in general a difficult problem --- see the article \cite{EichlerZagier} where the problem is solved in full generality -- but in the present case, one does not need to use these techniques. Indeed,  it is not too difficult, using elliptic integrals, to solve the particular equation ${\mathcal P}(u,0,4)=0$, \ie for $g_2=0,\; g_3=4$. One finds the solution $u= \tfrac{i}{6}\, B(\tfrac{1}{6}, \tfrac{1}{3})$, where $B$ is the Euler beta function. Since, by scaling (see Appendix),  ${\mathcal P}(u; g_2, g_3)=s^2 {\mathcal P}(u; g_2/s^4, g_3/s^6)$, we have ${\mathcal P}(u; 0, g_3)=s^2 {\mathcal P}(u; 0, g_3/s^6)$. If $g_3/s^6=4$, the solution  is known (cf supra), so we take $s=(g_3/4)^{1/6}$, and ${\mathcal P}(u; 0, g_3)$ will vanish for a value $u_f = \tfrac{i}{6} \, B(\tfrac{1}{6}, \tfrac{1}{3}) \, \tfrac{1}{(g_3/4)^{1/6}}$. We summarize the discussion as follows: 

Consider the equation ${\mathcal P}(u;0,g)=0$.  If $g>0$ there is no real solution. If $g < 0$, there are two real zeroes ($u_f$ and $u_g$) in each periodicity cell, $u_f+u_g$ being the smallest real period, and the position of the first is:
$$ 
u_f = \frac{\Gamma(1/6) \, \Gamma(1/3)}{2^{2/3} \; 3 \; (-g)^{1/6} \;Ê\sqrt{\pi}}
$$
Here $\Gamma$ is the Euler Gamma-function. 
In cosmology, with $k=0$ and $\lambda >0$, one has $g_3 = - \lambda/108$. Therefore the end of  time $u_f$ that  solves the equation $T(u_f) = 6 \, {\mathcal P}(u_f;0,g_3)=0$ is given by :
\begin{equation} 
u_f = \frac{ 2^{2/3} \; \sqrt{3} \; \Gamma(1/3) \, \Gamma(7/6)}{\lambda^{1/6}Ê\; \sqrt{\pi}}
\end{equation}
In the theory of elliptic functions, the case $g_2=0$ is often referred to as the {\sl equianharmonic case}. This name\footnote{which means ``equally anharmonic'' since it is related to a geometrical configuration where the anharmonic ratio (cross ratio) is a cube root or a sixth root of unity.} already appears in Abramowitz and Stegun \cite{AbramowitzStegun}, where it refers to the case $g_2=0$, $g_3=1$; the case where $g_2=0$ and $g_3$ is an arbitrary complex number can be obtained from the latter by scaling and it is therefore justified to keep the same terminology. 
For $g_2=0$, $g_3=1$, the equianharmonic half-periods are known \cite{AbramowitzStegun, NIST}: $(\tfrac{\Gamma(1/3)^3}{4\pi},  e^{i \pi/3} \; \tfrac{\Gamma(1/3)^3}{4\pi})$. The quantity called $\omega_r$ in this paper (half the smallest real period for the temperature, expressed as a function of conformal time) is then immediately obtained by scaling, one gets:
\begin{equation} 
 \omega_r = \frac{3 \; {\Gamma(1/3)^3}}{2\; 2^{2/3}\; \lambda^{1/6} \; \pi}
\end{equation}
This universe is therefore very special, not only geometrically, since we have set $k=0$, but also because of its dynamics. Indeed, when the first Weierstrass invariant vanishes ($g_2=0$),  the Klein $j$-invariant vanishes and
the modular parameter (ratio of the periods)  of the associated elliptic curve (or of the corresponding lattice, see our discussion in the Appendix) can be chosen at the corner\footnote{Choosing $e^{2 i \pi/3}$ or $e^{i \pi/3}$ is irrelevant: both values are equivalent under the modular transformation $\tau \mapsto \tau+1$.} ($e^{2 i \pi/3}$) of the boundary of the fundamental domain for the full modular group. 
The discussion carried out in the previous subsection holds: when $k=0$ the available experimental data does not single out a specific elliptic curve but an isomorphism class of elliptic curves specified by the j-invariant. If moreover one takes $\alpha = 0$, like in this section,  the $j$-invariant is zero (equianharmonic case).

\section{Miscellaneous (slightly eccentric) comments}

\paragraph{The elliptic curve associated with a given cosmology: is the universe special ?}
The flat case ($k=0$) with no radiation pressure ($\alpha = 0 \Leftrightarrow \Omega_r(u)= 0$) is clearly special since the associated modular parameter sits in a corner, $exp(2 i \pi/3)$, of the fundamental domain (equianharmonic case).
Whether or not $k=0$ is a debatable issue: it seems to be experimentally clear, nowadays, that $\Omega_K^o$ is ``small'', but $\Omega_K$ is a time-dependent function and one should not set it to zero at all times unless one can argue, on some theoretical grounds, that the constant $k$ should be $0$, rather than $1$ or $-1$ (see the comments in sec.~\ref{experimentalconstraints}). 
Assuming $k=0$ is certainly convenient from an experimental point of view and may be to the liking of partisans of the inflation mechanism (although widely accepted, one should remember that strong criticisms have been voiced against inflation). On the other hand, considering that space is flat, or asymptotically flat, and that it is filled in first approximation with an ideal fluid whose density goes to zero at infinity, may be perceived by some people 
%(including such people as the authors of \cite{MTW}) 
as a retrograde idea going against the evolution of the scientific concepts that were at the roots of General Relativity and Friedmann equations.
Any kind of hypothesis made on $k$ will also trigger epistemological or philosophical debates, for instance:  should one believe that the universe ``exists'' by  itself, independently of our more or less elaborated fits? \\
The no-radiation pressure hypothesis is of different nature, as it is clearly only an approximation: although small,  $\Omega_r^o$ is non-zero and it is well-measured. Whether or not $k=0$, one comment that one can make is the following: for the Weierstrass ${\mathcal P}$ function to have real values ---the CMB temperature is real--- on the real axis (conformal time), one needs the associated lattice to be identical with its conjugate, equivalently $g_2$ and $g_3$ should be real; this is automatic (see eqs.~\ref{eq: g2g3withradiation}) in the framework of Friedmann equations, then the $j$-invariant is real as well, so the modular parameter $\tau$, taken in the standard fundamental domain, should belong either to the boundary of the domain, or to the half-axis $i \, y$, with $y > 1$, since these geodesic arcs are precisely those subsets where $j$ takes real values. It seems unreasonable to extend this kind of discussion further... Elliptic curves can be ``special'' for a variety of arithmetical reasons, but it is clear that human experiments, in cosmology,  do not give exact results in the sense of arithmetics! 

\begin{figure} 
 \begin{center}
 \includegraphics*[scale=0.5]{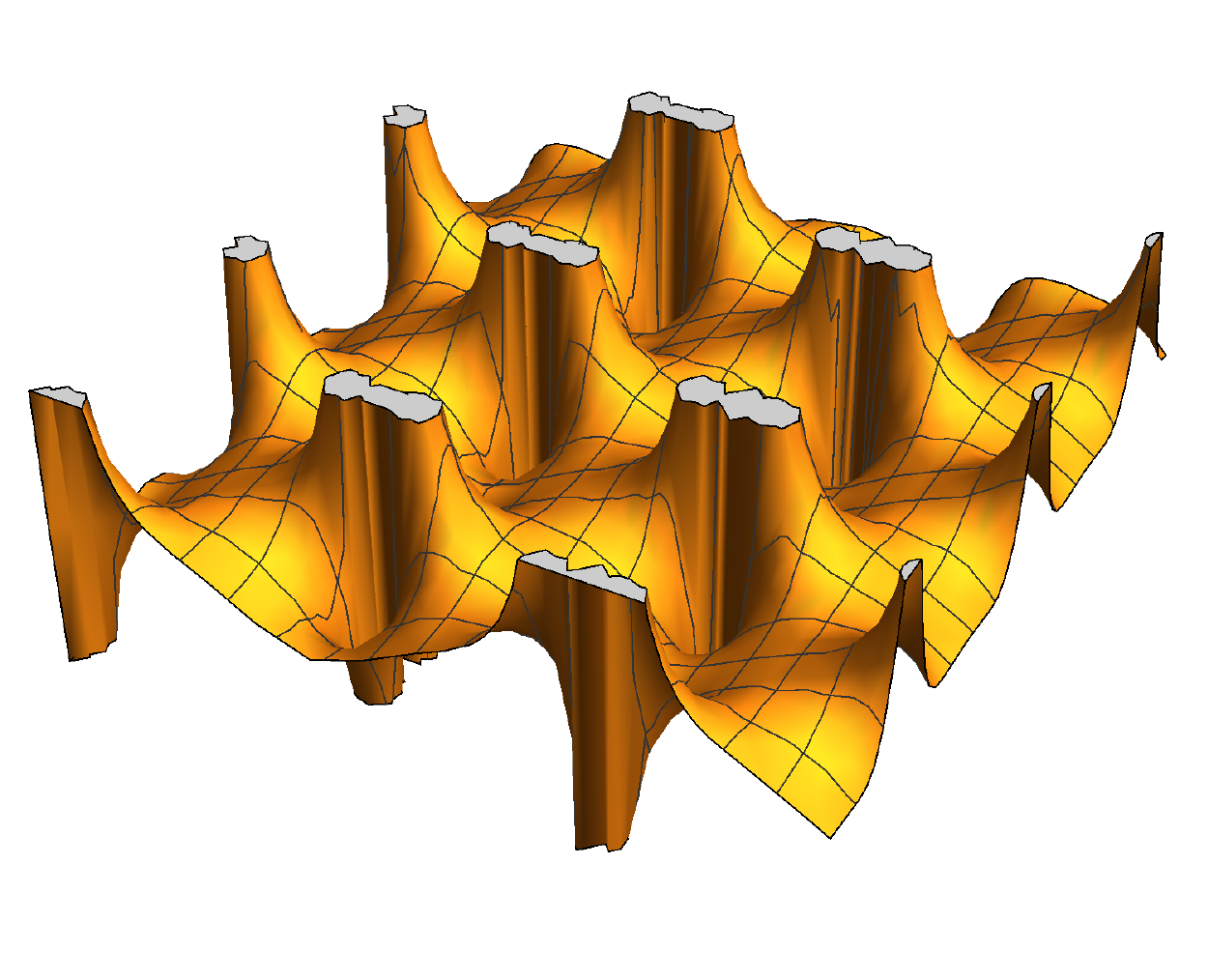} 
  \hspace{1.2cm}
  \includegraphics*[scale=0.5]{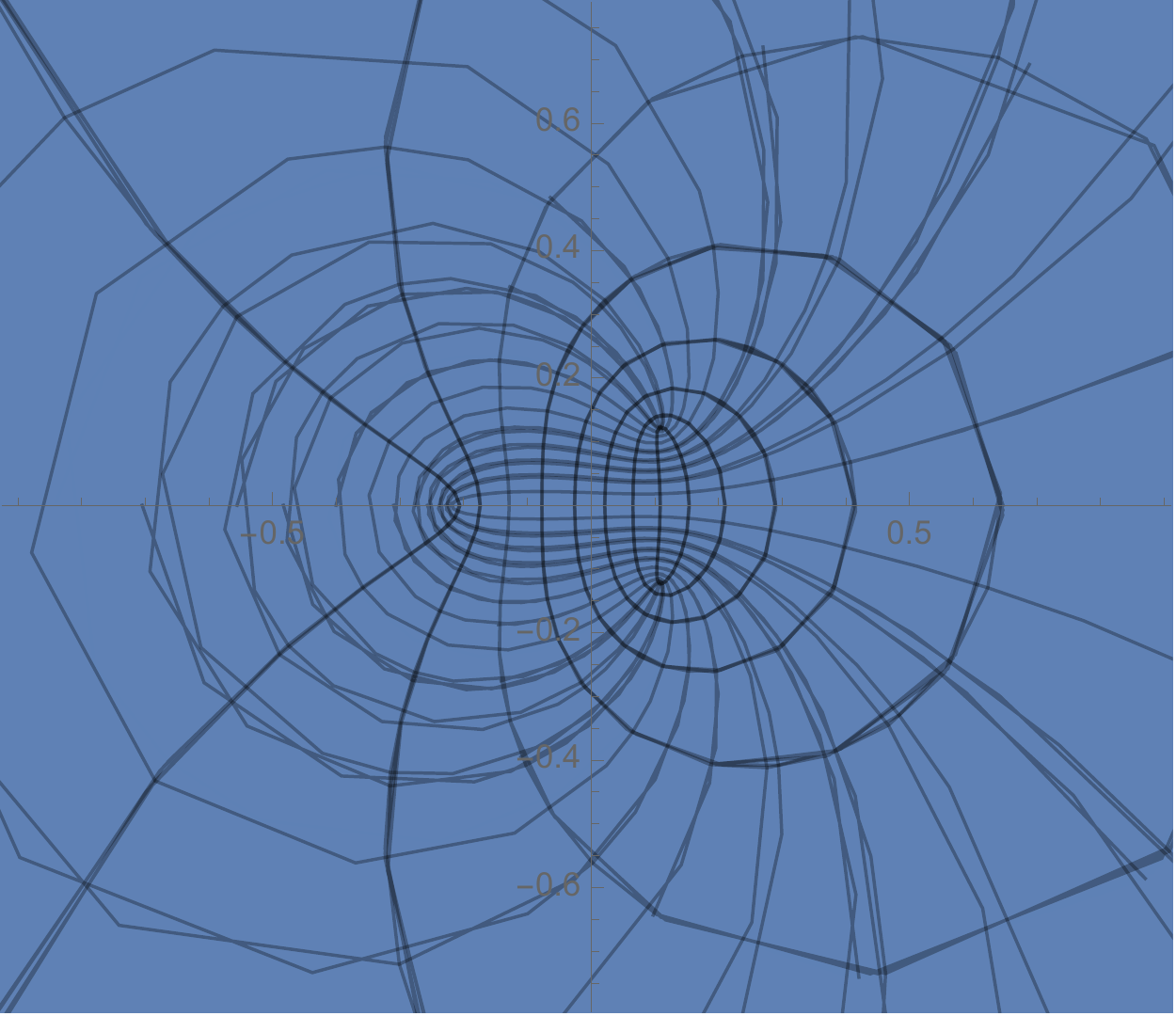}
\end{center}
\caption{The temperature $T$ as a function of the complex conformal time:
{\small 
(i)  The real part of $T(u)$ for an argument running in a domain containing several complex periods,
(ii) Plot of the parametric region defined by $X=Re(T(u))$, $Y=Re(T(u))$, with $u = x+i y$  running in the same set as in (i).}}
\label{fig:complexWeierstrass} 
\end{figure}

\paragraph{Aeons.}
The solutions of Friedmann equations, written as differential equations with respect to conformal time $u$, and almost all quantities of cosmological interest related to these solutions, are periodic functions of $u$, and are actually bi-periodic (elliptic functions) when the variable $u$ is extended to the complex plane. They automatically describe a denumerable infinity of identical cosmologies (depending upon the choice of the parameters, only one interval within each real period may describe a  possible physical history). For a wide range of parameters --- including the values that are experimentally measured ---  the temperature function, an elliptic function of $u$, goes to zero when the conformal time reaches a finite value $u_f$, corresponding, in terms of cosmic time,  to $t \rightarrow \infty$.  This happens, of course, in each period. In other words, Friedmann equations, from the very beginning, describe an infinite sequence of identical FLRW spacetimes, each one lasting an infinite number of years,  in terms of cosmic time. This observation (that was explicitly made in ref.~\cite{CoqueGros}) may inspire science-fiction writers and may trigger deep philosophical or metaphysical thoughts about the notion of reality or eternity, but it does not convey much physical meaning: a periodic function is a function on a circle, a bi-periodic function is a function on a torus, and each period of $T(u)$ describe the ``same'' universe. The fact of perceiving those universes as distinct does not mean much$\ldots$ unless, of course, one can introduce a way to make them (slightly) different, and a way to communicate ---\ie to send information--- from one period to the next.  In a slightly different framework, such a connection was proposed recently in \cite{Penrose:aeons} and popularized under the denomination ``conformal cyclic cosmology'', the name ``aeon'' was then chosen to denote each member of the infinite sequence of spacetimes. Along that same line of thought, and taking into account the fact that all the functions of interest are bi-periodic, one could go one step further and talk about ``complex aeons'' when considering the doubly periodic lattice of identical (or almost identical) spacetimes obtained by allowing the conformal time to be complex: see fig.~\ref{fig:complexWeierstrass}.
We did not try, in the present article, to suggest any perturbation to the background provided by homogeneous and isotropic cosmologies, or to suggest a mechanism that could connect the different aeons that are rooted in the structure of Friedmann-Lema\^{i}tre differential equations.

\section*{Appendix: a short memo on elliptic functions and elliptic curves}

\subsubsection*{Elliptic functions.}
\label{ellipticthings}
\small{
Given two independent vectors of the plane, \ie equivalently two non-zero complex numbers $\omega_1$ and $\omega_2$ such that $\omega_2/\omega_1$ is not real, one can build the lattice $L$ generated by $2 \omega_1$ and $2 \omega_2$ as the infinite set $\{m \, 2\omega_1 + n \,  2\omega_2 \}$,  where $m,n$ are arbitrary integers.
The two complex numbers $2 \omega_1$ and $2 \omega_1$ are called\footnote{About half the planet calls ($\omega_1, \omega_2$) what we call ($2\omega_1, 2\omega_2$). } the {\sl periods} of the lattice, and $\{\omega_1, \omega_2\}$ are the {\sl half-periods}.
An elliptic function with respect to $L$ is a non-constant meromorphic function of the complex variable $u$ that is periodic with respect to the given lattice, \ie doubly periodic. 

The Weierstrass elliptic function\footnote{Not to be confused with the pathological Weierstrass function that is continuous everywhere and differentiable nowhere.} ${\mathcal P}$ associated with the lattice $L$ is defined as 
  \begin{equation} 
{\mathcal P}(u) = \frac{1}{u^2} + \sum_{\omega \in L - \{ 0 \}} ( \frac{1}{(u-\omega)^2} - \frac{1}{\omega^2})
\label{defofP}
  \end{equation} 
From this definition it is clear that ${\mathcal P}$ is periodic with respect to $L$ (\ie doubly periodic), that it is even, that it has a pole of order two at the origin (and at all other vertices of $L$), that it is meromorphic, and that
  \begin{equation} 
{\mathcal P}^\prime(u)^2 = 4 {\mathcal P}(u)^3 - g_2 {\mathcal P}(u)  - g3 
\label{eqdifforP}
  \end{equation} 
where $g_2= 60 \sum_{\omega \in L - \{ 0 \}} \frac{1}{\omega^4}$, and $g_3=140 \sum_{\omega \in L - \{ 0 \}} \frac{1}{\omega^6}$. These two (complex in general) numbers are called the Weierstrass invariants of the lattice $L$.
Near the origin,  ${\mathcal P}(u)  \sim \dfrac{1}{u^2} + \frac{g_2}{20} u^2 + \frac{g_3}{28} u^4 + \ldots$ 

The Weierstrass elliptic function ${\mathcal P}$ is a function of the complex variable $u$, but it depends on the chosen lattice $L$. In order to make this dependence explicit one writes  ${\mathcal P}(u; L)$ although
it is quite standard to denote the same function by ${\mathcal P}(u; g_2, g_3)$ or ${\mathcal P}(u\vert \omega_1, \omega_2)$. }

\subsubsection*{Modular considerations}
\small{
A lattice $L$ can be specified by a pair of half-periods $(\omega_1, \omega_2)$, but given any $2 \times 2$ matrix with integer coefficients $a,b,c,d$ and determinant $1$ (\ie an element of the infinite discrete group $SL(2,\ZZ)$), one obtains a new pair of periods for the same lattice $L$ by setting $
 \begin{pmatrix}\omega_1^\prime \\ \omega_2^\prime \end{pmatrix} = \begin{pmatrix}a&b \\ c&d \end{pmatrix} .   \begin{pmatrix}\omega_1 \\ \omega_2 \end{pmatrix}$.
 The invariants $g_2$, $g_3$ only depend on the lattice: the invariants calculated from any two pairs of periods of the same lattice are the same.
  Conversely, given $g_2$ and $g_3$, one can determine a pair of periods for the lattice $L$, but the previous comment shows that this pair is not uniquely determined.
 
 Rather than choosing different pairs of periods for the same lattice, one can also rescale the lattice itself by an arbitrary (real or complex) non-zero number $s$, and in particular replace the pair of half-periods $(\omega_1, \omega_2)$ by $(s\,\omega_1, s\,\omega_2)$.  One gets another lattice, homothetic --by definition-- with the first.
  The lattice is different, its Weierstrass invariants are different (one finds $g_2(s\, \omega_1, s \, \omega_2)=s^{-4}\, g_2(\, \omega_1,\omega_2)$ and $g_3(s\, \omega_1, s \, \omega_2)=s^{-6}\, g_3(\, \omega_1,\omega_2)$) and the corresponding ${\mathcal P}$-function is also different, but the former is simply related to the latter: one has the homogeneity relation: ${\mathcal P}(s \, u \vert s\,\omega_1, s\,\omega_2) = s^{-2}\;  {\mathcal P}(u\vert \omega_1, \omega_2)$.
  In particular, choosing $s = 1/\omega_1$, one obtains:
  \begin{equation} 
{\mathcal P}(u\vert \omega_1, \omega_2) = \dfrac{1}{\omega_1^{2}} \; {\mathcal P}( \dfrac{u}{\omega_1} \vert 1 , \tau = \dfrac{\omega_2}{\omega_1})
\label{eq: homogeneity} 
\end{equation} 
Using an homothety with scale factor ${1}/{\omega_1}$,  one can replace the basis $\{\omega_1, \omega_2\}$ by $\{1, \tau\}$ with $\tau = \omega_2/\omega_1$.
 It does not cost anything to assume that $\tau = \omega_2/\omega_1$ belongs to the upper half-plane (if it does not, just permute the two periods).
 
The value of $\tau= \omega_2/\omega_1$, called the modular parameter, clearly does not change if one only rescales the periods; however if one uses another set of periods $\{\omega^{\prime}_1, \omega^{\prime}_2\}$ for the same lattice (rescaled or not), one will have in general $\tau^\prime \neq \tau$.  As the periods are only defined up to the action of $SL(2,\ZZ)$, the same is true for $\tau$. In other words, one can find four integers $a,b,c,d$, with $ad-bc = 1$, such that $\tau^\prime = (a \tau + b)/(c\tau +d)$.
Notice that $\tau^\prime$ also belongs to the upper half-plane.
Conversely, one proves that if $\tau$ and $\tau^\prime$ are related in this way, they are associated with the same lattice or with two homothetic lattices.\\
The $j$-invariant, or Klein invariant\footnote{It is convenient to set $J(\tau)=1728 \, j(\tau)$ but it is not uncommon to see $J$ called $j$ (or vice-versa) in the literature.}, of a lattice is defined as the complex number 
  \begin{equation} 
  j(L) =   \frac{g_2(L)^3}{\Delta} \quad \text{where} \quad \Delta={g_2(L)^3 - 27 g_3(L)^2}
  \label{eq:defKleinInvariantJ}
    \end{equation} 
From the homogeneity transformations of $g_2$ and $g_3$ one sees that $j(L)$ is invariant under lattice rescaling. Actually, a detailed analysis leads to the following result: 
Given $L$ and $L^\prime$, two lattices in $\CC$, then $j(L) = j(L^\prime)$ if and only if $L$ and $L^\prime$ are homothetic. 
As a function of $\tau$,  the meromorphic function $j$ is invariant with respect to the action of the modular group, and one has the following famous expansion (Fourier series), usually written in terms of $q=e^{2 i \pi \tau}$, where all the coefficients are positive integers:
  \begin{equation} 1728 \; j(\tau)= 1/q +744+196884 \, q+21493760 \, q^2 + \ldots \end{equation} 
} 
An arbitrary complex number $j$ being given, there are several ways to invert the Klein function, \ie to find a value of $\tau$ (up to a modular transformation) solving the equation $j(\tau) = j$. 
This can be done for instance as follows: solve the cubic equation $256 (1-x)^3/x^2 = 1728\, j$ in $x$, then write $x=\ell(1-\ell)$ and solve this quadratic equation in $\ell$. One obtains finally $\tau = i \, \tfrac{{}_2F_1(1/2,1/2,1,1-\ell)}{{}_2F_1(1/2,1/2,1,\ell)}$ where ${}_2F_1$ is the Gauss hypergeometric function. The first step of the method amounts to solve a sextic equation for $\ell$ and there are a priori six possible choices\footnote{The modular function $\ell$, as a function of $\tau$ is usually called $\lambda(\tau)$ in the literature, but we changed the notation for $\ell (\tau)$,  as $\lambda$ already refers to the reduced cosmological constant.} for the solution (six values for a cross-ratio), but $\tau$, up to a modular transformation, will not depend upon this choice.

\begin{figure} 
\begin{center}
 \includegraphics*[scale=0.60]{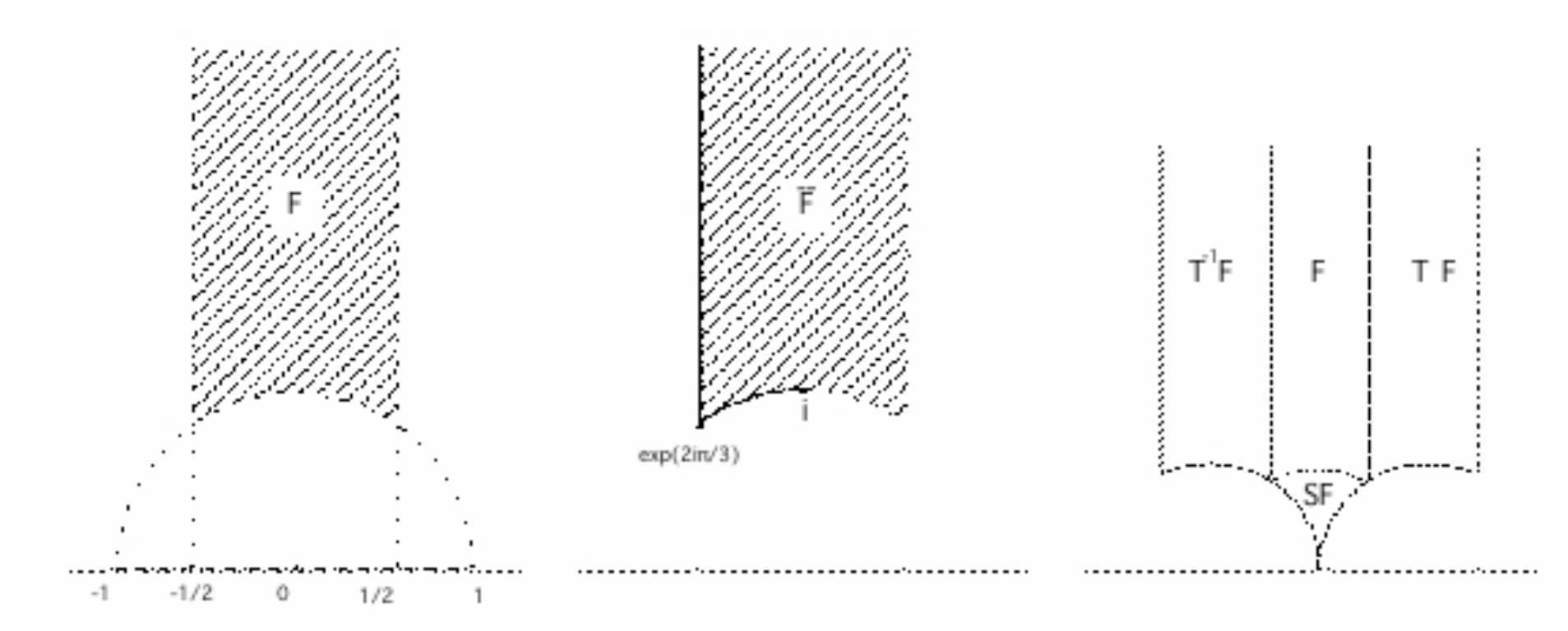}  
\end{center}
\caption{Standard fundamental domain for the modular group, and its neighbors. Each point of the domain specifies a lattice in the complex plane, up to homothety. 
The boundary point $e^{2 i \pi/3}$ describes a flat universe without radiation (equianharmonic elliptic curve).}
\label{fig:fundamentaldomain} 
\end{figure}

\subsubsection*{Elliptic curves}
\small{
Choose a lattice $L$. If we set $x={\mathcal P}(z)$ and  $y={\mathcal P}^\prime(z)$, we have  $y^2 = 4 x^3 - g_2  x - g_3$.
Using homogenous coordinates $x=X/U$, $y=Y/U$, one gets the equation of a cubic in the complex projective plane: $Y^2 U =   4 X^3 - g_2 X U^2 - g_3 U^3$.
The homothetic lattice $s L$ determines an isomorphic cubic (use the homogeneity relations for $g_2, g_3, {\mathcal P}, {\mathcal P}^\prime$, and set  $x = \tilde X/  U$, $y=\tilde Y/  U$, with $\tilde X = X/s^3$ and $\tilde Y = Y/s^3$).

An elliptic curve is a projective variety isomorphic to a non-singular curve of degree $3$ in the complex projective plane, together with a distinguished point. Every such cubic can be brought to the Weierstrass form (\ie to the previous form).
This (complex) curve is parametrized by setting $x={\mathcal P}(z)$ and  $y={\mathcal P}^\prime(z)$, for some lattice specified by $g_2$ and $g_3$.

A function (like ${\mathcal P}$) defined on the complex plane $\CC$ and periodic w.r.t. a lattice $L$ is, by definition, a function on $\CC/L$. The latter manifold is, topologically, and by construction, a torus (it is obtained by identifying the opposite sides of a period parallelogram).  All 2-tori are diffeomorphic as real manifolds, but a torus constructed as previously is also a complex manifold because the choice of a lattice specifies a complex structure (two proportional lattices determine the same complex structure).  Up to isomorphism, an elliptic curve can therefore also be defined as a 2-torus endowed with a complex structure.}

\subsubsection*{Elliptic functions (continuation).}
\small{
Taking into account equation (\ref{eqdifforP}), any elliptic function, for a given lattice, can be written as a rational function of ${\mathcal P}$ and its first derivative ${\mathcal P}^\prime$.
In particular, any rational function of an elliptic function $f$ is 
also elliptic (with respect to the same lattice) but its order 
will coincide with the order of $f$ only if the transformation is 
fractional linear, like the transformations (\ref{eq: TasPfunction}) or (\ref{eq:transformationT}). 
It is known, since Liouville, that if $a$ is an arbitrary 
complex number (including infinity), the number of solutions of the 
equation $f(u)=a$, called the {\sl order} of $f$,  is independent of $a$, if multiplicities are 
properly counted.
The order of an elliptic functions is 
at least two and the Weierstrass ${\cal P}$ function 
of a  lattice can be defined as {\sl the} elliptic 
function of order $2$ that has a pole of order $2$ at the origin and is such that 
$1/u^{2} - {\cal P}(u)$ vanishes at $u = 0$.

The Weierstrass $\zeta$ function is the primitive of ${-\mathcal P}$ which is such that $\zeta(u)-1/u$ vanishes at the origin. The odd meromorphic
function $\zeta$ has a pole of first order at all the vertices of the lattice defined by the invariants $(g_2,g_3)$. 
Warning: $\zeta$ is not periodic (hence not elliptic !) but it is quasi-periodic: if $a\neq b$ then $\zeta(u-a)-\zeta(u-b)$ is elliptic of order $2$ with poles at $u=a$ and $u=b$.

 The Weierstrass $\sigma$ function is defined as an entire function that vanishes at $u=0$ and whose logarithmic derivative is $\zeta(u)$.
 It is not elliptic, but if $a_1, a_2, b_1, b_2$ are complex numbers such that $a_1+a_2=b_1+b_2$, then ${\sigma(u-a_1)\sigma(u-a_2)}/{\sigma(u-b_1)\sigma(u-b_2)}$ is elliptic of order $2$ with poles at $b_1,b_2$ and zeroes at $a_1, a_2$. 
}

\section*{Acknowledgments}

This text is based, in parts, on talks or lectures given by the author in Gordon Mac Kay Laboratory (Harvard Univ., 1981), CERN (Geneva, 2000), CBPF (Rio de Janeiro, 2002), LUTH (Meudon, 2013), CPT (Marseille, 2013) and IAFE (Univ. Buenos Aires, 2014).
Hospitality of these research centers is greatly acknowledged.
\vfill \eject

 \end{document}